\documentclass[11pt]{article}
\usepackage{latexsym,epsfig,amssymb,amsmath,amsfonts}
\newcommand{\dotle} {\mbox{$\:\stackrel{\centerdot}{\le}\:$}}
\newcommand{\dotge} {\mbox{$\:\stackrel{\centerdot}{\ge}\:$}}
\newcommand{\one}{\mathbf{1}}
\newcommand{\calA}{{\cal A}}
\newcommand{\calB}{{\cal B}}
\newcommand{\calC}{{\cal C}}

\newcommand{\calE}{{\cal E}}

\newcommand{\calM}{{\cal M}}
\newcommand{\calP}{{\cal P}}
\newcommand{\calQ}{{\cal Q}}
\newcommand{\calR}{{\cal R}}
\newcommand{\calS}{{\cal S}}
\newcommand{\calT}{{\cal T}}
\newcommand{\calU}{{\cal U}}

\newcommand{\calX}{{\cal X}}
\newcommand{\calY}{{\cal Y}}

\newcommand{\balpha}{{\boldsymbol \alpha}}
\newcommand{\btheta}{{\boldsymbol \theta}}
\newcommand{\bk}{{\mathbf k}}

\newcommand{\bs}{{\mathbf s}}
\newcommand{\bS}{{\mathbf S}}
\newcommand{\bt}{{\mathbf t}}
\newcommand{\bu}{{\mathbf u}}
\newcommand{\bU}{{\mathbf U}}
\newcommand{\bv}{{\mathbf v}}

\newcommand{\bw}{{\mathbf w}}

\newcommand{\bx}{{\mathbf x}}
\newcommand{\bX}{{\mathbf X}}
\newcommand{\by}{{\mathbf y}}
\newcommand{\bY}{{\mathbf Y}}
\newcommand{\bz}{{\mathbf z}}

\newcommand{\tp}{\tilde{p}}

\newcommand{\tbs}{\tilde{\mathbf s}}

\newcommand{\tbu}{\tilde{\mathbf u}}

\newcommand{\tbx}{\tilde{\mathbf x}}

\newcommand{\tby}{\tilde{\mathbf y}}
\newcommand{\tbY}{\tilde{\mathbf Y}}
\newcommand{\pYXSa}{p_{\bY|\bX\bS^a}}

\def\eE{{\mathbb E}}
\def\iI{{\mathbb I}}

\def\rR{{\mathbb R}}
\def\uU{{\mathbb U}}

\def\argmax{\mathrm{argmax}}

\def\CDMC{\mathrm{CDMC}}
\def\CAM{\mathrm{CAM}}
\newtheorem{definition}{Definition}[section]
\newtheorem{remark}{Remark}[section]

\newtheorem{theorem}{Theorem}[section]
\newtheorem{lemma}[theorem]{Lemma}
\newtheorem{proposition}[theorem]{Proposition}
\newtheorem{corollary}[theorem]{Corollary}
\renewcommand{\theequation}{\arabic{section}.\arabic{equation}}
\newcommand{\Section}[1]{\section{#1}
\setcounter{equation}{0}}
\begin{document}
\title{Capacity and Random-Coding Exponents \\
    for Channel Coding with Side Information}
\author{Pierre Moulin and Ying Wang \\
    University of Illinois at Urbana-Champaign \\
    Beckman Institute, Coordinated Science Laboratory, \\
    and Dept of Electrical and Computer Engineering, \\
    Urbana, IL 61801, USA \\
    {\tt \{moulin,ywang11\}@ifp.uiuc.edu}
    \thanks{This research was supported by NSF under ITR grants
    CCR 00-81268 and CCR 03-25924.}
    }
\date{September 30, 2004. Revised August 25, 2006 and December 19, 2006}
\maketitle

\begin{abstract}
Capacity formulas and random-coding exponents are derived for a
generalized family of Gel'fand-Pinsker coding problems. These
exponents yield asymptotic upper bounds on the achievable log
probability of error. In our model, information is to be reliably
transmitted through a noisy channel with finite input and output
alphabets and random state sequence, and the channel is selected
by a hypothetical adversary. Partial information about the state
sequence is available to the encoder, adversary, and decoder. The
design of the transmitter is subject to a cost constraint. Two
families of channels are considered: 1) compound discrete
memoryless channels (CDMC), and 2) channels with arbitrary memory,
subject to an additive cost constraint, or more generally to a
hard constraint on the conditional type of the channel output
given the input. Both problems are closely connected. The
random-coding exponent is achieved using a stacked binning scheme
and a maximum {\em penalized} mutual information decoder, which
may be thought of as an empirical generalized Maximum a Posteriori
decoder. For channels with arbitrary memory, the random-coding
exponents are larger than their CDMC counterparts. Applications of
this study include watermarking, data hiding, communication in
presence of partially known interferers, and problems such as
broadcast channels, all of which involve the fundamental idea of
binning.
\end{abstract}

{\bf Index terms}: channel coding with side information, error exponents,
arbitrarily varying channels, universal coding and decoding,
randomized codes, MAP decoding, random binning, capacity, reliability function,
method of types, watermarking, data hiding, broadcast channels.

\newpage

\Section{Introduction}
In 1980, Gel'fand and Pinsker studied the problem of coding
for a discrete memoryless channel (DMC) $p(y|x,s)$ with random states
$S$ that are observed by the encoder but not by the decoder \cite{Gel80}.
They derived the capacity of this channel and showed it is achievable
by a random binning scheme and a joint-typicality decoder.
Applications of their work include computer memories with defects
\cite{Hee83}, writing on dirty paper, and communication in presence of
a known interference \cite{Cos83,Ere01,Ere04,Liu06}.
Duality with source coding problems with side information
was explored in \cite{Cov02,Bar03,Pra03}.
In the late 1990's, it was discovered that the problems of
embedding and hiding information in cover signals are closely related
to the Gel'fand-Pinsker problem: the cover signal plays
the role of the state sequence in the Gel'fand-Pinsker problem
\cite{Wil00,Che01,Mou03}.
Capacity expressions were derived under expected distortion constraints for
the transmitter and a memoryless adversary \cite{Mou03}.
One difference between the basic Gel'fand-Pinsker problem and
the various formulations of data-hiding and watermarking problems
resides in the amount of side information available to the encoder,
channel designer (adversary), and decoder.
A unified framework for studying such problems is considered
in this paper. The encoder, adversary and decoder have access
to degraded versions $\bs^e, \bs^a, \bs^d$, respectively,
of a state sequence $\bs$. Capacity is obtained as the solution
to a mutual-information game:
\[ C = \sup_{p_{XU|S^e}} \min_{p_{Y|XS^a}} [I(U;YS^d) - I(U;S^e)] , \]
where $U$ is an auxiliary random variable, and the sup and min
are subject to appropriate constraints.

In problems such as data hiding, the assumption of a fixed channel
is untenable when the channel is under partial control of an
adversary. This motivated the game-theoretic approach of
\cite{Mou03}, where the worst channel in a class of memoryless
channels was derived, and capacity is the solution to a maxmin
mutual-information game. This game-theoretic approach was recently
extended by Cohen and Lapidoth \cite{Coh02} and Somekh-Baruch and
Merhav \cite{Som03,Som04}, who considered a class of channels with
arbitrary memory, subject to almost-sure distortion constraints.
In the special case of {\em private data hiding}, in which the
cover signal is known to both the encoder and the decoder,
Somekh-Baruch and Merhav also derived random-coding and
sphere-packing exponents \cite{Som03}. Binning is not needed in
this scenario. The channel model of \cite{Coh02,Som03,Som04} is
different from but reminiscent of the classical memoryless
arbitrary varying channel (AVC) \cite{Csi81,Csi88,Lap98} which is
often used to analyze jamming problems. In the classical AVC
model, no side information is available to the encoder or decoder.
Error exponents for this problem were derived by Ericson
\cite{Eri85} and Hughes and Thomas \cite{Hug96}. The capacity of
the AVC with side information at the encoder was derived by
Ahlswede \cite{Ahl86}.

The coding problems considered in this paper are motivated by data
hiding applications in which the decoder has partial\footnote{For
instance, the decoder may have access to a noisy, compressed
version of the original cover signal.} or no knowledge of the
cover signal. In all cases capacity is achievable by
random-binning schemes. Roughly speaking, the encoder designs a
codebook for the auxiliary $U$. The selected sequence $\bU$ plays
the role of input to a fictitious channel and conveys information
about both the encoder's state sequence $\bS^e$ and the message
$M$ to the decoder. Finding the best error exponents for such
schemes is challenging. Initial attempts in this direction for the
Gel'fand-Pinsker DMC have been reported by Haroutunian {\em et
al.} \cite{Har88,Har91}, but errors were discovered later
\cite{Har00,Har01}. Very recently, random-coding exponents have
been independently obtained by Haroutunian and Tonoyan
\cite{Har04} and Somekh-Baruch and Merhav \cite{Som04b}. Their
results and ours \cite{Mou04} were presented at the 2004 ISIT
conference.

The random-coding exponents we have derived cannot be achieved by
standard binning schemes and standard maximum mutual information
(MMI) decoders \cite{Csi81,Lap98}. Instead we use a {\em stack} of
variable-size codeword-arrays indexed by the type of the encoder's
state sequence $\bS^e$. The appropriate decoder is a maximum
penalized mutual information (MPMI) decoder, where the penalty is
a function of the encoder's state sequence type. The MPMI decoder
may be thought of as an empirical generalized MAP decoder, just
like the conventional MMI decoder may be thought of as an
empirical MAP decoder.

This paper is organized as follows. A statement of the problem is
given in Sec.~\ref{sec:statement}, together with basic
definitions. Our main results are stated in Sec.~\ref{sec:main} in
the form of four theorems. An application to binary alphabets
under Hamming cost constraints for the transmitter and adversary
is given in Sec.~\ref{sec:binary}. Proofs of the theorems appear
in Secs.~\ref{sec:thm1}---\ref{sec:thm4}. All derivations are
based on the method of types \cite{Csi98}. The paper concludes
with a discussion in Sec.~\ref{sec:discussion} and appendices.
\subsection{Notation}
We use uppercase letters for random variables,
lowercase letters for individual values, and boldface fonts for sequences.
The p.m.f. of a random variable $X \in \calX$ is denoted by
$p_X = \{ p_X(x), \, x \in \calX \}$, and the probability
of a set $\Omega$ under $p_X$ is denoted by $P_X(\Omega)$.
Entropy of a random variable $X$ is denoted by $H(X)$,
and mutual information between two random variables $X$ and $Y$
is denoted by $I(X;Y) = H(X) - H(X|Y)$, or by $\tilde{I}_{XY}(p_{XY})$
when the dependency on $p_{XY}$ should be explicit; similarly we sometimes
use the notation $\tilde{I}_{XY|Z}(p_{XYZ})$.
The Kullback-Leibler divergence between two p.m.f.'s $p$ and $q$
is denoted by $D(p||q)$. We denote by
$D(p_{Y|X} || q_{Y|X} | p_X) = D(p_{Y|X} p_X || q_{Y|X} p_X)$
the conditional Kullback-Leibler divergence of $p_{Y|X}$ and
$q_{Y|X}$ with respect to $p_X$. The base-2 logarithm of $x$ is
denoted by $\log x$, and the natural logarithm is denoted by $\ln
x$.

Following the notation in Csisz\'{a}r and K\"{o}rner \cite{Csi81},
let $p_\bx$ denote the type of a sequence $\bx \in \calX^N$
($p_\bx$ is an empirical p.m.f. over $\calX$) and $T_\bx$ the type class
associated with $p_\bx$, i.e., the set of all sequences of type $p_\bx$.
Likewise, we define the joint type $p_{\bx\by}$ of a pair of sequences
$(\bx, \by) \in \calX^N \times \calY^N$
(a p.m.f. over $\calX \times \calY$) and $T_{\bx\by}$ the type class
associated with $p_{\bx\by}$, i.e., the set of all sequences
of type $p_{\bx\by}$.
We define the conditional type $p_{\by|\bx}$ of a pair of sequences
($\bx, \by$) as $\frac{p_{\bx\by}(x,y)}{p_{\bx}(x)}$ for all $x \in \calX$
such that $p_{\bx}(x) > 0$.
The conditional type class $T_{\by|\bx}$ is the set of
all sequences $\tby$ such that $(\bx, \tby) \in T_{\bx\by}$.
We denote by $H(\bx)$ the entropy of the p.m.f. $p_{\bx}$
and by $I(\bx;\by)$ the mutual information for the joint p.m.f. $p_{\bx\by}$.
Recall that
\begin{equation}
   (N+1)^{- |\calX|} \,2^{NH(\bx)} \le |T_{\bx}| \le 2^{NH(\bx)}
\label{eq:type-size1}
\end{equation}
and
\begin{equation}
   (N+1)^{- |\calX| \,|\calY|} \,2^{NH(\by|\bx)}
    \le |T_{\by|\bx}| \le 2^{NH(\by|\bx)} .
\label{eq:type-size2}
\end{equation}

We let $\calP_X$ and $\calP_X^{[N]}$ represent the set of all
p.m.f.'s and empirical p.m.f.'s, respectively, for a random variable $X$.
Likewise, $\calP_{Y|X}$ and $\calP_{Y|X}^{[N]}$ denote
the set of all conditional p.m.f.'s and all empirical conditional p.m.f.'s,
respectively, for a random variable $Y$ given $X$.
The notations $f(N) \ll g(N)$, $f(N) = O(g(N))$, and $f(N) \gg g(N)$ indicate that
$\lim_{N \rightarrow \infty} \left| \frac{f(N)}{g(N)} \right|$ is zero,
finite but nonzero, and infinite, respectively.
The shorthands $f(N) \doteq g(N)$, $f(N) \dotle g(N)$ and $f(N) \dotge g(N)$
denote equality and inequality on the exponential scale:
$\lim_{N \rightarrow \infty} \frac{1}{N} \ln \frac{f(N)}{g(N)} = 0$,
$\lim_{N \rightarrow \infty} \frac{1}{N} \ln \frac{f(N)}{g(N)} \le 0$, and
$\lim_{N \rightarrow \infty} \frac{1}{N} \ln \frac{f(N)}{g(N)} \ge 0$,
respectively.
We let $\iI\{ x \in \Omega \}$ denote the indicator function
of a set $\Omega$,
and $\uU(\Omega)$ denote the uniform p.m.f. over a finite set $\Omega$.
We define $|t|^+ \triangleq \max(0,t)$, $\exp_2(t) \triangleq 2^t$, and
$\overline{h}(t) \triangleq -t \log t - (1-t) \log (1-t)$
(the binary entropy function). We adopt the notational convention
that the minimum of a function over an empty set is $+ \infty$.

\Section{Statement of the Problem}
\label{sec:statement}

\begin{figure}[h]
\centering
\setlength{\epsfxsize}{6in} \epsffile{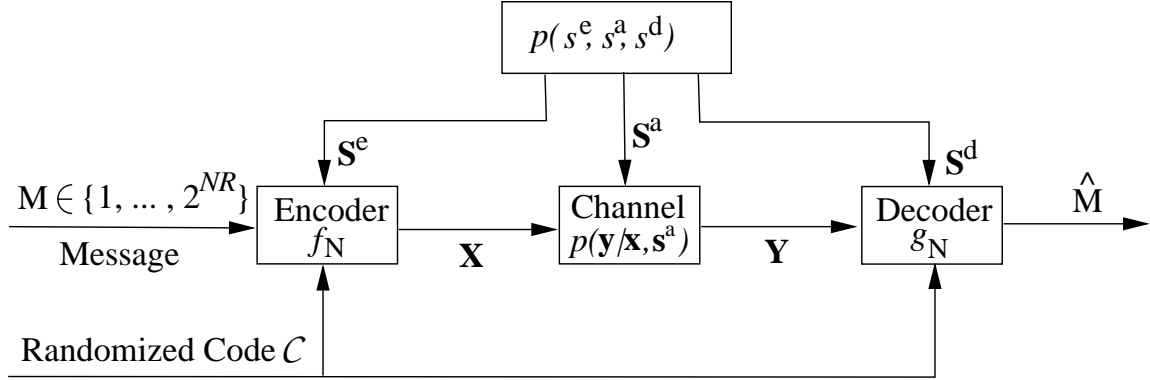}
\caption{Communication with side information at the encoder and decoder.
    Cost constraints are imposed on the encoder and channel.}
\label{fig:unified}
\end{figure}

Our generic problem of communication with side information at the encoder
and decoder is diagrammed in Fig.~\ref{fig:unified}.
There three versions $\bS^e$, $\bS^a$, and $\bS^d$ of a state sequence
are available to the encoder, adversary and decoder, respectively.
We use the short hand $\bS$ to denote the joint state sequence
$(\bS^e, \bS^a, \bS^d)$. This sequence consists of independent and identically
distributed (i.i.d.) samples drawn from a p.m.f. $p(s^e, s^a, s^d)$.
The individual sequences $\bS^e$, $\bS^a$, $\bS^d$ are available noncausally
to the encoder, adversary and decoder, respectively.
The adversary's channel is of the form $p_{\bY|\bX\bS^a}(\by|\bx,\bs^a)$.
This includes the problems listed in Table~1 
as special cases.
The alphabets $\calS^e$, $\calS^a$, $\calS^d$, $\calX$ and $\calY$ are finite.

\begin{table}[h]
\centering
\begin{tabular}{|l|ccc|c|} \hline
Problem & $\bS^a$ & $\bS^d$ & Binning? \\ \hline
Gel'fand-Pinsker \cite{Gel80} & $\bS^e$ & $\emptyset$ & yes \\
Public Watermarking \cite{Mou03,Som04} & $\emptyset$ & $\emptyset$ & yes \\
Semiblind Watermarking \cite{Mou03} & $\emptyset$ & $\bS^d \ne \bS^e$ & yes \\
Cover-Chiang \cite{Cov02} & $(\bS^e, \bS^d)$ & $\bS^d$ & yes \\ \hline
Private Watermarking \cite{Mou03,Som03} & $\emptyset$ & $\bS^e$ & no \\ \hline
\end{tabular}
\label{tab:problems}
\caption{Relation between $\bS^e$, $\bS^a$, and $\bS^d$
    for various coding problems with side information.}
\end{table}

A message $M$ is to be transmitted to a decoder; $M$ is uniformly
distributed over the message set $\calM$. The transmitter produces
a sequence $\bX = f_N(\bS^e, M)$. The adversary passes $\bX$
through the channel $\pYXSa(\by|\bx,\bs^a)$ to produce corrupted
data $\bY$. The decoder does not know $\pYXSa$ selected by the
adversary and has access to $\bS^d$. The decoder produces an
estimate $\hat{M} = g_N(\bY,\bS^d) \in \calM$ of the transmitted
message.\footnote{
   At first sight the problem setup could be simplified by eliminating
   the variable $\bS^a$ and considering the ``average channel''
   $p_{\bY|\bX\bS^e}(\by|\bx,\bs^e) = \sum_{\bs^a} p_{\bY|\bX\bS^a}(\by|\bx,\bs^a)
   p_{S^a|S^e}^N(\bs^a|\bs^e)$.
   We do not follow this approach because $p_{S^a|S^e}$ is fixed and
   $p_{\bY|\bX\bS^a}$ is optimized by the adversary; hence these p.m.f.'s
   appear separately in the problem formulation and its solution.
   A similar comment applies to $\bS^d$.
}

We allow the encoder/decoder pair $(f_N, g_N)$
to be {\em randomized}, i.e., the choice of $(f_N, g_N)$
is a function of a random variable known to the encoder and
decoder but not to the adversary. This random variable is independent
of all other random variables and plays the role of a secret key.
The randomized code will be denoted by $(F_N, G_N)$.

To summarize, the random variables $M$, $F_N$, $G_N$, $\bS^e$,
$\bS^a$, $\bS^d$, $\bX$ and $\bY$ have joint p.m.f.
\[ p_M(m) p_{F_N G_N}(f_N,g_N) \left[
    \prod_{i=1}^N p_{S^eS^aS^d}(s_i^e, s_i^a, s_i^d)
    \right] \iI\{ \bx = f_N(\bs^e,m)\} \pYXSa(\by|\bx,\bs^a) .
\]
\subsection{Constrained Side-Information Codes}

A cost function $\Gamma ~:~\calS^e \times \calX \to \rR^+$
is defined to quantify the cost $\Gamma(s^e,x)$ of transmitting symbol
$x$ when the channel state at the encoder is $s^e$.
This definition is extended to $N$-vectors using
$\Gamma^N(\bs^e,\bx)=\frac{1}{N}\sum_{i=1}^N \Gamma(s_i^e,x_i)$.
In information embedding applications, $\Gamma$ is a distortion function
measuring the distortion between host signal and marked signal.

We now define a class of codes satisfying maximum-cost constraints
(Def.~\ref{def:code-as}) and a class of codes satisfying average-cost
constraints (Def.~\ref{def:code-avg}). The latter class is of course
larger than the former. We also define a class of randomly-modulated (RM)
codes (Def.~\ref{def:code-RM}), adopting terminology from \cite{Hug96}.

Def.~\ref{def:code-avg} is analogous to the definition of a
length-$N$ information hiding code in \cite{Mou03}. The common
source of randomness between encoder and decoder appears via the
distribution $p_{F_N G_N}(f_N,g_N)$ whereas in \cite{Mou03} it
appears via a cryptographic key sequence $\bk$ with finite entropy
rate.
\begin{definition}
A length-$N$, rate-$R$, randomized code with side information and
{\bf maximum cost} $D_1$ is a triple $(\calM, F_N, G_N)$, where
\begin{itemize}
\item $\calM$ is the message set of cardinality
    $|\calM| = \lceil 2^{NR} \rceil$;
\item $(F_N, G_N)$ has joint distribution $p_{F_N G_N}(f_N,g_N)$;
\item $f_N~:~(\calS^e)^N \times \calM \to \calX^N$
    is the encoder mapping the state sequence $\bs^e$ and message $m$
    to the transmitted sequence $\bx=f_N(\bs^e,m)$.
    The mapping is subject to the cost constraint
    \begin{equation}
       \Gamma^N(\bs^e,f_N(\bs^e,m)) \le D_1 \quad
        \mbox{almost~surely}~(p_{S^e} p_{F_N} p_M) ;
    \label{eq:D1-as}
    \end{equation}
    \item $g_N~:~\calY^N \times (\calS^d)^N \to \calM \cup \{ e \}$
    is the decoder mapping the received sequence $\by$ and channel state
    sequence $\bs^d$ to a decoded message $\hat m = g_N(\by,\bs^d)$.
    The decision $\hat{m} = e$ is a declaration of error.
    \end{itemize}
\label{def:code-as}
\end{definition}
\begin{definition}
A length-$N$, rate-$R$, randomized code with side information and
{\bf expected cost} $D_1$ is a triple $(\calM, F_N, G_N)$
which satisfies the same conditions as in Def.~\ref{def:code-as},
except that (\ref{eq:D1-as}) is replaced with the weaker constraint
\begin{equation}
   \sum_{\bs^e} p_{S^e}^N(\bs^e) \sum_{f_N} p_{F_N}(f_N) \sum_{m \in \calM}
   \frac{1}{|\calM|} \Gamma^N(\bs^e,f_N(\bs^e,m)) \le D_1 .
\label{eq:D1-avg}
\end{equation}
\label{def:code-avg}
\end{definition}
\begin{definition}
A {\bf randomly modulated} (RM) code with side information is a randomized code
defined via permutations of a prototype $(f_N,g_N)$. Such codes are of the form
\begin{eqnarray*}
   \bx = f_N^{\pi}(\bs^e,m) & \triangleq & \pi^{-1} f_N (\pi \bs^e, m) \\
     g_N^{\pi}(\by,\bs^d) & \triangleq & g_N(\pi \by, \pi \bs^d)
\end{eqnarray*}
where $\pi$ is chosen uniformly from the set of all $N!$ permutations
and is not revealed to the adversary.
The sequence $\pi \bx$ is obtained by applying $\pi$ to the elements of $\bx$.
\label{def:code-RM}
\end{definition}

\subsection{Constrained Attack Channels}

Next we define a class $\calA$ of DMC's (Def.~\ref{def:CDMC})
and a corresponding class $\calP_{\bY|\bX\bS^a}[\calA]$
of channels with arbitrary memory (CAM) in which the conditional type
of $\by$ given $(\bx,\bs^a)$ is constrained (Def.~\ref{def:CAM}).
\begin{definition}
A compound DMC (CDMC) class $\calA$ is any compact (under $L_1$ norm)
subset of $\calP_{Y|XS^a}$.
\label{def:CDMC}
\end{definition}
For CDMC's, we have
$\pYXSa(\by|\bx,\bs^a) = \prod_{i=1}^N p_{Y|XS^a}(y_i|x_i,s_i^a)$,
where $p_{Y|XS^a} \in \calA$.
The set $\calA$ is defined according to the application.
\begin{enumerate}
\item In the case of a known channel \cite{Gel80}, $\calA$ is a singleton.
\item In information hiding problems \cite{Mou03}, $\calA$ is the class
    of DMC's that introduce expected distortion between $X$ and $Y$
    at most equal to $D_2$:
    \begin{equation}
       \sum_{s_a,x,y} p_{XS^a}(x,s^a) p_{Y|XS^a}(y|x,s^a) d(x,y) \le D_2 ,
    \label{eq:D2-avg}
    \end{equation}
    where $d~:~\calX \times \calY \to \rR^+$ is a distortion function.
    $\calA$ can also be defined to be a subset of the above class.
\item In some applications, $\calA$ could be defined via multiple
    cost constraints.
\end{enumerate}
Given a p.m.f. $p_{XUS^e}$, we denote by $\calP_{YS^aS^d|XUS^e}[\calA, p_{XUS^e}]$
the class of DMC's $p_{YS^aS^d|XUS^e}$ whose conditional marginal $p_{Y|XS^a}$
is in the CDMC class $\calA$.
\begin{definition}
The CAM class $\calP_{\bY|\bX\bS^a}[\calA]$ is the set of channels such that
for any channel input $(\bx,\bs^a)$ and output $\by$,
the conditional type $p_{\by|\bx\bs^a}$ belongs to
$\calA \bigcap \calP_{Y|XS^a}^{[N]}$ with probability 1:
\begin{equation}
   Pr[p_{\by|\bx\bs^a} \in \calA] = 1.
\label{eq:CAM}
\end{equation}
\label{def:CAM}
\end{definition}
If $\calA$ is defined via the distortion constraint (\ref{eq:D2-avg}),
let $d^N(\bx,\by) = \frac{1}{N} \sum_{i=1}^N d(x_i,y_i)$.
Condition (\ref{eq:CAM}) may then be rewritten as
\begin{equation}
   Pr[ d^N(\bx,\by) \le D_2] = 1 ,
\label{eq:D2-as}
\end{equation}
i.e., feasible channels have total distortion bounded by $N D_2$
and arbitrary memory.\footnote{ The case of channels with
arbitrary memory subject to expected-distortion constraints admits
a trivial solution: the adversary ``obliterates'' $\bX$ with a
fixed, nonzero probability that depends on $D_2$ but not on $N$,
and therefore no reliable communication is possible in the sense
of Def.~\ref{def:ach} below. } Comparing the CDMC class $\calA$
and the CAM class $\calP_{\bY|\bX\bS^a}[\calA]$, we see that 1)
for $(\bX,\bS^a,\bY)$ in any given type class, the conditional
p.m.f. of $\bY$ given $(\bX,\bS^a)$ is uniform in the CDMC case
but not necessarily so in the CAM case, and 2) while conditional
types $p_{\by|\bx\bs^a} \notin \calA$ may have exponentially
vanishing probability under the CDMC model, such types are
prohibited in the CAM case. One may expect that both factors have
an effect on capacity and random-coding exponents. As we shall
see, only the latter factor does have an effect on random-coding
exponents.

The relation between the CAM class $\calP_{\bY|\bX\bS^a}[\calA]$
in (\ref{eq:D2-as}) and the classical AVC model \cite{Csi81} is
detailed in Appendix~\ref{sec:CAM}. The class (\ref{eq:D2-as}) is
not a special case of the classical AVC model because arbitrary
memory is allowed.

We also introduce the following class of attack channels, which turn out
to be the worst CAM channels for the problems considered in this paper.
\begin{definition}
An attack channel $p_{\bY|\bX\bS^a}$
{\bf uniform over single conditional types}
is defined via a mapping $\Lambda~:~ \calP_{XS^a}^{[N]} \to \calP_{Y|XS^a}^{[N]}$
such that with probability 1, the channel output
$\by$ has conditional type $p_{\by|\bx\bs^a} = \Lambda(p_{\bx\bs^a})$.
Moreover, $\bY$ is uniformly distributed over the corresponding
conditional type class.
\label{def:attack-CC}
\end{definition}

Lastly, given a type $p_{\bx\bu\bs^e}$, we denote by
$\calP_{YS^aS^d|XUS^e}^{[N]}[\calA, p_{\bx\bu\bs^e}]$
the class of conditional types $p_{\by\bs^a\bs^d|\bx\bu\bs^e}$
such that $p_{\by|\bx\bs^a}$ is in the CAM class $\calP_{\bY|\bX\bS^a}[\calA]$.

\subsection{Probability of Error}

The average probability of error for a deterministic code
$(f_N, g_N)$ when channel $p_{\bY|\bX\bS^a}$ is in effect
is given by
\begin{eqnarray}
   \lefteqn{P_e(f_N,g_N,p_{\bY|\bX\bS^a})}      \nonumber \\
    & = & \mbox{Pr}(\hat M \not = M)            \nonumber \\
    & = & \frac{1}{|\calM|} \sum_m \sum_{\bs^e,\bs^a,\bx}
        \,\sum_{(\by,\bs^d) \notin g_N^{-1}(m)}
        p_{\bY|\bX\bS^a}(\by|\bx,\bs^a)
            \,\iI\{ \bx=f_N(\bs^e,m)\}
        p_{S^eS^aS^d}^N(\bs^e,\bs^a,\bs^d) .
\label{eq:Pe-det-code}
\end{eqnarray}
For a randomized code the expression above is averaged
with respect to $p_{F_N G_N}(f_N,g_N)$; this average is denoted by
$P_e(F_N,G_N,p_{\bY|\bX\bS^a})$.
The minmax probability of error for the class of randomized codes
and the class of attack channels considered is given by
\begin{equation}
   P_{e,N}^* = \min_{p_{F_N G_N}} \max_{p_{\bY|\bX\bS^a}}
    \sum_{f_N,g_N} p_{F_N G_N}(f_N,g_N) P_e(f_N, g_N, p_{\bY|\bX\bS^a}) .
\label{eq:Pe-minmax}
\end{equation}
\bigskip
\begin{definition}
A rate $R$ is said to be achievable if $P_{e,N}^* \to 0$ as $N \to \infty$.
\label{def:ach}
\end{definition}
\begin{definition}
The capacity $C(D_1, \calA)$ is the supremum of all achievable rates.
\end{definition}
\begin{definition}
The reliability function of the class of attack channels considered is
\begin{equation}
   E(R) = \liminf_{N \to \infty} \left[ - \frac{1}{N} \log P_{e,N}^* \right] .
\label{eq:E}
\end{equation}
\end{definition}

There are four combinations of maximum/expected cost constraints
for the transmitter and CDMC/CAM designs for the adversary
(four flavors of the generalized Gel'fand-Pinsker problem), and
a question is whether same capacity and error exponents will be obtained
in all four cases. We now define {\em transmit channels}, which
play a crucial role in deriving capacity and error-exponents.
\begin{definition}
Given alphabets $\calX$, $\calU$ and $\calS^e$, a {\bf transmit channel}
$p_{XU|S^e}$ is a conditional p.m.f. that satisfies the following
distortion constraint on the conditional marginal $p_{X|S^e}$:
\[ \sum_{u,s^e,x} p_{XU|S^e}(x,u|s^e) p_{S^e}(s^e) \Gamma(s^e,x) \le D_1 . \]
\end{definition}

Given an alphabet $\calU$ of cardinality $L$,
we denote by $\calP_{XU|S^e}(L,D_1)$ the set of feasible transmit channels.
Note that transmit channels have been termed covert channels \cite{Mou03}
and watermarking channels \cite{Som03,Som04} in the context of information
hiding. In those papers, the channel $p_{Y|X}$ was termed
attack channel; we retain this terminology for $p_{Y|XS^a}$ in this paper.

\subsection{Preliminaries}
\label{sec:preliminaries}

Consider a sextuple of random variables $(S^e,S^a,S^d,U,X,Y)$ with joint p.m.f.
$p_{S^eS^aS^dUXY}$, where $U$ is an auxiliary random variable taking values
in $\calU \triangleq \{ 1, 2, \cdots, L \}$.
The following difference of mutual informations plays a
fundamental role in capacity analysis \cite{Gel80}---\cite{Som04}
of channels with side information. It plays a central role in the
analysis of error exponents as well:
\begin{equation}
   J_L(p_{S^eS^aS^dUXY}) \triangleq I(U;YS^d) - I(U;S^e) .
\label{eq:J}
\end{equation}
Note that $J_L$ depends on $p_{S^eS^aS^dUXY}$ only via the marginal
$p_{US^eS^dY}$; moreover, the cardinality $L$ of the alphabet $\calU$
has been made explicit in the definition (\ref{eq:J}).
%

Channel capacity for the problems studied in \cite{Gel80}---\cite{Som04}
is given by
\begin{equation}
    C(D_1, \calA) = \lim_{L \to \infty} \,\max_{p_{XU|S^e}} \min_{p_{Y|XS^a}}
    J_L(p_{S^eS^aS^d} \,p_{XU|S^e} \,p_{Y|XS^a})
\label{eq:C0}
\end{equation}
where restrictions are imposed on the joint distribution
of $(S^e,S^a,S^d)$ (including the absence of some of these variables,
see Table~1),   
and the maximization over $p_{XU|S^e}$ and minimization over $p_{Y|XS^a}$
are possibly subject to cost constraints.

The cardinality of the alphabet $\calU$ may be unbounded \cite[p.~514]{Som04}
\footnote{
   The capacity formula in \cite[Corollary~1, p.~514]{Som04} was obtained
   under the restriction of constant composition codes.}
, hence the infinite range for $L$ in (\ref{eq:C0}).
To evaluate (\ref{eq:C0}) in the case $S^a=\emptyset$,
Moulin and O'Sullivan \cite{Mou03} claimed that one
can choose $L = |\calS^e| \, |\calX| + 1$ without loss of optimality.
The proof is based on Caratheodory's theorem, as suggested in \cite{Gel80}.
However the proof in \cite{Mou03} applies only to the fixed-channel case
\footnote{Equation (A7) in the proof of \cite[Prop.~4.1(iv)]{Mou03}
is associated with a fixed DMC.}.

The use of alphabets with unbounded cardinality introduces some
technical subtleties. The following two lemmas are straightforward
but will be useful. The proof of the first one is based on the nested nature
of the feasible sets $\calP_{XU|S^e}(L,D_1), \, 1 \le L < \infty$.
\begin{lemma}
Let $\calU = \{ 1, 2, \cdots, L \}$ and $\psi_L$ a functional defined over
$\calP_{XU|S^e}(L,D_1)$. Then
\begin{equation}
   \psi_L^* \triangleq \max_{p_{XU|S^e} \in \calP_{XU|S^e}(L,D_1)}
    \psi_L(p_{XU|S^e})
\label{eq:psiL-star}
\end{equation}
is a nondecreasing function of $L$.
\label{lem:monotonic}
\end{lemma}
{\em Proof}. We need to prove that $\psi_L^* \le \psi_{L+1}^*$ for any $L$.
Let $p_{XU|S^e}^*$ achieve the maximum defining $\psi_L^*$ and
define the extended p.m.f. $p_{XU|S^e}^{ex}$ over $\{ 1, 2, \cdots, L+1 \}$
as follows:
\[ p_{XU|S^e}^{ex}(x,u|s^e) =  \left\{ \begin{array}{ll}
    p_{XU|S^e}^*(x,u|s^e) & u=1,2,\cdots,L \\ 0 & u=L+1
    \end{array} \right. \quad \forall x, s^e .
\]
Since $p_{XU|S^e}^{ex}$ and $p_{XU|S^e}^*$ have the same conditional marginal
$p_{X|S^e}$, we have $p_{XU|S^e}^{ex} \in \calP_{XU|S^e}(L+1,D_1)$, and
\[ \psi_{L+1}(p_{XU|S^e}^{ex}) = \psi_L(p_{XU|S^e}^*) . \]
Therefore
\[ \psi_L^* = \psi_{L+1}(p_{XU|S^e}^{ex}) \le \psi_{L+1}^* . \]
\hfill $\Box$
\begin{lemma}
    Given three compact sets $\calP$, $\calQ$, $\calR$ and
    a functional $\phi ~:~ \calP \times \calQ \times \calR \to \rR$,
    let $(p^*,q^*,r^*)$ achieve the min max min in
    \begin{equation}
       \min_{p \in \calP} \,\max_{q \in \calQ} \,\min_{r \in \calR} \phi(p,q,r) .
    \label{eq:minmaxmin}
    \end{equation}
    It is assumed that $\phi$ is continuous in an $L_1$ neighborhood of
    $(p^*,q^*,r^*)$. Then, given three sequences of subsets
    $\calP_n, \calQ_n, \calR_n, \,1 \le n < \infty$ respectively
    dense in $\calP$, $\calQ$ and $\calR$ under the $L_1$ norm,
    we have the following property:
    \begin{eqnarray}
       \min_{p \in \calP} \,\max_{q \in \calQ} \,\min_{r \in \calR} \phi(p,q,r)
        & = & \lim_{n \to \infty} \,\min_{p \in \calP_n}
            \,\max_{q \in \calQ_n} \,\min_{r \in \calR_n} \phi(p,q,r) .
                            \label{eq:l1-property}
    \end{eqnarray}
\label{lem:compact}
\end{lemma}
{\em Proof}: Denote the left side of (\ref{eq:l1-property}) by $\phi^{***}$
    and the argument of the limit in the right side by $A_n$. We have
    $\underline{A}_n \le A_n \le \overline{A}_n$ where
    \begin{eqnarray*}
       \underline{A}_n & = & \min_{p \in \calP}
            \max_{q \in \calQ_n} \min_{r \in \calR} \phi(p,q,r) , \\
       \overline{A}_n & = & \min_{p \in \calP_n} \max_{q \in \calQ}
            \min_{r \in \calR_n} \phi(p,q,r) .
    \end{eqnarray*}
    Since the maximization (resp. minimizations) defining $\underline{A}_n$
    (resp. $\overline{A}_n$) is over a dense subset of $\calQ$
    (resp. $\calP \times \calR$), we have
    \[ \lim_{n \to \infty} \underline{A}_n = \lim_{n \to \infty} \overline{A}_n
        = \phi^{***} .
    \]
    Hence $\lim_{n \to \infty} A_n = \phi^{***}$.
    \hfill $\Box$

Finally, recall that the Kullback-Leibler divergence \cite{Csi81} and related
functionals (including mutual information $\tilde{I}(p_{XY}) = D(p_{XY}||p_X p_Y)$
and $J_L$ functionals) are continuous with respect to $L_1$ norm.
For instance, for any $L$, any p.m.f.'s $p$ and $p'$ with finite values
of $J_L$, and any $\epsilon > 0$, there exists $\delta$ such that
\[ \|p-p'\| < \delta
    \quad \Rightarrow \quad |J_L(p)-J_L(p')| < \epsilon ,
\]
where the norm on $p-p'$ is the $L_1$ norm.

\Section{Main Results}
\label{sec:main}

The main tool used to prove the coding theorems in this paper is
the method of types \cite{Csi98}. Our random-coding schemes are
binning schemes in which the auxiliary random variable $U$ is
input to a fictitious channel.

In all derivations, optimal types for sextuples
$(\bs^e,\bs^a,\bs^d,\bu,\bx,\by)$ are obtained as solutions to maxmin problems.
Two key facts used to prove the theorems are:
1) the number of conditional types is polynomial in $N$, and
2) in the CAM case, the worst attacks are uniform over conditional types,
as in Somekh-Baruch and Merhav's watermarking capacity game \cite{Som04}.
Proof of the theorems appears in Secs.~\ref{sec:thm1}---\ref{sec:thm4}.
Related, known results for CDMC's without side information
are summarized in Appendix~\ref{sec:DMC}.

The expression (\ref{eq:C0}), restated below in a slightly
different form, turns out to be a capacity expression for the
problems considered in this paper (Theorems~\ref{thm:C-CDMC} and
\ref{thm:C-CAM}):
\begin{equation}
   C = C(D_1, \calA) = \lim_{L \to \infty} C_L
\label{eq:C}
\end{equation}
where
\begin{equation}
   C_L \triangleq \max_{p_{XU|S^e} \in \calP_{XU|S^e}(L,D_1)}
    \min_{p_{Y|XS^a} \in \calA} J_L(p_{S^eS^aS^d} p_{XU|S^e} p_{Y|XS^a}) .
\label{eq:CL}
\end{equation}
By application of Lemma~\ref{lem:monotonic}, the sequence $C_L$
is nondecreasing in $L$.

In the special case of degenerate $p_{S^eS^d}$ (no side
information at the encoder and decoder), it is known that the
maximum above is achieved by $U=X$, and capacity reduces to the
standard formula $C = \max_{p_X} \min_{p_{Y|XS^a}}
\tilde{I}_{XY}(p_X \,p_{Y|XS^a} \,p_{S^a})$. If $S^e = S^d = S$
and $S^a = \emptyset$ (private watermarking), the optimal choice
is again $U=X$, and $C = \max_{p_{X|S}} \min_{p_{Y|X}}
\tilde{I}_{XY|S}(p_{X|S} \,p_{Y|X} \,p_S)$.

\subsection{Random-Coding Exponents for CDMC Model}
\label{sec:RandomCoding}

\begin{lemma}
The function
\begin{eqnarray}
   E_{r,L}^{\CDMC}(R) & \triangleq & \min_{\tp_{S^e} \in \calP_{S^e}}
        \,\max_{p_{XU|S^e} \in \calP_{XU|S^e}(L,D_1)}
        \,\min_{\tp_{YS^aS^d|XUS^e} \in \calP_{YS^aS^d|XUS^e}}
        \,\min_{p_{Y|XS^a} \in \calA} \nonumber \\
    & & \mbox{\hspace*{0.25in}}
        \left[ D(\tp_{S^e} \,p_{XU|S^e} \,\tp_{YS^aS^d|XUS^e}
        ||p_{S^eS^aS^d} \,p_{XU|S^e}\,p_{Y|XS^a}) \right.
                                        \nonumber \\
    & & \mbox{\hspace{0.25in}} \left.
        + |J_L(\tp_{S^e} \,p_{XU|S^e} \,\tp_{YS^aS^d|XUS^e}) - R|^+ \right]
\label{eq:ErL-CDMC}
\end{eqnarray}
satisfies the following properties:
\begin{description}
\item[(i)] $E_{r,L}^{\CDMC}(R) = 0$ if and only if $R \ge C_L$;
\item[(ii)] $E_{r,L}^{\CDMC}(R) \le |C_L - R|^+ $; \item[(iii)]
$E_{r,L}^{\CDMC}(R) \le E_{r,L+1}^{\CDMC}(R)$ (nondecreasing in
$L$).
\end{description}
\label{lem:ErL-CDMC}
\end{lemma}
{\em Proof}.\\
(i) Clearly $E_{r,L}^{\CDMC}(R) \ge 0$,
with equality if and only if the following three conditions are met:
\begin{enumerate}
\item the minimizing $\tp_{S^e}$ in (\ref{eq:ErL-CDMC}) is equal to $p_{S^e}$,
\item the minimizing $\tp_{YS^aS^d|XUS^e}$ in (\ref{eq:ErL-CDMC})
    is equal to $p_{Y|XS^a} \,p_{S^aS^d|S^e}$, and
\item $R \ge C_L$.
\end{enumerate}

\noindent
(ii) This upper bound on (\ref{eq:ErL-CDMC}) is obtained by fixing
$\tp_{S^e} = p_{S^e}$ and $\tp_{YS^aS^d|XUS^e} = p_{Y|XS^a} \,p_{S^aS^d|S^e}$.
The upper bound is achieved if the minimizing $\tp_{S^e}$ and
$\tp_{YS^aS^d|XUS^e}$ in (\ref{eq:ErL-CDMC}) are equal to $p_{S^e}$ and
$p_{Y|XS^a} \,p_{S^aS^d|S^e}$, respectively.

\noindent
(iii) This is a direct consequence of Lemma~\ref{lem:monotonic}.

\begin{theorem}
For the CDMC case (Def.~\ref{def:CDMC}) with maximum-cost constraint
(\ref{eq:D1-as}) or expected-cost constraint (\ref{eq:D1-avg})
on the transmitter, the reliability function is lower-bounded
by the random-coding error exponent
\begin{eqnarray}
   E_r^{\CDMC}(R) & = & \lim_{L \to \infty} E_{r,L}^{\CDMC}(R) .
\label{eq:Er-CDMC}
\end{eqnarray}
Moreover, $E_r^{\CDMC}(R) = 0$ if and only if $R \ge C$.
\label{thm:ach-CDMC}
\end{theorem}

For any value of $L$, the random-coding error exponent $E_{r,L}^{\CDMC}(R)$
of (\ref{eq:ErL-CDMC}) is achieved by a binning code with conditionally
constant composition and the MPMI decoder.
We now present a brief overview of this scheme and an interpretation
for the MPMI decoder.

\begin{figure}
\centering
\setlength{\epsfxsize}{5in} \epsffile{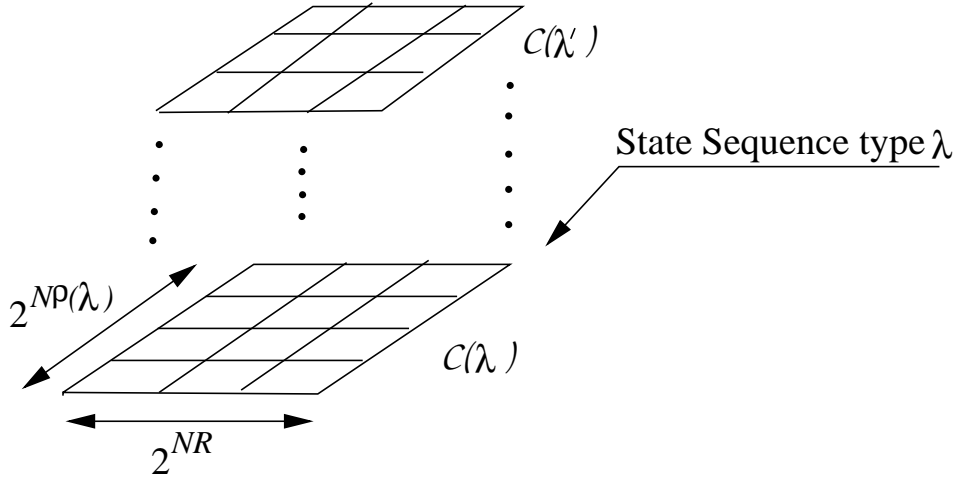}
\caption{Representation of binning scheme as a stack of
    arrays indexed by the encoder's state sequence type $\lambda$.
    The arrays have $2^{NR}$ columns and $2^{N\rho(\lambda)}$ rows,
    and the random-coding exponent is optimized by choosing
    $\rho(\lambda) = I_{US^e}^*(\lambda) + \epsilon$.}
\label{fig:BinStack}
\end{figure}

For notational simplicity, here we use the shorthand $\lambda$ to denote
the type of the encoder's state sequence (recall there is a polynomial
number of such types). Let $\calU = \{ 1, 2, \cdots, L \}$, where the
value of $L$ is arbitrary. Referring to Fig.~\ref{fig:BinStack},
to each value of $\lambda$ corresponds an array
\begin{equation}
   \calC(\lambda) = \{ \bu^{lm|\lambda},
    1 \le l \le 2^{N\rho(\lambda)} ,\, 1 \le m \le |\calM| \}
\label{eq:code}
\end{equation}
of codewords, drawn uniformly from some optimized type class. We refer to
$\rho(\lambda)$ as the {\em depth parameter} of the array $\calC(\lambda)$.
The codebook $\calC$ is the union of these arrays. Each array has
exponential size, but the number of arrays is polynomial in $N$.

The array depth parameter $\rho(\lambda)$ is designed to optimally balance
the probability of encoding error and the probability of decoding error,
conditioned on the encoder's state sequence type $\lambda$.
Upon seeing $m$ and $\bs^e$, the encoder evaluates the type $\lambda$ of
$\bs^e$ and seeks a codeword $\bu^{lm|\lambda}$
that belongs to some optimized conditional type class $T_{\bu|\bs^e}$.
Let $I_{US^e}^*(\lambda)$ denote the empirical mutual information
associated with $T_{\bu|\bs^e}$.
An encoding error arises when no codeword can be found in the conditional
type class $T_{\bu|\bs^e}$.
The probability of that event does not vanish when
$\rho(\lambda) \le I_{US^e}^*(\lambda)$ but vanishes at a double-exponential
rate when $\rho(\lambda) > I_{US^e}^*(\lambda) + \epsilon$.
The probability of decoding error increases exponentially with $\rho(\lambda)$.
Therefore the optimal tradeoff is given by
$\rho(\lambda) = I_{US^e}^*(\lambda) + \epsilon$.

Instead of choosing $\rho(\lambda)$ as a function of $\lambda$, a suboptimal
design would be to fix the value of $\rho$ and draw all the codewords uniformly
and i.i.d. from a single type class $T_{\bu}$. The scheme would then be
more akin to the original Gel'fand-Pinsker binning scheme,
which uses a single array of codewords (drawn i.i.d. from a p.m.f. $p_U$).
When $\rho$ is fixed, the fact that a polynomial number
of equal-size arrays is used rather than a single array is inconsequential
as far as error exponents are concerned.

The MPMI decoder is matched to the selected random binning scheme.
Given $(\by,\bs^d)$, the MPMI decoder seeks the codeword in
$\calC = \bigcup_\lambda \calC(\lambda)$ that achieves the maximum of
the {\bf penalized empirical mutual information} criterion
\begin{equation}
   \hat{m}_{\mathrm{MPMI}} = \argmax_m \max_{l,\lambda}
            [I(\bu^{lm|\lambda};\by\bs^d) - \rho(\lambda)] .
\label{eq:MPMI-0}
\end{equation}
As the proof of Theorem~\ref{thm:ach-CDMC} indicates, the penalty $\rho(\lambda)$
is optimal among all functions of $\lambda$; the optimal penalty is thus matched
to the array depth parameter.

The MPMI decoder may be thought of as an {\em empirical generalized MAP decoder}.
Indeed, all messages are equiprobable, and the encoding procedure ensures that
for any given type $\lambda$, all bins are equiprobable as well.
The probability of the pair $(m,l)$ is thus equal to $1/|\calC(\lambda)|$
for all $l,m$. Hence, given $\calC(\lambda)$, the {\em a priori}
distribution of the codewords is uniform:
$p(\bu^{lm|\lambda}) = 1/|\calC(\lambda)| = 2^{-N[R+\rho(\lambda)]}$. Therefore
\begin{equation}
   \rho(\lambda) = - R - \frac{1}{N} \log p(\bu^{lm|\lambda}) \quad
    \forall l,m .
\label{eq:rho-0}
\end{equation}
We may write
\begin{equation}
   I(\bu^{lm|\lambda};\by\bs^d) = \frac{1}{N} \log
    \frac{\hat{p}(\by,\bs^d|\bu^{lm|\lambda})}{\hat{p}(\by,\bs^d)}
\label{eq:MI-0}
\end{equation}
where $\hat{p}$ denotes an empirical p.m.f. or empirical conditional p.m.f..
Substituting (\ref{eq:rho-0}) and (\ref{eq:MI-0}) into (\ref{eq:MPMI-0}),
we obtain
\begin{eqnarray}
   \hat{m}_{\mathrm{MPMI}}
    & = & \argmax_m \max_{l,\lambda} \left[
            \frac{1}{N} \log \frac{\hat{p}(\by,\bs^d|\bu^{lm|\lambda})}
                {\hat{p}(\by,\bs^d)}
            + \frac{1}{N} \log p(\bu^{lm|\lambda}) \right]  \nonumber \\
    & = & \argmax_m \max_{l,\lambda} \hat{p}(\bu^{lm|\lambda}|\by\bs^d) .
\label{eq:empirical-GMAP}
\end{eqnarray}
This may be thought of as an empirical version of the generalized MAP decoder
\begin{eqnarray}
   \hat{m}_{\mathrm{GMAP}}
    & \triangleq & \argmax_m \max_{l,\lambda} p(\bu^{lm|\lambda}|\by\bs^d)
                                            \nonumber \\
    & = & \argmax_m \max_{l,\lambda} \left[
            \frac{1}{N} \log \frac{p(\by,\bs^d|\bu^{lm|\lambda})}{p(\by,\bs^d)}
            + \frac{1}{N} \log p(\bu^{lm|\lambda}) \right]
\label{eq:GMAP}
\end{eqnarray}
which requires knowledge of the channel from $\bu$ to $(\by,\bs^d)$.
We do not know whether the GMAP decoder is as good (on the exponential scale)
as the optimal MAP decoder
\begin{equation}
   \hat{m}_{\mathrm{MAP}} \triangleq \argmax_m \,p(m|\by,\bs^d)
    = \argmax_m \eE_{l,\lambda|\by\bs^d} \,p(\bu^{lm|\lambda}|\by,\bs^d)
\label{eq:MAP}
\end{equation}
which averages out the nuisance parameters $(l,\lambda)$
and is more difficult to analyze.

The MPMI decoder is matched to the encoding scheme in that the same function
$\rho(\lambda)$ is used as the depth parameter of the array $\calC(\lambda)$
and as the penalty in the decoding function. As the proof of
Theorem~\ref{thm:ach-CDMC} indicates, any other choice of the penalty
function would in general result in a lower error exponent.
This is not surprising in view of the above generalized MAP interpretation.

\subsection{Random-Coding Exponents for CAM Model}
\label{sec:RandomCoding-CAM}

We now turn our attention to the CAM channel model.
First we state the following lemma, which is analogous to Lemma~\ref{lem:ErL-CDMC}.

\begin{lemma}
The function
\begin{eqnarray}
   E_{r,L}^{\CAM}(R) & \triangleq & \min_{\tp_{S^e} \in \calP_{S^e}}
        \,\max_{p_{XU|S^e} \in \calP_{XU|S^e}(L,D_1)}
        \,\min_{\tp_{YS^aS^d|XUS^e} \in
            \calP_{YS^aS^d|XUS^e}[\calA,p_{XU|S^e}\tp_{S^e}]}
                                        \nonumber \\
    & & \mbox{\hspace*{0.25in}} \left[ D(\tp_{S^eS^aS^d}||p_{S^eS^aS^d})
        + \tilde{I}_{Y;US^eS^d|XS^a}(\tp_{S^e} \,p_{XU|S^e} \,\tp_{YS^aS^d|XUS^e})
                                \right. \nonumber \\
    & & \mbox{\hspace*{0.25in}} \left.
        +  |J_L(\tp_{S^e} \,p_{XU|S^e} \,\tp_{YS^aS^d|XUS^e}) - R|^+ \right]
\label{eq:ErL-CAM}
\end{eqnarray}
satisfies the following properties:
\begin{description}
\item[(i)] $E_{r,L}^{\CAM}(R) = 0$ if and only if $R \ge C_L$;
\item[(ii)] $E_{r,L}^{\CAM}(R) \le |C_L - R|^+ $; \item[(iii)]
$E_{r,L}^{\CAM}(R) \le E_{r,L+1}^{\CAM}(R)$ (monotonicity in $L$).
\end{description}
\label{lem:ErL-CAM}
\end{lemma}

\begin{theorem}
For the CAM case (Def.~\ref{def:CAM}) with maximum-cost constraint
(\ref{eq:D1-as}) or expected-cost constraint (\ref{eq:D1-avg})
on the transmitter, the reliability function is lower-bounded by
the random-coding error exponent
\begin{eqnarray}
   E_r^{\CAM}(R) & = & \lim_{L \to \infty} E_{r,L}^{\CAM}(R)
\label{eq:Er-CAM}
\end{eqnarray}
Moreover, $E_r^{\CAM}(R) = 0$ if and only if $R \ge C$.
\label{thm:ach-CAM}
\end{theorem}

For any value of $L$, the random-coding error exponent (\ref{eq:ErL-CAM})
is achieved by a randomly-modulated code with conditionally constant composition,
stacked binning, and a MPMI decoder. The worst attack channel
is uniform over single conditional types (Def.~\ref{def:attack-CC}).

\subsection{Comparison of Random-Coding Exponents for CDMC and CAM Models}
\label{sec:comparison}

For both the CDMC and the CAM models, it should be noted that:
\begin{enumerate}
\item the worst type classes $T_{\bs^e}$, $T_{\by\bs^a\bs^d|\bx\bu\bs^e}$,
    and best type class $T_{\bx\bu|\bs^e}$
    (in an appropriate min max min sense) determine the error exponents;
\item the order of the min, max and min is determined by the knowledge
    available to the encoder. The encoder knows $\bs^e$ and can optimize
    $T_{\bx\bu|\bs^e}$, but has no control over $T_{\by\bs^a\bs^d|\bx\bu\bs^e}$;
\item the straight-line part of $E_r(R)$ results from the union bound;
\item random codes are generally suboptimal at low rates.
\end{enumerate}

Theorems~\ref{thm:ach-CDMC} and \ref{thm:ach-CAM} imply the following
relationship between error exponents in the CDMC and CAM cases.

\begin{corollary}
$E_r^{\CDMC}(R) \le E_r^{\CAM}(R) \le |C-R|^+$ for all $R$.
\label{cor:Er}
\end{corollary}
{\em Proof}.
Fix $L$. Using the relation
\[ I(Y;US^eS^d|XS^a) = D(p_{YXUS^eS^aS^d} || p_{Y|XS^a} p_{XUS^eS^aS^d}) , \]
we write
\[ \tilde{I}_{Y;US^eS^d|XS^a}(\tp_{S^e} \,p_{XU|S^e} \,\tp_{YS^aS^d|XUS^e})
    = D(\tp_{S^e} \,p_{XU|S^e} \,\tp_{YS^aS^d|XUS^e}
    || \tp_S \,p_{XU|S^e} \,p_{Y|XS^a})
\]
where we have defined the marginal conditional p.m.f.
\[ p_{Y|XS^a}(y|x,s^a) = \frac{\sum_{us^es^d} \tp_{YS^aS^d|XUS^e}(y,s^a,s^d|x,u,s^e)
    \,p_{XU|S^e}(x,u|s^e) \,\tp_{S^e}(s^e)}
    {\sum_{yus^es^d} \tp_{YS^aS^d|XUS^e}(y,s^a,s^d|x,u,s^e)
    \,p_{XU|S^e}(x,u|s^e) \,\tp_{S^e}(s^e)} .
\]
Since $\tp_{YS^aS^d|XUS^e}$ is an element of
$\calP_{YS^aS^d|XUS^e}[\calA,p_{XU|S^e}\tp_{S^e}]$ in (\ref{eq:ErL-CAM}),
$p_{Y|XS^a}$ defined above is an element of $\calA$ and may be viewed as
a functional of $\tp_{YS^aS^d|XUS^e}$ (for fixed $p_{XU|S^e}$ and $\tp_{S^e}$).
Hence the cost function in (\ref{eq:ErL-CAM}) may be written as
\begin{eqnarray*}
   \lefteqn{D(\tp_S || p_S) + D(\tp_{S^e} \,p_{XU|S^e} \,\tp_{YS^aS^d|XUS^e}
    || \tp_S \,p_{XU|S^e} \,p_{Y|XS^a}) + |J-R|^+} \\
    & = & D(\tp_{S^e} \,p_{XU|S^e} \,\tp_{YS^aS^d|XUS^e}
        ||p_{S^eS^aS^d} \,p_{XU|S^e}\,p_{Y|XS^a}) + |J-R|^+
\end{eqnarray*}
where the equality follows from the chain rule for divergence.
Thus the cost functions in (\ref{eq:ErL-CAM}) and (\ref{eq:ErL-CDMC})
are identical; the only difference is the domain over which
the minimizations are performed. In (\ref{eq:ErL-CDMC}),
the minimization over $\tp_{YS^aS^d|XUS^e}$ is unconstrained, and
the minimization over $p_{Y|XS^a}$ is over $\calA$.
In (\ref{eq:ErL-CAM}), the minimization over $\tp_{YS^aS^d|XUS^e}$ is
constrained to the set $\calP_{YS^aS^d|XUS^e}[\calA,p_{XU|S^e}\tp_{S^e}]$,
and $p_{Y|XS^a}$ is a fixed element of $\calA$ once $\tp_{YS^aS^d|XUS^e}$
is fixed. In other words the minimization in (\ref{eq:ErL-CDMC}) is over
a larger set, and we have $E_{r,L}^{\CDMC}(R) \le E_{r,L}^{\CAM}(R)$.
Taking the limits of both sides of this inequality as $L \to \infty$,
we obtain $E_r^{\CDMC}(R) \le E_r^{\CAM}(R)$.

Similarly, from Lemma~\ref{lem:ErL-CAM} we have
$E_{r,L}^{\CAM}(R) \le |C_L - R|^+$; taking limits as $L \to \infty$,
we obtain $E_r^{\CAM}(R) \le |C-R|^+$.
\hfill $\Box$

The inequality $E_r^{\CDMC}(R) \le E_r^{\CAM}(R)$
is not as surprising as it initially seems, because the proof of
Theorem~\ref{thm:ach-CAM} shows there is no loss in optimality
in considering CAM's that are uniform over conditional types,
and there are more conditional types to choose from under
the CDMC model. Generally that additional flexibility is
beneficial for the adversary, and the worst conditional type
does not satisfy the hard constraint (\ref{eq:CAM}).
See Sec.~\ref{sec:binary} for an example.

\begin{remark}
In the absence of side information (degenerate $p_{S^eS^d}$),
the optimal $U=X$, and (\ref{eq:Er-CAM}) becomes
$E_r^{\CAM}(R) = |C-R|^+$. The expression for $E_r(R)$
derived by Hughes and Thomas \cite{Hug96} (Eqns (9), (6),
also see the observation on top of p.~96) is upper-bounded by
$|C-R|^+$; they also provide a binary-Hamming example in which
equality is achieved. Our result implies that the upper bound
$|C-R|^+$ is in fact achieved for any problem without side
information in which there exists a hard constraint on the conditional
type of the channel output given the input.
\end{remark}

\subsection{Capacity}
\label{sec:converse}

As discussed in Sec.~\ref{sec:preliminaries}, Gel'fand and Pinsker's proof
of the converse theorem in \cite{Gel80} can be extended to more complex problems
such as compound Gel'fand-Pinsker channels \cite{Mou03,Som04}.
The capacity for the generalized Gel'fand-Pinsker problem is given
in Theorems~\ref{thm:C-CDMC} and \ref{thm:C-CAM}, respectively.
Achievability of $C$ follows from Theorems~\ref{thm:ach-CDMC} and
\ref{thm:ach-CAM}. Indeed, for any $\epsilon > 0$, there exists
$L(\epsilon)$ such that $C_L \ge C-\epsilon$.
The proof of the converses appear in Sec.~\ref{sec:thm3} and \ref{sec:thm4}.

\begin{theorem}
Under the CDMC model (Def.~\ref{def:CDMC}) for the adversary,
capacity for the generalized Gel'fand-Pinsker problem
is given by (\ref{eq:C}) for both combinations
of maximum-cost constraints (\ref{eq:D1-as}) and expected-cost constraints
(\ref{eq:D1-avg}) on the transmitter.
\label{thm:C-CDMC}
\end{theorem}

\begin{theorem}
Under the CAM model (Def.~\ref{def:CAM}) for the adversary,
capacity for the generalized Gel'fand-Pinsker problem
is given by (\ref{eq:C}) for both combinations
of maximum-cost constraints (\ref{eq:D1-as}) and expected-cost constraints
(\ref{eq:D1-avg}) on the transmitter.
\label{thm:C-CAM}
\end{theorem}

The proof of the CDMC converse is similar to that in \cite{Mou03};
the proof in the CAM case exploits the close connection between
the CAM and CDMC problems.

\subsection{Remarks on Cardinality of $\calU$}

The sequence $C_L$ defined in (3.2) is nondecreasing and converges to the
capacity limit $C$, but one may ask at what rate. When the feasible set
$\calA$ has finite cardinality, by application of Caratheodory's
theorem, it suffices to select $L = |\calX|\,|\calS^e| + |\calA| - 1$
(see \cite{Som04,Mou04b} for related problems).
When $\calA$ is a compact set, one may construct a sequence
$\epsilon_L \downarrow 0$ and a sequence $\{\calA_L\}$ of subsets of $\calA$
that is dense in the $L_1$ norm:
\[ \forall\, p_{Y|XS^a} \in \calA, \exists\, \hat{p}_{Y|XS^a} \in \calA_L
    ~:~ \max_{x,s^a} \| \hat{p}_{Y|XS^a}(\cdot|x,s^a) - p_{Y|XS^a}(\cdot|x,s^a) \|
        < \epsilon_L . \]
This may be done, for instance, by applying a uniform quantizer to each
$p_{Y|XS^a}(y|x,s^a)$ to obtain $\hat{p}_{Y|XS^a}(y|x,s^a)$.
By continuity of the functional $J_L$, the effect of this quantization
on $J_L$ can be made arbitrarily small by letting $L \to \infty$.
Finally, Caratheodory's theorem can be applied to the set of $|\calA_L|$ attack
channels so that $\max_{p_{XU|S^e}} \min_{p_{Y|XS^a} \in \calA_L}
J_L(p_{S^e} \,p_{XU|S^e} \,p_{Y|XS^a})$
is achieved using $L = |\calX|\,|\calS^e| + |\calA_L| - 1$.
Proposition~\ref{prop:cara-C} below formally states this result
when the feasible set of attack channels
is defined by the distortion constraint (2.3).

\begin{proposition}
Consider the subsequence $\{C_L\}$ indexed by
\begin{equation}
   L = |\calX|\,|\calS^e| + (l+1)^{|\calY|\,|\calX|\,|\calS^a|} - 1 ,
    \quad l = 1, 2, \cdots
\label{eq:L-cara-capacity}
\end{equation}
Then
\[ C - 2 |\calY| \,\frac{\log l}{l} \le C_L \le C . \]
\label{prop:cara-C}
\end{proposition}
{\em Proof}: See Appendix~\ref{sec:cara-C}.

For the random-coding exponent $E_r(R)$, the idea is similar but
the derivations are more involved because Kullback-Leibler
divergence is not absolutely continuous with respect to its
arguments; attack channels that lie on the boundary of the
probability simplex require a special treatment.

In Proposition~\ref{prop:cara-Er} below, the random-coding
exponent is viewed as a function of $D_2$ and, with a little abuse
of notation, written as $E_r(D_2)$. The random-coding exponent
when the alphabet $\calU$ has size $L$ is similarly denoted by
$E_{r,L}(D_2)$.

\begin{lemma}
The function $E_r(D_2)$ is continuous and nonincreasing in $D_2$.
\label{lem:Er-continuous}
\end{lemma}

The above statement is a consequence of the fact that the Kullback-Leibler and
mutual-information functionals are continuous in their arguments, and that
the sets $\{\calA(D_2)\}$ are continuously nested.

\begin{proposition}
Denote by $c \le \frac{1}{|\calS^a|\,|\calS^d|}$ the minimum of
$p_{S^aS^d|S^e}$ over its support set. Define the constants
\[ l_{\min} = \max \left\{ |\calY|\,|\calS^a|\,|\calS^d| ,
    \,\exp_2 \left[ 1 + \sqrt{ \left| - \frac{1}{2}
        \log (8 |\calS^a|^5 |\calS^d| c) \right|^+ } \,\right] \right\}
\]
and $\overline{D} = \max_x \frac{1}{|\calY|} \sum_y d(x,y)$.
Consider the subsequence $\{E_{r,L}(D_2)\}$ indexed by
\begin{equation}
   L = |\calX|\,|\calS^e| + l^{|\calY|\,|\calX|\,(|\calS| + |\calS^a|)} - 1,
    \quad l = l_{\min}, \,l_{\min} + 1, \cdots
\label{eq:L-cara}
\end{equation}
Then
\begin{equation}
   E_{r,L}(D_2) \le E_r(D_2)
    \le E_{r,L} \left( D_2 - \frac{|\calY|\,|\calS^a|\,|\calS^d| \overline{D}
        + D_2 \ln l}{l} \right)
        + 7 \,|\calY|\,|\calS^a|\,|\calS^d| \frac{\log^2 l}{l} .
\label{eq:ErL-bounds}
\end{equation}
\label{prop:cara-Er}
\end{proposition}
The gap between the lower and upper bounds in (\ref{eq:ErL-bounds})
is $O(\frac{\log^2 l}{l})$ as $l \to \infty$.

{\em Proof}: See Appendix~\ref{sec:cara-Er}.

\Section{Binary-Hamming Case}
\label{sec:binary}

\begin{figure}[h]
\centering
\setlength{\epsfxsize}{4in} \epsffile{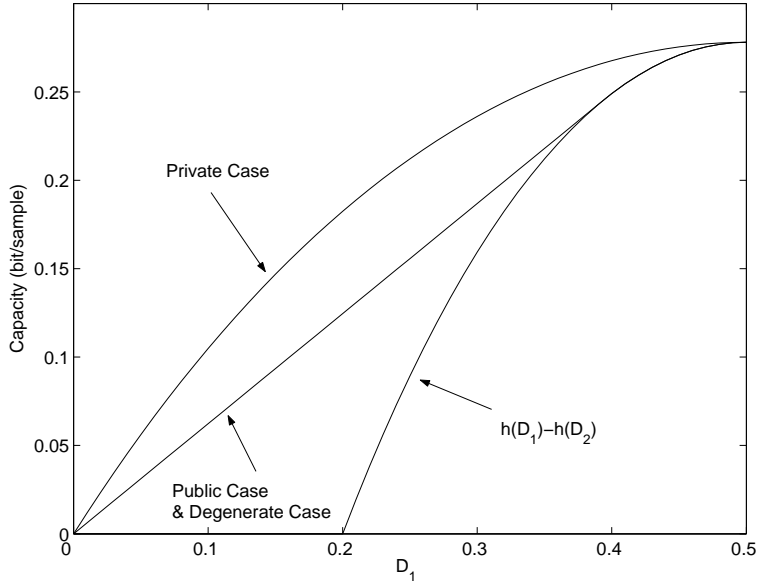}
\caption{Capacity functions for Cases I--III when $D_2 = 0.2$.}
\label{fig:BinaryCapacity}
\end{figure}

In this section, we consider a problem of theoretical and practical interest
where $\calS^e = \{ 0, 1 \}$, $\bS^e$ is a Bernoulli sequence with
$Pr[S^e=1] = p^e = 1 - Pr[S^e=0]$,
transmission is subject to the cost constraint (\ref{eq:D1-as})
in which $\Gamma$ is Hamming distance, and the adversary is
subject to the expected-distortion constraint (\ref{eq:D2-avg})
or to the maximum-distortion constraint (\ref{eq:D2-as}),
in which $d$ is also Hamming distance.
In both cases the set $\calA$ is given by (\ref{eq:D2-avg}).
We study three cases:
\begin{description}
\item[Case I:] $p^e = \frac{1}{2}$, $S^a = S^d = \emptyset$.
    This problem is analogous to the public watermarking problem
    of \cite{Bar03,Pra03,Mou03}.
\item[Case II:] $p^e = \frac{1}{2}$, $S^a = \emptyset$, $S^d = S^e$.
    This is the private watermarking problem of \cite{Mou03}.
    The CAM version of this problem is closely related to a problem
    studied by Csisz\'{a}r and Narayan \cite{Csi88} and
    Hughes and Thomas \cite{Hug96}.
\item[Case III:] Degenerate side information:
    $p^e = 0$, $S^e = S^a = S^d = \emptyset$.
    Unlike \cite{Csi88,Hug96}, the attacker's noise may depend on $X$.
\end{description}
In all three cases, we were able to derive some analytical results
and to numerically evaluate error exponents.
Capacity formulas for these problems are given below
and illustrated in Fig.~\ref{fig:BinaryCapacity}.

In this section we use the notation $p \star q \triangleq p(1-q) + (1-p)q$.

\subsection{Case I: Public Watermarking}
Here $p^e = \frac{1}{2}$ and $S^a = S^d = \emptyset$, so we have $S = S^e$.
Capacity for a fixed-DMC problem
(adversary implements a binary symmetric channel (BSC) with crossover
probability $D_2$) is given in Barron {\em et al.} \cite{Bar03}
and Pradhan {\em et al.} \cite{Pra03}:
\begin{equation}
   C^{\mathrm{\mathrm{pub}}} = g^*(D_1,D_2) \triangleq \left \{
   \begin{array}{ll}
   \frac{D_1}{\delta_2} [\overline{h}(\delta_2)-\overline{h}(D_2)],
    & \mbox {if} \; 0 \le D_1 < \delta_2 ; \\
   \overline{h}(\delta_2)-\overline{h}(D_2),
    &\mbox {if} \; \delta_2 \le D_1 \le 1/2;\\
   1-\overline{h}(D_2), &\mbox {if} \; D_1 > 1/2,
   \end{array} \right.
\label{eq:C-pub}
\end{equation}
where $\delta_2 = 1-2^{-\overline{h}(D_2)}$ and
$\overline{h}(\cdot)$ is the binary entropy function.
The straight-line portion of the capacity function is achieved
by time-sharing.
Proposition~\ref{prop:pub-C} shows that the BSC is the worst
channel for the CDMC and CAM classes considered.

\begin{proposition}
Capacity under the CDMC and CAM models defined by the distortion constraints
(\ref{eq:D2-avg}) and (\ref{eq:D2-as}), respectively, is equal to $C^{\mathrm{\mathrm{pub}}}$
and is achieved for $|\calU| = 2$.
\label{prop:pub-C}
\end{proposition}

\noindent \emph{Proof}: See Appendix~\ref{sec:pub-C}.

\begin{proposition}
The random-coding error exponent is a straight line in the CAM case:
$E_r^{\mathrm{CAM,pub}}(R) = |C^{\mathrm{\mathrm{pub}}}-R|^+$ for all $R$.
The minimizing $\tp_S$ in (\ref{eq:ErL-CAM}) coincides with $p_S$,
the maximizing $L = |\calU| = 2$,
and the minimizing $p_{Y|XUS}$ is the BSC $p_{Y|X}$ with crossover
probability $D_2$.
\label{prop:pub-Er-CAM}
\end{proposition}

\noindent \emph{Proof}: See Appendix~\ref{sec:pub-Er-CAM}.

Unlike the CAM case, in the CDMC case we have no guarantee that $L=2$ is optimal
for random-coding exponents.
The exponents $E_{r,L}^{\CDMC}(R)$ and $E_{r,L}^{\CAM}(R)$ are shown
in Fig.~\ref{fig:ErrExpoPublic} for the case $D_1 = 0.4$, $D_2 = 0.2$, and $L=2$;
see Sec.~\ref{sec:binary-discuss} for details of these calculations.
For the CDMC case, we have found numerically that
the worst attack channel $p_{Y|X}$ is the BSC with crossover
probability $D_2$, and that the worst-case $\tp_S$
in (\ref{eq:Er-CDMC}) coincides with $p_S$.

\subsection{Case II: Private Watermarking}

Here $p^e = \frac{1}{2}$, $S^a = \emptyset$, $S^d = S^e = S$.

\begin{proposition} \cite{Mou03}.
Capacity is given by
\begin{equation}
   C^{\mathrm{priv}} = \overline{h}(D_1 \star D_2) - \overline{h}(D_2)
\label{eq:C-priv}
\end{equation}
and is achieved when $U=X$ ($L=2$).
\end{proposition}

\begin{figure}
\centering
\setlength{\epsfxsize}{4in} \epsffile{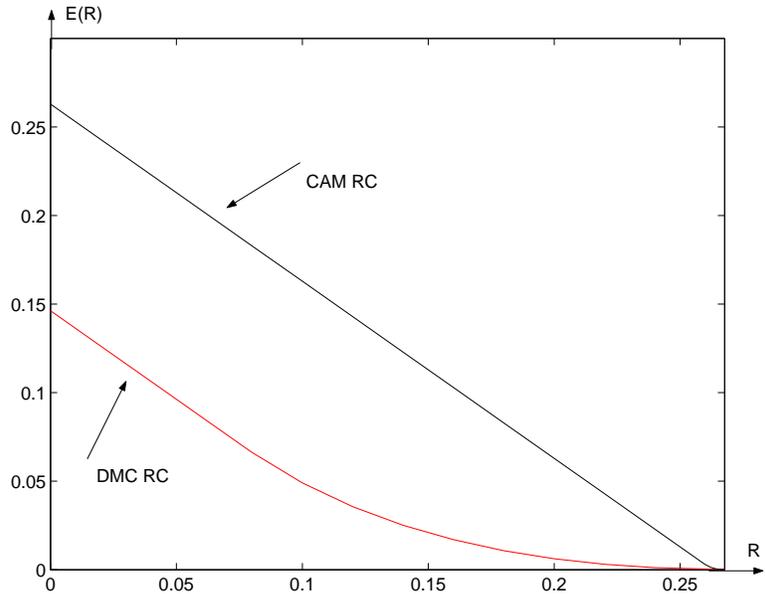}
\caption{Error exponents when $p^e=\frac{1}{2}$, $S^a = \emptyset$,
    $S^e = S^d$ (private watermarking).}
\label{fig:ErrExpoPrivate}
\end{figure}

For the random-coding exponents, we have no guarantee that $L=2$
is an optimal choice. The exponents $E_{r,L}^{\CDMC}(R)$ and
$E_{r,L}^{\CAM}(R)$ in that case are shown in
Fig.~\ref{fig:ErrExpoPrivate} for the case $D_1 = 0.4$, $D_2 =
0.2$. As in Case~I, for both the CAM and CDMC cases, the
worst-case $\tp_S$ in (\ref{eq:ErL-CDMC}) and (\ref{eq:ErL-CAM})
coincides with $p_S$.

The capacity expression (\ref{eq:C-priv}) was also derived for the AVC
problem of Csisz\'{a}r and Narayan \cite{Csi88}, albeit with different
assumptions ($p^e=0$, i.e., degenerate side information, and channel state
$\theta$ selected independently of $X$, see Appendix~\ref{sec:CAM}).
Error exponents for the latter problem were derived by Hughes and Thomas
\cite{Hug96}. They obtained
$E_r(R) = |C-R|^+$ at all rates below capacity.

\subsection{Case III: Degenerate side information}

Here $p^e = 0$, $S^e = S^a = S^d = \emptyset$.

\begin{proposition}
Capacity is the same as in the public watermarking game: $C^{\mathrm{deg}} = C^{\mathrm{\mathrm{pub}}}$.
\label{prop:deg-C}
\end{proposition}

\noindent \emph{Proof}: See Appendix~\ref{sec:deg-C}.

\begin{proposition}
$E_r^{\mathrm{CAM,deg}}(R) = |C^{\mathrm{deg}} - R|^+$ for all $R < C^{\mathrm{deg}}$.
\label{prop:deg-Er-CAM}
\end{proposition}

\noindent \emph{Proof}: Follows from Remark~3.1.

\begin{figure}
\centering
\setlength{\epsfxsize}{4in} \epsffile{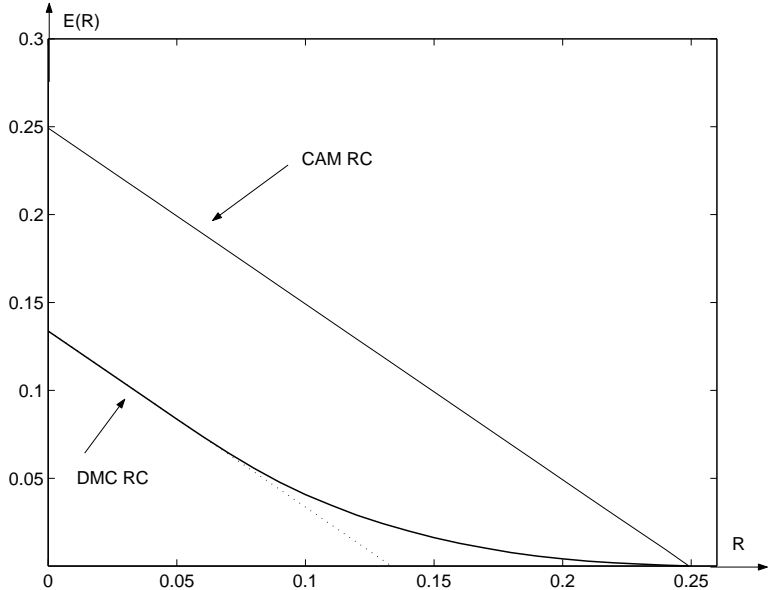}
\caption{Error exponents when $S^e = S^a = S^d = \emptyset$
    ($p^e=0$, no side information).}
\label{fig:ErrExpoDegen}
\end{figure}

Unlike Case~I and Case~II, the worst attack is an asymmetric
binary channel, favoring outputs with low Hamming weight. Error
exponents in the case $D_1 = 0.4$, $D_2 = 0.2$, are given in
Fig.~\ref{fig:ErrExpoDegen}.
\begin{figure}
\centering
\setlength{\epsfxsize}{4in} \epsffile{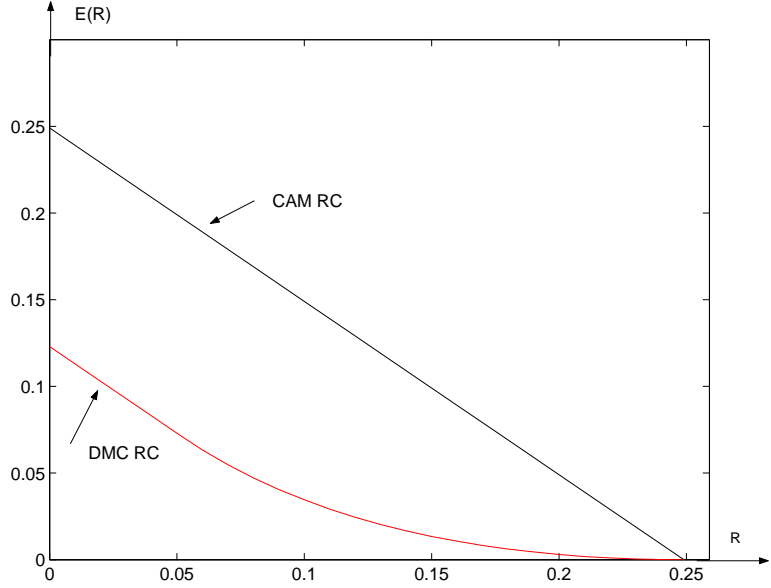}
\caption{Error exponents when $p^e=\frac{1}{2}$, $S^a = S^d = \emptyset$
    (public watermarking).}
\label{fig:ErrExpoPublic}
\end{figure}

\subsection{Discussion}
\label{sec:binary-discuss}

Comparing Figs~\ref{fig:ErrExpoPublic} and \ref{fig:ErrExpoPrivate},
we see that the random-coding error exponents for $L=2$ are only slightly
larger when side information is available to the decoder. For instance, the
zero-rate exponents are 0.123 and 0.146 at rate zero in the CDMC case;
and 0.249 and 0.263 in the CAM case.

Some practical comments about the optimization problems solved in
this section are in order. Among these problems, the calculations of
random-coding exponents for the CDMC / public
watermarking scenario are the most complicated ones, both of
which have four layers of minimization or maximization. The
number of the parameters to be optimized is $8|\calU|+1$
(1 for $\tp_S$, $4|\calU| - 2$ for $p_{XU|S}$, 2 for $p_{Y|X}$
and $4|\calU|$ for $\tp_{Y|XUS}$). Other difficulties arise due
to the lack of nice properties such as everywhere differentiability
and convexity. There appears to be a substantial increase
of computational difficulty going from $|\calU| = 2$ to larger $\calU$.
Based on the analytical results derived above, it is tempting
to conjecture that $|\calU| = 2$ is an sufficient choice
for optimality; unfortunately at this time we are unable
to validate that conjecture analytically or numerically.

We have used a genetic algorithm \cite{Michalewicz} to numerically solve
the above-mentioned optimization problems. Advantages of genetic
algorithms include easy implementation,
robustness with respect to selection of starting points,
no need for evaluation of function derivatives, and ability to
handle high-dimensional problems. The parameters of
a genetic algorithm may be selected to ensure that the algorithm
is globally convergent. In particular, we have used
an ``elitist'' genetic algorithm, in which the value of the best
individual in each iteration is nondecreasing for a maximization
problem (or nonincreasing for a minimization problem). The
sequence of the best solutions in each iteration is guaranteed
to converge to the global optimum almost surely
\cite{Michalewicz,Rudolph}.

\Section{Proof of Theorem~\ref{thm:ach-CDMC}}
\label{sec:thm1}

Owing to Lemma~\ref{lem:ErL-CDMC}, for any $\epsilon > 0$, chosen independently
of $N$, there exists $L(\epsilon)$ such that
\begin{equation}
   E_{r,L}^{\CDMC}(R) \ge E_r^{\CDMC}(R) - \epsilon ,
    \qquad \forall L > L(\epsilon) .
\label{eq:ErL-eps}
\end{equation}

We shall prove the existence of a sequence of codes $(f_N, g_N)$ such that
\[ \lim_{N \to \infty} \left[ - \frac{1}{N} \log \max_{p_{Y|XS^a} \in \calA}
    \,P_e(f_N, g_N, p_{Y|XS^a}) \right] = E_{r,L}^{\CDMC}(R) . \]

The proof is given for the maximum-cost constraint (\ref{eq:D1-as})
on the transmitter. Any code that achieves the error exponent $E_{r,L}^{\CDMC}(R)$
is therefore also feasible under the weaker average-cost constraint
(\ref{eq:D1-avg}). A random ensemble of binning codes is constructed,
and it is shown that the error probability averaged over this ensemble
vanishes exponentially with $N$ at rate $E_{r,L}^{\CDMC}(R)$.
Since the error probability functional $P_e(f_N, g_N, p_{Y|XS^a})$
is continuous in $p_{Y|XS^a}$ (by (\ref{eq:Pe-det-code}))
and the feasible set $\calA$ of attack channels
can be approximated with arbitrary precision (in the $L_1$ norm)
by a subset whose cardinality is polynomial in $N$,
there exists a code $(f_N,g_N)$ from the ensemble that achieves
$E_{r,L}^{\CDMC}(R)$ uniformly over $\calA$.
It is therefore sufficient to prove that
\begin{equation}
   \lim_{N \to \infty} \left[ - \frac{1}{N} \log \max_{p_{Y|XS^a} \in \calA}
    \,P_e(F_N, G_N, p_{Y|XS^a}) \right] = E_{r,L}^{\CDMC}(R)
\label{eq:PeFG-exp}
\end{equation}
for the random ensemble considered.
Combining (\ref{eq:PeFG-exp}) and (\ref{eq:ErL-eps}) then proves the claim.

The maximum-cost constraint (\ref{eq:D1-as}) may be written as
\begin{equation}
    \sum_{s^e,x} p_{\bs^e \bx}(s^e,x) \Gamma(s^e,x) \le D_1 \quad a.s.
\label{eq:code-D1}
\end{equation}

Assume $R < C_L - \epsilon$.
Define the function
\begin{eqnarray}
   \widetilde{E}_{r,L,N}^{\CDMC}(R, p_{\bs^e}, p_{\bx\bu|\bs^e})
    & \triangleq &
        \min_{p_{\by\bs^a\bs^d|\bx\bu\bs^e} \in \calP_{YS^aS^d|XUS^e}^{[N]}}
        \,\min_{p_{Y|XS^a} \in \calA}
                                    \nonumber \\
    & & \left[ D(p_{\bs^e} \,p_{\bx\bu|\bs^e}
        \,p_{\by\bs^a\bs^d|\bx\bu\bs^e} ||p_{S^eS^aS^d} p_{XU|S^e} p_{Y|XS^a})
                            \right. \nonumber \\
    & & \left. + | J_L(p_{\bs^e} p_{\bx\bu|\bs^e}
        p_{\by\bs^a\bs^d|\bx\bu\bs^e}) - \epsilon - R|^+ \right] ,  \nonumber \\
    & & \hspace*{1in} p_{\bs^e} \in \calP_{S^e}^{[N]}, \,
            p_{\bx\bu|\bs^e} \in \calP_{XU|S^e}^{[N]}(L,D_1) .
\label{eq:tilde-ErN-CDMC}
\end{eqnarray}
Let
\begin{eqnarray}
   E_{r,L,N}^{\CDMC}(R) & \triangleq & \min_{p_{\bs^e} \in \calP_{S^e}^{[N]}}
        \,\max_{p_{\bx\bu|\bs^e} \in \calP_{XU|S^e}^{[N]}(L,D_1)}
        \widetilde{E}_{r,L,N}^{\CDMC}(R, p_{\bs^e}, p_{\bx\bu|\bs^e}) ,
\label{eq:ErN-CDMC}
\end{eqnarray}
which differs from (\ref{eq:ErL-CDMC}) in that the optimizations
are performed over empirical p.m.f.'s instead of arbitrary p.m.f.'s.

Consider the maximization over the conditional type $p_{\bx\bu|\bs^e}$
(viewed as a function of $p_{\bs^e}$) in (\ref{eq:ErN-CDMC}).
As a result of this optimization, we may associate the following:
\begin{itemize}
\item to any type $p_{\bs^e}$, \\
    a type class $T_U^*(p_{\bs^e}) \triangleq T_{\bu}$
    and a mutual information $I_{US^e}^*(p_{\bs^e}) \triangleq
    \tilde{I}_{US^e}(p_{\bu|\bs^e} p_{\bs^e})$;
\item to any sequence $\bs^e$, a conditional type class
    $T_{U|S^e}^*(\bs^e) \triangleq T_{\bu|\bs^e}$;
\item to any sequences $\bs^e$ and $\bu \in T_{U|S^e}^*(\bs^e)$,
    a conditional type class
    $T_{X|US^e}^*(\bu,\bs^e) \triangleq T_{\bx|\bu\bs^e}$.
\end{itemize}

A random codebook $\calC$ for $\bU$ is the union of codebooks
$\calC(p_{\bs^e})$ indexed by the state sequence type $p_{\bs^e}$
(recall there is a polynomial number of types).
The codebook $\calC(p_{\bs^e})$ is obtained by
a) drawing $2^{N(R+\rho(p_{\bs^e}))}$ random vectors
independently from the uniform distribution over $T_U^*(p_{\bs^e})$, and
b) arranging them in an array with $2^{NR}$ columns and
$2^{N\rho(p_{\bs^e})}$ rows.
The design of the function $\rho(p_{\bs^e})$ is arbitrary at this point
but will be optimized later.

\underline{\bf Encoder}.
The encoding (given $\bs^e$ and $m$) proceeds in two steps.
\begin{enumerate}
\item Find $l$ such that $\bu(l,m) \in \calC(p_{\bs^e}) \bigcap T_{U|S^e}^*(\bs^e)$.
    If more than one such $l$ exists, pick one of them randomly (with uniform
    distribution). Let $\bu = \bu(l,m)$.
    If no such $l$ can be found, generate $\bu$ uniformly
    from the conditional type class $T_{U|S^e}^*(\bs^e)$.
\item   Generate $\bX$ uniformly distributed over the conditional type class
    $T_{X|US^e}^*(\bu,\bs^e)$.
\end{enumerate}
Clearly, the p.m.f. of $(\bS^e,\bU,\bX)$, conditioned on its joint type,
is uniform, and the encoder's maximum-cost constraint is satisfied.

\underline{\bf Decoder.}
Given $(\by,\bs^d)$, the decoder seeks
$\hat{\bu} \in \calC = \bigcup_{p_{\bs^e}} \calC(p_{\bs^e})$ that
maximizes the {\bf penalized empirical mutual information} criterion
\begin{equation}
   \max_{p_{\bs^e}} \max_{\bu \in \calC(p_{\bs^e})}
    [I(\bu;\by\bs^d) - \psi(p_{\bs^e})] .
\label{eq:MPMI}
\end{equation}
The decoder declares an error if maximizers with different column
indices are found.
Otherwise the decoder outputs the column index of $\hat{\bu}$.
The penalty function $\psi(\cdot)$ in (\ref{eq:MPMI}) will soon be optimized,
resulting in the ``matched design'' $\psi = \rho$.

We now analyze the probability of error
\[ P_e \triangleq \max_{p_{Y|XS^a} \in \calA} P_e(F_N,G_N,p_{Y|XS^a}^N) \]
of the decoder.

{\bf Step 1.}
An encoding error arises under the following event:
\begin{eqnarray}
  \calE_m & = & \{ (\calC, \bs^e) ~:~ (\bu(l,m) \in \calC
    ~\mathrm{and}~ \bu(l,m) \notin T_{U|S^e}^*(\bs^e))
    ~\mathrm{for}~ 1 \le l \le 2^{N\rho(p_{\bs^e})} \}
\label{eq:E1}
\end{eqnarray}
conditioned on message $m$ being selected. The probability
that a vector $\bU$ uniformly distributed over $T_U^*(p_{\bs^e})$ also belongs
to $T_{U|S^e}^*(\bs^e)$ is equal to $\exp_2 \{ -N I_{US^e}^*(p_{\bs^e}) \}$
on the exponential scale. Therefore
\begin{eqnarray}
   Pr[\calE_m | T_{\bs^e}]
    & = & ( 1 - Pr[\bU \in T_{U|S^e}^*(\bS^e) ~|~ \bU \sim \uU(T_U^*(p_{\tbs^e}))] )
        ^{2^{N\rho(p_{\bs^e})}} \nonumber \\
    & \doteq & ( 1 - 2^{-N I_{US^e}^*(p_{\bs^e})} )^{2^{N\rho(p_{\bs^e})}} \nonumber \\
    & \le & \exp \{ - 2^{N (\rho(p_{\bs^e}) - I_{US^e}^*(p_{\bs^e}))} \}
                                    \label{eq:double-exp0} \\
    & \le & \left\{ \begin{array}{ll}
            \exp \{ - 2^{N \epsilon} \} & :~\mathrm{if~}
                \rho(p_{\bs^e}) \ge I_{US^e}^*(p_{\bs^e}) + \epsilon    \\
            1 & :~\mathrm{else.}
        \end{array} \right.
\label{eq:double-exp}
\end{eqnarray}
The inequality (\ref{eq:double-exp0}) follows from $1+a \le e^a$.
The double-exponential term in (\ref{eq:double-exp})
vanishes faster than any exponential function.

{\bf Step 2.}
We have a decoding error under the following event $\calE_m'$:
conditioned on message $m$ being selected, there exists $\bu'$
not in column $m$ of an array $\calC(p_{\bs^e}')$ such that
\[ I(\bu';\by\bs^d) - \psi(p_{\bs^e}') \ge I(\bu;\by\bs^d) - \psi(p_{\bs^e}) . \]
Therefore
\begin{eqnarray}
   P_e & = & \max_{p_{Y|XS^a}} Pr[error~|~ m=1, p_{Y|XS^a}] \nonumber \\
    & = & \max_{p_{Y|XS^a}} \sum_{T_{\bs\bu\bx\by}} Pr[T_{\bs\bu\bx\by}]
        \,Pr[error~|~ T_{\bs\bu\bx\by}, m=1]
                                \nonumber \\
    & \le &  \max_{p_{Y|XS^a}} \sum_{T_{\bs\bu\bx\by}} Pr[T_{\bs\bu\bx\by}]
        \,\left( Pr[\calE_1 | T_{\bs\bu\bx\by}]
        + Pr[\calE_1'~|~ T_{\bs\bu\bx\by}, \calE_1^c] \right)
                                \nonumber \\
    & = &  \max_{p_{Y|XS^a}} \sum_{T_{\bs\bu\bx\by}} Pr[T_{\bs\bu\bx\by}]
        \,\left( Pr[\calE_1 | T_{\bs^e}]
        + Pr[\calE_1'~|~ T_{\bs\bu\bx\by}, \calE_1^c] \right)
                                \label{eq:Pe1-DMC} \\
    & \le &  \sum_{T_{\bs\bu\bx\by}} \max_{p_{Y|XS^a}} Pr[T_{\bs\bu\bx\by}]
        \,\left( Pr[\calE_1 | T_{\bs^e}]
        + Pr[\calE_1'~|~ T_{\bs\bu\bx\by}, \calE_1^c] \right) .
                                \label{eq:Pe1b-DMC}
\end{eqnarray}
We will see in Step~3 that $Pr[\calE_1'~|~ T_{\bs\bu\bx\by}, \calE_1^c]$
does not depend on $p_{Y|XS^a}$ for the MPMI decoder.
Using the asymptotic relations
$P_Z^N(T_{\bz}) \doteq \exp_2 \{-ND(p_{\bz}||p_Z)\}$
and $P_{Z|V}^N(T_{\bz|\bv}) \doteq \exp_2 \{-ND(p_{\bz|\bv}||p_{Z|V}|p_{\bv})\}$
\cite{Csi81}, we derive
\begin{eqnarray}
  Pr[T_{\bs\bu\bx\by}]
    & = & P_S^N P_{\bX\bU|\bS^e} P_{Y|XS^a}^N (T_{\bs\bu\bx\by})
                                        \nonumber \\
    & \doteq & \exp_2 \{ - N D(p_{\bs\bu\bx\by}
            || p_S p_{\bx\bu|\bs^e} p_{Y|XS^a}) \}  \nonumber \\
    & = & \exp_2 \{ - N D(p_{\bs^e} p_{\bx\bu|\bs^e}
        p_{\by\bs^a\bs^d|\bx\bu\bs^e}
            || p_S p_{\bx\bu|\bs^e} p_{Y|XS^a}) \} .
\label{eq:PrTsuxy-DMC}
\end{eqnarray}

{\bf Step 3.}
Next we evaluate $Pr[\calE_1'~|~ T_{\bs\bu\bx\by}, \calE_1^c]$,
which can be written as
$Pr[\calE_1'~|~ T_{\bs\bu\bx\by}, \bu, \by, \bs^d, \calE_1^c]$,
where $\bu,\by,\bs^d$ is an arbitrary member of the conditional type class
$T_{\bu\by\bs^d|\bx\bs^e\bs^a}$.

Denote by $p_e(\bu, \by, \bs^d, p_{\bs^e}', T_{\bs\bu\bx\by})$
the probability that the decoder outputs the codeword in row $l'$ and column
$m' \ne 1$ of the array $\calC(p_{\bs^e}')$, conditioned on $\bu, \by, \bs^d$,
and $T_{\bs\bu\bx\by}$.
This conditional error probability is independent of $(l',m')$. We have
\begin{equation}
   Pr[\calE_1'~|~ T_{\bs\bu\bx\by}, \bu, \by, \bs^d, \calE_1^c]
    = 1 - \prod_{p_{\bs^e}'}
        [1 - p_e(\bu, \by, \bs^d, p_{\bs^e}', T_{\bs\bu\bx\by})]
        ^{2^{N\rho(p_{\bs^e}')} (2^{NR}-1)}
\label{eq:pe-su}
\end{equation}
where
\begin{eqnarray}
   p_e(\bu, \by, \bs^d, p_{\bs^e}', T_{\bs\bu\bx\by})
    & = & \sum_{\bu' \in \calU_e(\bu, \by, \bs^d,
        p_{\bs^e}', p_{\bs\bu\bx\by})} p(\bu'|p_{\bs^e}')
                                        \nonumber \\
    & = & \sum_{\bu' \in \calU_e(\bu, \by, \bs^d, p_{\bs^e}',
        p_{\bs\bu\bx\by})} \frac{1}{|T_U^*(p_{\bs^e}')|}
\label{eq:pe-s'su}
\end{eqnarray}
and
\begin{eqnarray}
   \calU_e(\bu, \by, \bs^d, p_{\bs^e}', p_{\bs\bu\bx\by})
    & = & \left\{ \bu' \in T_U^*(p_{\bs^e}')
        ~:~ I(\bu';\by\bs^d) - \psi(p_{\bs^e}')
        \ge I(\bu;\by\bs^d) - \psi(p_{\bs^e}) \right\}
\end{eqnarray}
is the set of codewords $\bu'$ in the array indexed by $p_{\bs^e}'$,
that cause a decoding error, conditioned on $\bu, \by, \bs^d$,
and $T_{\bu\bs\bx\by}$. Also define the corresponding set of
conditional types
\begin{eqnarray}
   \calT_e(\bu, \by, \bs^d, p_{\bs^e}', p_{\bs\bu\bx\by})
    & = & \left\{ T_{\bu'|\by\bs^d} ~:~   T_{\bu'} = T_U^*(p_{\bs^e}'),
        \,I(\bu';\by\bs^d) - \psi(p_{\bs^e}')
        \ge I(\bu;\by\bs^d) - \psi(p_{\bs^e}) \right\}      \nonumber \\
    & \subseteq & \left\{ T_{\bu'|\by\bs^d} ~:~
        I(\bu';\by\bs^d) - \psi(p_{\bs^e}')
        \ge I(\bu;\by\bs^d) - \psi(p_{\bs^e}) \right\} .
\end{eqnarray}
Therefore
\begin{eqnarray}
   p_e(\bu, \by, \bs^d, p_{\bs^e}', T_{\bs\bu\bx\by})
    & = & \sum_{T_{\bu'|\by\bs^d} \in
        \calT_e(\bu, \by, \bs^d, p_{\bs^e}', p_{\bs\bu\bx\by})}
        \frac{|T_{\bu'|\by\bs^d}|}{|T_{\bu'}|}      \nonumber \\
    & \doteq & \sum_{T_{\bu'|\by\bs^d} \in
        \calT_e(\bu, \by, \bs^d, p_{\bs^e}', p_{\bs\bu\bx\by})}
        2^{-N I(\bu';\by\bs^d)}             \nonumber \\
    & \dotle & 2^{-N [I(\bu;\by\bs^d) - \psi(p_{\bs^e}) + \psi(p_{\bs^e}')] }
\label{eq:pe-s'us-bound}
\end{eqnarray}
because $\frac{|T_{\bu'|\by\bs^d}|}{|T_{\bu'}|} \doteq 2^{-N I(\bu';\by\bs^d)}$,
and the number of conditional types $T_{\bu'|\by\bs^d}$ is polynomial in $N$.

Next we use the following inequality, which is proved in Appendix~\ref{sec:bound2}.
\begin{eqnarray}
   1 - \prod_i (1-\alpha_i)^{t_i} & \le & \min \left(
        1, \sum_i \alpha_i t_i \right) ,
            \quad 0 \le \alpha_i \le 1, \,t_i \ge 1 .
\label{eq:bound2}
\end{eqnarray}

Applying (\ref{eq:bound2}) and (\ref{eq:pe-s'us-bound})
successively to (\ref{eq:pe-su}), we obtain
\begin{eqnarray}
   Pr[\calE_1'~|~ T_{\bs\bu\bx\by}, \calE_1^c]
      & = & Pr[\calE_1'~|~ T_{\bs\bu\bx\by}, \bu, \by, \bs^d, \calE_1^c]
                                                \nonumber \\
    & \le & \min \left\{ 1 , \sum_{p_{\bs^e}'}
        p_e(\bu, \by, \bs^d, p_{\bs^e}', T_{\bs\bu\bx\by})
        \,2^{N\rho(p_{\bs^e}')} (2^{NR}-1) \right\}         \nonumber \\
    & \dotle & \min \left\{ 1 , \sum_{p_{\bs^e}'}
        2^{-N[I(\bu;\by\bs^d) - \psi(p_{\bs^e}) + \psi(p_{\bs^e}')
                - \rho(p_{\bs^e}') - R]} \right\}           \nonumber \\
    & \doteq & \exp_2 \left\{ -N |I(\bu;\by\bs^d) - \psi(p_{\bs^e})
            + \min_{p_{\bs^e}'} [\psi(p_{\bs^e}')-\rho(p_{\bs^e}')] - R|^+
            \right\} .
\label{eq:pe-su-bound}
\end{eqnarray}

{\bf Step 4.}
Combining (\ref{eq:pe-su-bound}) and (\ref{eq:double-exp}),
we obtain
\begin{equation}
   Pr[\calE_1 | T_{\bs^e}] + Pr[\calE_1'~|~ T_{\bs\bu\bx\by}, \calE_1^c]
    \dotle \exp_2 \left\{ - N \,\Gamma(R, \rho, \psi, p_{\bs^e}, p_{\bx\bu|\bs^e},
        p_{\by\bs^a\bs^d|\bx\bu\bs^e}) \right\}
\label{eq:Pr-condl-error}
\end{equation}
where we have defined the function
\begin{eqnarray}
   \lefteqn{\Gamma(R, \rho, \psi, p_{\bs^e}, p_{\bx\bu|\bs^e},
        p_{\by\bs^a\bs^d|\bx\bu\bs^e})}         \nonumber \\
    & \triangleq & \left\{ \begin{array}{ll}
        | I(\bu;\by\bs^d) - \psi(p_{\bs^e})
            + \min_{p_{\bs^e}'} [\psi(p_{\bs^e}')-\rho(p_{\bs^e}')] - R|^+
            & :~ \rho(p_{\bs^e}) \ge I_{US^e}^*(p_{\bs^e}) + \epsilon \\
        0 & :~\mbox{else.}
        \end{array} \right.
\label{eq:Gamma}
\end{eqnarray}
Applying the inequality $\min_{p_{\bs^e}'} F(p_{\bs^e}') \le F(p_{\bs^e})$
to the function $F=\psi-\rho$, we obtain
\begin{eqnarray*}
   \lefteqn{\Gamma(R, \rho, \psi, p_{\bs^e}, p_{\bx\bu|\bs^e},
        p_{\by\bs^a\bs^d|\bx\bu\bs^e})}         \nonumber \\
    & \le & \left\{ \begin{array}{ll}
        | I(\bu;\by\bs^d) - \rho(p_{\bs^e}) - R|^+
            & :~ \rho(p_{\bs^e}) \ge I_{US^e}^*(p_{\bs^e}) + \epsilon \\
        0 & :~\mbox{else.}
        \end{array} \right.
\end{eqnarray*}
and thus
\begin{equation}
  \Gamma(R, \rho, \psi, p_{\bs^e}, p_{\bx\bu|\bs^e}, p_{\by\bs^a\bs^d|\bx\bu\bs^e})
    \le |J_L(p_{\bs^e} p_{\bx\bu|\bs^e} p_{\by\bs^a\bs^d|\bx\bu\bs^e})
        - \epsilon - R|^+
\label{eq:Gamma-ineq}
\end{equation}
with equality when
\begin{equation}
   \psi(p_{\bs^e}) = \rho(p_{\bs^e}) = I_{US^e}^*(p_{\bs^e}) + \epsilon .
\label{eq:rho}
\end{equation}

Combining (\ref{eq:Pe1b-DMC}), (\ref{eq:PrTsuxy-DMC}), (\ref{eq:Pr-condl-error}),
and (\ref{eq:Gamma-ineq}), we obtain
\begin{eqnarray}
   P_e & \le & \sum_{T_{\bs\bu\bx\by}} \max_{p_{Y|XS^a}} Pr[T_{\bs\bu\bx\by}]
        \left( Pr[\calE_1 | T_{\bs^e}]
            + Pr[\calE_1'~|~ T_{\bs\bu\bx\by}, \calE_1^c] \right)
                                        \nonumber \\
    & \dotle & \max_{p_{\bs^e}} \min_{p_{\bx\bu|\bs^e}} \min_{\rho, \psi}
        \max_{p_{\by\bs^a\bs^d|\bx\bs^e}} \max_{p_{Y|XS^a}}
        \exp_2 \left\{ -N [D(p_{\bs^e} p_{\bx\bu|\bs^e}
            p_{\by\bs^a\bs^d|\bx\bu\bs^e} || p_S p_{\bx\bu|\bs^e} p_{Y|XS^a})
            \right.
                                        \nonumber \\
    & & \mbox{\hspace*{2in}} \left. + \Gamma(R, \rho, \psi, p_{\bs^e},
            p_{\bx\bu|\bs^e}, p_{\by\bs^a\bs^d|\bx\bu\bs^e}) \right\}
                                        \label{eq:Pe-UB-DMC-0} \\
    & = & \exp_2 \left\{ -N E_{r,L,N}^{\CDMC}(R) \right\}
                                        \label{eq:Pe-UB-DMC}
\end{eqnarray}
where (\ref{eq:Pe-UB-DMC-0}) holds because $p_{\bx\bu|\bs^e}$ and $(\rho,\psi)$
can be optimized to achieve the exponent $E_{r,L,N}^{\CDMC}(R)$ in (\ref{eq:ErN-CDMC}).

{\bf Step 5.}
By Lemma~\ref{lem:ErL-CDMC}, the function $E_{r,L}^{\CDMC}(R)$
is nonnegative and upper bounded by $|C_L-R|^+$.
Applying (\ref{eq:l1-property}) with $\tp_{S^e}$, $p_{XU|S^e}$,
$(\tp_{YS^aS^d|XUS^e}, p_{Y|XS^a})$, and $D+|J-R|^+$
in the roles of the variables $p$, $q$, $r$, and the functional $\phi$,
respectively, we conclude that
the exponent $E_{r,L,N}^{\CDMC}(R)$ in (\ref{eq:Pe-UB-DMC}) converges to
the limit $E_{r,L}^{\CDMC}(R)$ in (\ref{eq:ErL-CDMC}) as $N \rightarrow \infty$.
Since $E_{r,L}^{\CDMC}(R) > 0$ for all $R < C_L$,
the probability of error vanishes if $R < C_L$.
The claim follows from the fact that we can choose $L$ such that
$C_L > C - \epsilon$, for any arbitrarily small $\epsilon$.

\Section{Proof of Theorem~\ref{thm:ach-CAM}}
\label{sec:thm2}

The proof is similar to the proof of Theorem~\ref{thm:ach-CDMC}
and is again given for the maximum-cost constraint (\ref{eq:D1-as})
on the transmitter.
A random ensemble $\calE$ of binning codes $(f_N,g_N)$ with fixed $|\calU|=L$
is constructed. This ensemble may also be viewed as a random ensemble of RM codes
(Def.~\ref{def:code-RM}). RM codes are obtained by selecting a prototype
$(f_N,g_N)$ from $\calE$ and generating the RM family
$\{ f_N^{\pi}, g_N^{\pi} \}$ according to Def.~\ref{def:code-RM}.
For RM codes there is no loss of optimality in restricting
the attack channel to a class of channels that are uniform
over conditional types (see Step~2 below).
It is shown that the error probability averaged over the ensemble
$\calE$ vanishes exponentially with $N$ at the rate $E_{r,L}^{\CAM}(R)$
given in (\ref{eq:ErL-CAM}).
Since the class of attack channels considered in Step~2 has
polynomial complexity, there exists a RM code that achieves $E_{r,L}^{\CAM}(R)$
for all attack channels in $\calP_{\bY|\bX\bS^a}[\calA]$.

The codebook-generation, encoding and decoding procedures are the same
as those in the CDMC case, with the difference that the types and conditional
types generated/selected by the encoder are obtained by optimizing a slightly
different payoff function. The probability of error analysis
is similar as well.

Assume $R < C_L - \epsilon$.
Define
\begin{eqnarray}
   \widetilde{E}_{r,L,N}^{\CAM}(R, p_{\bs^e}, p_{\bx\bu|\bs^e}) & \triangleq &
        \min_{p_{\by\bs^a\bs^d|\bx\bu\bs^e}
            \in \calP_{YS^aS^d|XUS^e}^{[N]}[\calA, p_{\bx\bu|\bs^e} p_{\bs^e}]}
                                    \nonumber \\
    & & \left[ D(p_{\bs^e\bs^a\bs^d}||p_{S^eS^aS^d})
        + \tilde{I}_{Y;US^eS^d|XS^a}
        (p_{\bs^e} p_{\bx\bu|\bs^e} p_{\by\bs^a\bs^d|\bx\bu\bs^e})
                            \right. \nonumber \\
    & & \left. + |J_L(p_{\bs^e} p_{\bx\bu|\bs^e} p_{\by\bs^a\bs^d|\bx\bu\bs^e})
        - \epsilon - R|^+ \right]
\label{eq:tilde-ErLN-CAM}
\end{eqnarray}
for all $p_{\bs^e} \in \calP_{S^e}^{[N]}$ and
$p_{\bx\bu|\bs^e} \in \calP_{XU|S^e}^{[N]}(L,D_1)$.
Let
\begin{eqnarray}
   E_{r,L,N}^{\CAM}(R) & \triangleq & \min_{p_{\bs^e} \in \calP_{S^e}^{[N]}}
        \,\max_{p_{\bx\bu|\bs^e} \in \calP_{XU|S^e}^{[N]}(D_1)}
        \widetilde{E}_{r,N}^{\CAM}(R, p_{\bs^e}, p_{\bx\bu|\bs^e}) .
\label{eq:ErLN-CAM}
\end{eqnarray}
which differs from (\ref{eq:ErL-CAM}) in that the optimizations
are performed over empirical p.m.f.'s instead of arbitrary p.m.f.'s.
Consider the maximization over $p_{\bx\bu|\bs^e}$ (viewed as a function of
$p_{\bs^e}$) in (\ref{eq:ErLN-CAM}). As in the proof of
Theorem~\ref{thm:ach-CDMC}, to the resulting optimal $p_{\bx\bu|\bs^e}$
we can associate a type class $T_U^*(p_{\bs^e})$,
conditional type classes $T_{U|S^e}^*(\bs^e)$ and $T_{X|US^e}^*(\bu,\bs^e)$,
and a mutual information $I_{US^e}^*(p_{\bs^e} p_{\bu|\bs^e})$.

Define $\rho(p_{\bs^e})$ and $\psi(p_{\bs^e})$ as in (\ref{eq:rho}).
The random codebook $\calC$ is a stack of codebooks $\calC(p_{\bs^e})$,
each of which is obtained by
a) drawing $2^{N(R+\rho(p_{\bs^e}))}$ independent random vectors
whose components are uniformly distributed in $T_U^*(p_{\bs^e})$, and
b) arranging them in an array with $2^{NR}$ columns and
$2^{N\rho(p_{\bs^e})}$ rows.

\underline{\bf Encoder}.
The encoding (given $\bs^e$ and $m$) proceeds exactly as in the CDMC case:
\begin{enumerate}
\item Find $l$ such that $\bu(l,m) \in \calC(p_{\bs^e}) \bigcap T_{U|S^e}^*(\bs^e)$.
    If more than one such $l$ exists, pick one of them randomly (with uniform
    distribution). Let $\bu = \bu(l,m)$.
    If no such $l$ can be found, generate $\bu$ uniformly
    from the conditional type class $T_{U|S^e}^*(\bs^e)$.
\item   Generate $\bX$ uniformly distributed over the conditional type class
    $T_{X|US^e}^*(\bu\bs^e)$.
\end{enumerate}

\underline{\bf Decoder.}
The decoder is the MPMI decoder of (\ref{eq:MPMI}).
We now analyze its probability of error
\[ P_e \triangleq \max_{p_{\bY|\bX\bS^a} \in \calP_{\bY|\bX\bS^a}[\calA]}
    P_e(F_N,G_N,p_{\bY|\bX\bS^a}) . \]

{\bf Step 1}.
An encoding error arises when no codeword with the appropriate type
can be found. The probability $Pr[\calE_m | T_{\bs^e}]$ of this event
is given by (\ref{eq:double-exp}).

{\bf Step 2}.
We have a decoding error under the following event $\calE_m'$: there
exists $\bu'$ not in column $m$ of an array $\calC(p_{\bs^e}')$ such that
$I(\bu';\by\bs^d) - \rho(p_{\bs^e}') \ge I(\bu;\by\bs^d) - \rho(p_{\bs^e})$.
Therefore
\begin{eqnarray}
   P_e & = & \max_{p_{\bY|\bX\bS^a}} Pr[error~|~ m=1] \nonumber \\
    & = & \max_{p_{\bY|\bX\bS^a}} \sum_{T_{\bs\bu\bx\by}} Pr[T_{\bs\bu\bx\by}]
        \,Pr[error~|~ T_{\bs\bu\bx\by}, m=1]
                                \nonumber \\
    & = &  \max_{p_{\bY|\bX\bS^a}}
        \sum_{T_{\bs\bu\bx\by}} Pr[T_{\bs\bu\bx\by}] \,\left(
        Pr[\calE_1 | T_{\bs^e}]
        + Pr[\calE_1'~|~ T_{\bs\bu\bx\by}, \calE_1^c] \right) .
                                \label{eq:Pe1}
\end{eqnarray}
Unlike (\ref{eq:Pe1b-DMC}), no dependency on a DMC $p_{Y|XS^a}$ appears here.
Observe that
\begin{equation}
   Pr[\calE_1'~|~ T_{\bs\bu\bx\by}, \calE_1^c]
    = \sum_{(\tbs\tbu\tbx\tby) \in T_{\bs\bu\bx\by}}
        p(\tbs|T_{\bs}) \,p(\tbx\tbu|\tbs^e)
        p_{\bY|\bX\bS^a}(\tby|\tbx\tbs^a)
        Pr[\calE_1'~|~ \tbu,\tby,\tbs^d,T_{\bs\bu\bx\by}, \calE_1^c] .
\label{eq:Pe2}
\end{equation}

Here we can apply the following argument from \cite{Som04}.
From (\ref{eq:Pe1}) (\ref{eq:Pe2}), we see that $P_e(F_N,G_N,p_{\bY|\bX\bS^a})$
is an affine functional of $p_{\bY|\bX\bS^a}$. Moreover, it can be verified
that $P_e(F_N,G_N,p_{\bY|\bX\bS^a}) = P_e(F_N,G_N,p_{\bY|\bX\bS^a}^{\pi})$
where $\pi$ is a permutation operator, and
$p_{\bY|\bX\bS^a}^{\pi}(\by|\bx\bs^a)
\triangleq p_{\bY|\bX\bS^a}(\pi \by| \pi \bx, \pi \bs^a)$.
By uniform averaging over all permutations $\pi$, we obtain an attack channel
$\overline{p}_{\bY|\bX\bS^a} = \frac{1}{N!} \sum_{\pi} p_{\bY|\bX\bS^a}^{\pi}$
which is {\em strongly exchangeable}: if $(\bX,\bS^a)$ is uniformly distributed
over a type class, then $\bY$ is uniformly distributed over conditional
class types.
So without loss of optimality for the adversary, we can consider only
strongly exchangeable channels in the analysis, for which
$p_{\bY|\bX\bS^a}(\by|\bx,\bs^a)$ is given by
\begin{equation}
   p_{\bY|\bX\bS^a}(\by|\bx,\bs^a)
    = \frac{Pr[T_{\by|\bx\bs^a}]}{|T_{\by|\bx\bs^a}|} ,
\label{eq:pYXSa}
\end{equation}
with $T_{\by|\bx\bs^a}$ to be optimized.
Using the upper bound
\begin{equation}
   Pr[T_{\by|\bx\bs^a}] \le \iI\{ p_{\by|\bx\bs^a} \in \calA \}
\label{eq:bound1}
\end{equation}
and the asymptotic relations
$Pr[T_{\bs}] \doteq \exp_2 \{ - N D(p_{\bs} || p_S) \}$ and
$\frac{|T_{\by|\bz\bw}|}{|T_{\by|\bz}|} \doteq
\exp_2 \{ - N I(\by;\bw|\bz) \}$, we obtain
\begin{eqnarray}
  Pr[T_{\bs\bu\bx\by}]
    & = & P_S^N P_{\bX\bU|\bS^e} P_{\bY|\bX\bS^a} (T_{\bs\bu\bx\by})
                                        \nonumber \\
    & = & |T_{\bs\bu\bx\by}| \frac{Pr[T_{\bs}] \,Pr[T_{\by|\bx\bs^a}]}
        {|T_{\bs}| \,|T_{\bx\bu|\bs^e}| \,|T_{\by|\bx\bs^a}|}
                                        \nonumber \\
    & \le & Pr[T_{\bs}] \frac{|T_{\bs\bu\bx\by}|}
        {|T_{\bs}| \,|T_{\bx\bu|\bs^e}| \,|T_{\by|\bx\bs^a}|}
        \iI\{ p_{\by|\bx\bs^a} \in \calA \}
                                        \nonumber \\
    & = & Pr[T_{\bs}] \frac{|T_{\by|\bx\bu\bs}|}{|T_{\by|\bx\bs^a}|}
        \iI\{ p_{\by|\bx\bs^a} \in \calA \}
                                        \nonumber \\
    & \doteq & \exp_2 \left\{ - N [ D(p_{\bs} || p_S)
        + \tilde{I}_{Y;US^eS^d|XS^a}(p_{\bs^e} p_{\bx\bu|\bs^e}
        p_{\by\bs^a\bs^d|\bx\bu\bs^e}) ] \right\}
        \,\iI\{ p_{\by|\bx\bs^a} \in \calA \} .
\label{eq:PrTsuxy-CAM}
\end{eqnarray}

{\bf Step 3.} This step is identical to the corresponding step in the DMC case
and yields
\begin{equation}
   Pr[\calE_1'~|~ T_{\bs\bu\bx\by}, \calE_1^c]
    \le \exp_2 \left\{ -N |J_L(p_{\bs^e} \,p_{\bx\bu|\bs^e}
        \,p_{\by\bs^a\bs^d|\bx\bu\bs^e}) - \epsilon - R|^+ ] \right\}
\label{eq:pe-su-bound-CAM}
\end{equation}

{\bf Step 4.}
Combining (\ref{eq:Pe1}), (\ref{eq:PrTsuxy-CAM}), (\ref{eq:double-exp}),
and (\ref{eq:pe-su-bound-CAM}), we obtain
\begin{eqnarray}
   P_e & \dotle & \sum_{T_{\bs\bu\bx\by}}
        Pr[T_{\bs\bu\bx\by}] \,Pr[\calE_1'~|~ T_{\bs\bu\bx\by}, \calE_1^c]
                                    \nonumber \\
    & \dotle & \max_{p_{\bs^e}} \min_{p_{\bx\bu|\bs^e}}
        \max_{p_{\by\bs^a\bs^d|\bx\bu\bs^e}}
        \exp_2 \left\{ -N [D(p_{\bs^e\bs^a\bs^d}||p_{S^eS^aS^d})
        + \tilde{I}_{Y;US^eS^d|XS^a}
        (p_{\bs^e} p_{\bx\bu|\bs^e} p_{\by\bs^a\bs^d|\bx\bu\bs^e}) \right.
                                    \nonumber \\
    & & \mbox{\hspace*{2in}} \left.
        + |J_L(p_{\bs^e} \,p_{\bx\bu|\bs^e} \,p_{\by\bs^a\bs^d|\bx\bu\bs^e})
            - \epsilon - R|^+ ] \right\}
                                    \nonumber \\
    & = & \exp_2 \{ -N E_{r,L,N}^{\CAM}(R) \}
                                    \label{eq:Pe-UB-CAM}
\end{eqnarray}
because $p_{\bx\bu|\bs^e}$ was optimized to achieve the exponent
$E_{r,L,N}^{\CAM}(R)$ in (\ref{eq:ErLN-CAM}).

{\bf Step 5.}
The last step is identical to that in the DMC case:
the exponent $E_{r,L,N}^{\CAM}(R)$ in (\ref{eq:Pe-UB-CAM}) converges to
the limit $E_{r,L}^{\CAM}(R)$ in (\ref{eq:ErL-CAM}) as $N \rightarrow \infty$,
and all rates below $C_L$ are achievable. By choosing $L$ large enough,
$C_L$ can be made arbitrarily close to $C$.
\hfill $\Box$

\Section{Proof of Converse of Theorem~\ref{thm:C-CDMC}}
\label{sec:thm3}

The proof of the converse theorem is an extension of \cite[Prop.~4.3]{Mou03}.
To prove the claim (derive an upper bound on capacity), we only need
to consider the expected-cost constraint (\ref{eq:D1-avg}) for the transmitter.
Indeed replacing (\ref{eq:D1-avg}) with the stronger maximum-cost constraint
(\ref{eq:D1-as}) cannot increase capacity, so the same upper bound applies.
Likewise, we assume as in \cite{Mou03} that the decoder knows the
attack channel $p_{Y|XS^a}$, because the resulting upper bound
on capacity applies to an uninformed decoder as well.

{\bf Step~1}.
Choose an arbitrary small $\eta > 0$.
For any rate-$R$ encoder $f_N$ and attack channel $p_{Y|XS^a} \in \calA$ such that
\begin{equation}
   I(M,\bY\bS^d) \le N(R-\eta) ,
\label{eq:IMY-R}
\end{equation}
we have
\begin{eqnarray*}
   NR = H(M)
    & = & H(M|\bY\bS^d) + I(M;\bY\bS^d) \\
    & \le & 1 + P_e(f_N,g_N, p_{Y|XS^a}^N)  \,NR + I(M;\bY\bS^d) \\
    & \le & 1 + P_e(f_N,g_N, p_{Y|XS^a}^N)  \,NR + N(R-\eta)
\end{eqnarray*}
where the first inequality is due to Fano's inequality,
and the second is due to (\ref{eq:IMY-R}). Hence
\[ P_e(f_N,g_N, p_{Y|XS^a}^N)  \ge \frac{N \eta -1}{NR} . \]
We conclude that the probability of error is bounded away from zero:
\begin{equation}
   P_e(f_N, g_N, p_{Y|XS^a}^N) \ge \frac{\eta}{2R}
\label{eq:strong-converse1}
\end{equation}
for all $N > \frac{2}{\eta}$.
Therefore rate $R$ is not achievable if
\begin{equation}
   \min_{p_{Y|XS^a} \in \calA} I(M;\bY\bS^d) \le N(R-\eta) .
\label{eq:cond1}
\end{equation}

{\bf Step~2}.
The joint p.m.f. of $(M,\bS,\bX,\bY)$ is given by
\begin{equation}
   p_{M\bS\bX\bY} = p_M p_{\bS^e} \iI\{ \bX = f_N(\bS^e,M) \}
    \prod_{i=1}^N p_{Y|XS^a}(y_i|x_i,s_i^a) p_{S^aS^d|S^e}(s_i^a,s_i^d|s_i^e) .
\label{eq:converse-joint-pmf}
\end{equation}
Define the random variables
\begin{equation}
   W_i = (M, S_{i+1}^e, \cdots, S_N^e, S_1^d, \cdots, S_{i-1}^d,
        Y_1, \cdots, Y_{i-1}) ,
        \quad 1 \le i \le N .
\label{eq:Vi}
\end{equation}
Since $(M, \{ S_j^e, S_j^d, Y_j \}_{j \ne i}) \to X_i S_i^e \to Y_i S_i^a S_i^d$
forms a Markov chain for any $1 \le i \le N$,
so does
\begin{equation}
   W_i \to X_i S_i^e \to Y_i S_i^a S_i^d .
\label{eq:markov-Wi}
\end{equation}

Also define the quadruple of random variables $(W,S,X,Y)$ as $(W_T,S_T,X_T,Y_T)$,
where $T$ is a time-sharing random variable, uniformly distributed
over $\{ 1, \cdots, N \}$ and independent of all the other random variables.
The random variable $W$ is defined over an alphabet of cardinality
$\exp_2 \{ N[R + \log \max(|\calS^e|, |\calY| \,|\calS^d|)] \}$.
Due to (\ref{eq:converse-joint-pmf}) and (\ref{eq:markov-Wi}),
$W \to XS^e \to YS^aS^d$ forms a Markov chain.

Using the same inequalities as in \cite[Lemma~4]{Gel80}
(with $(Y_i, S_i^d)$ and $S_i^e$ playing the roles of $Y_i$ and $S_i$,
respectively), we obtain
\begin{equation}
   I(M;\bY\bS^d) \le \sum_{i=1}^N [I(W_i;Y_i S_i^d) - I(W_i;S_i^e)] .
\label{eq:ineq-GP}
\end{equation}
Using the definition of $(W,S,X,Y)$ above and the same inequalities as in
\cite[(C16)]{Mou03}, we obtain
\begin{eqnarray}
   \sum_{i=1}^N [I(W_i;Y_i S_i^d) - I(W_i;S_i^e)]
    & = & N [I(W;YS^d|T) - I(W;S^e|T)] \nonumber \\
    & \le & N [I(WT;YS^d) - I(WT;S^e)] \nonumber \\
    & = & N [I(U;YS^d) - I(U;S^e)]
\label{eq:TimeShare}
\end{eqnarray}
where $U = (W,T)$ is defined over an alphabet of cardinality
\[ L(N) \triangleq N \,\exp_2 \{ N[R + \log \max(|\calS^e|, |\calY| \,|\calS^d|)] \} . \]
Therefore
\begin{eqnarray}
   I(M;\bY\bS^d) & \le & N [I(U;YS^d) - I(U;S^e)] \nonumber \\
            & = & N J_{L(N)}(p_{USXY}) \nonumber \\
   \min_{p_{Y|XS^a} \in \calA} I(M;\bY\bS^d)
    & \le & N \min_{p_{Y|XS^a} \in \calA} J_{L(N)}(p_{USXY}) \nonumber \\
    & \le & N \sup_L \max_{p_{XU|S^e}}
        \min_{p_{Y|XS^a} \in \calA} J_L(p_{USXY})       \nonumber \\
    & = & N \lim_{L \to \infty} \,\max_{p_{XU|S^e}}
        \min_{p_{Y|XS^a} \in \calA} J_L(p_{USXY})
\label{eq:cond2}
\end{eqnarray}
where the last equality follows from Lemma~\ref{lem:monotonic}.
Combining (\ref{eq:cond1}) and (\ref{eq:cond2}), we conclude
that $R$ is not achievable if
\[ \lim_{L \to \infty} \,\max_{p_{XU|S^e}} \min_{p_{Y|XS^a} \in \calA}
    J_L(p_{USXY}) \le R - \eta, \]
which proves the claim, because $\eta$ is arbitrarily small.
\hfill $\Box$

\Section{Proof of Converse of Theorem~\ref{thm:C-CAM}}
\label{sec:thm4}

The proof of the converse theorem builds on the proof for the C-DMC case.

{\bf Step~1}.
Consider an attack channel $p_{Y|XS^a}^*$ that achieves $C(D_1,\calA)$
in (\ref{eq:C}). Without loss of generality, assume that $p_{S^a}(s^a) > 0$
for all $s^a \in \calS^a$. For any positive $\epsilon$, consider
the following $L_1$ neighborhood of $p_{Y|XS^a}^*$:
\[ {\calB}(\epsilon) = \left\{ p_{Y|XS^a} ~:~
    \sum_{y,x,s^a} | p_{Y|XS^a}(y|x,s^a) - p_{Y|XS^a}^*(y|x,s^a)|
    \le \epsilon \right\} .
\]
We have $\lim_{\epsilon \to 0} C(D_1, \calB(\epsilon)) = C(D_1, \calA)$.
For any arbitrarily small $\eta$, there exists $\epsilon$ such that
\[ C(D_1, \calA) - \eta \le C(D_1, \calB(\epsilon)) \le C(D_1, \calA) . \]

In order to prove the converse theorem, it is sufficient to show that
reliable communication at rates
$R > C(D_1, \calB(\epsilon)) + 2\eta \ge C(D_1, \calA) + \eta$
is impossible for a particular attack channel
$p_{\bY|\bX\bS^a} \in \calP_{\bY|\bX\bS^a}[\calB(\epsilon)]$.
The channel we select is ``nearly memoryless''.
Given any rate-$R$ randomized code $(\calM, F_N, G_N)$, we show that
$\lim_{N \to \infty} P_{e,N}(F_N,G_N,p_{\bY|\bX\bS^a}) \ge \frac{\eta}{4R}$
hence is nonzero.

{\bf Step~2: Construction of $p_{\bY|\bX\bS^a}$}. Consider any
rate-$R$ deterministic code ($\calM, f_N, g_N$) where $R >
C(D_1,\calA)$. From Theorem~\ref{thm:C-CDMC}, we know that
$\min_{f_N,g_N} P_{e,N}(f_N,g_N,(p_{Y|XS^a}^*)^N) \not \to 0$ as
$N \to \infty$. Define an arbitrary mapping $\Lambda~:~\calX^N
\times (\calS^a)^N \to \calY^N$ such that
$p_{\Lambda(\bx,\bs^a)|\bx\bs^a} \in \calB(\epsilon)$ for all
$(\bx,\bs^a)$. Denote by $\tbY$ the output of $(p_{Y|XS^a}^*)^N$.
Define the following functions of $(\tby, \bx, \bs^a)$: the binary
quantity
\[
   B = 1 \quad \Leftrightarrow \quad p_{\tby|\bx\bs^a} \notin \calB(\epsilon)
\]
and the sequence
\begin{equation}
   \by = \left\{ \begin{array}{lll}
    \tby & \mathrm{:~if~} p_{\tby|\bx\bs^a} \in \calB(\epsilon) & (B=0) \\
    \Lambda(\bx,\bs^a) & \mathrm{:~else} & (B=1).
    \end{array} \right.
\label{eq:y-CAM}
\end{equation}
Therefore the p.m.f.
\[
    \pYXSa(\by|\bx,\bs^a) = \left[ (p_{Y|XS^a}^*)^N(\by|\bx,\bs^a)
    Pr(B=0) + \iI\{\by = \Lambda(\bx,\bs^a)\} Pr(B=1) \right]
    \iI\{ p_{\by|\bx,\bs^a} \in \calB(\epsilon) \}
\]
belongs to $\calB(\epsilon)$.

{\bf Step~3}. Now we seek an upper bound on $Pr[B=1]$. Define the
binary random variable $A$ such that
\begin{equation}
   A = 1 \quad \Leftrightarrow \quad \min_{x,s^a} p_{\bx\bs^a}(x,s^a) \le \epsilon^2 .
\label{eq:A=1}
\end{equation}
The probability that $A=1$ is a function of the code $f_N$.
We assume momentarily that $f_N$ is such that
\begin{equation}
   Pr[A=1] < \frac{\eta}{9R} .
\label{eq:Pr-A=1}
\end{equation}
In Step~5 we show this assumption causes no loss of generality.

Define the shorthand
\[ \tilde{\epsilon} = \frac{\epsilon^6}{2 \ln 2} \]
and the class of types
\begin{equation}
   \calP_{YXS^a}^{[N]}(\tilde{\epsilon}) \triangleq \left\{ p_{\tby\bx\bs^a}
    ~:~ D(p_{\tby|\bx\bs^a} || p_{Y|XS^a}^* |p_{\bx\bs^a}) > \tilde{\epsilon} ,
        \,\,\underbrace{\min_{x,s^a} p_{\bx\bs^a}(x,s^a) > \epsilon^2}_{A=0}
        \right\} .
\label{eq:type-epsilon}
\end{equation}

With this notation, we have
\begin{eqnarray}
  B=1 & \Rightarrow & \sum_{y,x,s^a} |p_{\by|\bx\bs^a}(y|x,s^a)
        - p_{Y|XS^a}^*(y|x,s^a)| > \epsilon             \nonumber \\
  B=1, A=0 & \Rightarrow &
        \|p_{\tby|\bx\bs^a} p_{\bx\bs^a} - p_{Y|XS^a}^* p_{\bx\bs^a} \|
            > \epsilon^2 \sum_{y,x,s^a} |p_{\tby|\bx\bs^a}(y|x,s^a)
                - p_{Y|XS^a}^*(y|x,s^a)| > \epsilon^3   \nonumber \\
    & \Rightarrow &
        D(p_{\tby|\bx\bs^a} || p_{Y|XS^a}^* |p_{\bx\bs^a}) > \tilde{\epsilon}
\label{eq:B=1,A=0}
\end{eqnarray}
where the first line follows from the definition of $\calB(\epsilon)$,
the second line from (\ref{eq:Pr-A=1}), and the third line from Pinsker's
inequality \cite[p.~58]{Csi81}: $D(p||p') \ge \|p-p'\|^2 / (2 \ln 2)$.

We have
\begin{equation}
   \mbox{Pr}[B=1] \le \mbox{Pr}[B=1, A=0] + \mbox{Pr}[A=1] .
\label{eq:B=1}
\end{equation}
Due to (\ref{eq:type-epsilon}) and (\ref{eq:B=1,A=0}),
the first term in the right side is bounded as follows:
\begin{eqnarray}
   \mbox{Pr}[B=1, A=0]
    & \le & \mbox{Pr} \left[ p_{\tby\bx\bs^a} \in
        \calP_{YXS^a}^{[N]}(\tilde{\epsilon}) \right]
                                            \nonumber \\
    & = & \sum_{p_{\tby\bx\bs^a} \in \calP_{YXS^a}^{[N]}(\tilde{\epsilon})}
        \mbox{Pr} \left[ T_{\tby\bx\bs^a} \right]
                                            \nonumber \\
    & \le & (N+1)^{|\calY| \,|\calX| \,|\calS^a|}
        \max_{p_{\tby\bx\bs^a} \in \calP_{YXS^a}^{[N]}(\tilde{\epsilon})}
        \mbox{Pr} \left[ T_{\tby\bx\bs^a} \right]
                                            \nonumber \\
    & \le & (N+1)^{|\calY| \,|\calX| \,|\calS^a|}
        \max_{p_{\tby\bx\bs^a} \in \calP_{YXS^a}^{[N]}(\tilde{\epsilon})}
        \exp_2 \left\{ - N
            D(p_{\tby|\bx\bs^a} || p_{Y|XS^a}^* | p_{\bx\bs^a}) \right\}
                                            \nonumber \\
    & \le & (N+1)^{|\calY| \,|\calX| \,|\calS^a|}
        \exp_2 \left\{ - N \tilde{\epsilon} \right\}
\label{eq:Pr-B=1}
\end{eqnarray}
which vanishes as $N \to \infty$. Combining (\ref{eq:Pr-A=1}), (\ref{eq:B=1})
and (\ref{eq:Pr-B=1}), we obtain
\begin{eqnarray*}
   Pr[B=1] & \le & (N+1)^{|\calY| \,|\calX| \,|\calS^a|}
            \,2^{- N \tilde{\epsilon}} + \frac{\eta}{9R} \\
    & \le & \frac{\eta}{8R}
\end{eqnarray*}
for $N$ large enough.

{\bf Step~4}. For any $(f_N, g_N)$, we have
\begin{eqnarray}
  \lefteqn{P_{e,N}(f_N, g_N, (p_{Y|XS^a}^*)^N)} \nonumber \\
    & = & Pr[ \hat{M} \ne M | \tbY, \bS^d] \nonumber \\
    & = & Pr[ \hat{M} \ne M | \tbY, \bS^d, B=0] Pr(B=0)
        + Pr[ \hat{M} \ne M | \tbY, \bS^d, B=1] Pr(B=1) .
                            \label{eq:RHS1}
\end{eqnarray}
Likewise,
\begin{eqnarray}
  \lefteqn{P_{e,N}(f_N, g_N, \pYXSa)} \nonumber \\
    & = & Pr[ \hat{M} \ne M | \bY, \bS^d] \nonumber \\
    & = & Pr[ \hat{M} \ne M | \bY, \bS^d, B=0] Pr(B=0)
        + Pr[ \hat{M} \ne M | \bY, \bS^d, B=1] Pr(B=1) .
                            \label{eq:RHS2}
\end{eqnarray}
Noting that the terms multiplying $Pr[B=0]$ in (\ref{eq:RHS1})
and (\ref{eq:RHS2}) are identical by construction of $\by$ in (\ref{eq:y-CAM})
and that $Pr[B=1]$ is upper bounded by $\frac{\eta}{8R}$, we obtain
\[ |P_{e,N}(f_N, g_N, \pYXSa) - P_{e,N}(f_N, g_N, (p_{Y|XS^a}^*)^N)|
    \le  2 Pr[B=1] \le \frac{\eta}{4R} ,
    \quad \forall f_N, g_N .
\]
Since $R > C(D_1, \calA) + \eta$, Theorem~\ref{thm:C-CDMC} implies that
\[ \lim_{N \to \infty} \min_{f_N,g_N}
    P_{e,N}(f_N, g_N, (p_{Y|XS^a}^*)^N) \ge \frac{\eta}{2R} .
\]
Hence
\[ \lim_{N \to \infty} \min_{f_N,g_N} P_{e,N}(f_N, g_N, \pYXSa)
    \ge \frac{\eta}{2R} - \frac{\eta}{4R} = \frac{\eta}{4R} .
\]
Therefore
\[ \lim_{N \to \infty} \min_{f_N,g_N}
    P_{e,N}(F_N, G_N, \pYXSa) \ge \frac{\eta}{4R} .
\]
for any randomized code $(F_N, G_N)$.

{\bf Step~5}. It remains to prove there was no loss of generality
in making the assumption (\ref{eq:Pr-A=1}). This is done as
follows. Given any code $f_N$ (that may not satisfy
(\ref{eq:Pr-A=1})), we can extend the code by appending $N
\epsilon |\calX|^{-1}$ letters $x$ at the end of the sequence, for
each $x \in \calX$. The resulting code has length $(1+\epsilon)N$
and will be denoted by $\tilde{f}_{(1+\epsilon)N}$. To this code
we can associate a decoding function $\tilde{g}_{(1+\epsilon)N}$
that ignores the last $\epsilon N$ letters of the received
sequence and outputs the same decision as $g_N$ based on the first
$N$ received letters. Hence
\begin{equation}
   P_{e,N}(f_N, g_N, \pYXSa) = P_{e,N}(\tilde{f}_{(1+\epsilon)N},
    \tilde{g}_{(1+\epsilon)N}, \pYXSa) , \quad \forall \pYXSa .
\label{eq:samePe}
\end{equation}
If $\tilde{f}_{(1+\epsilon)N}$ satisfies (\ref{eq:Pr-A=1}),
it follows from Step~4 that reliable communication is impossible using
such codes, and from (\ref{eq:samePe}) the same conclusion applies to $f_N$.

We now show that $\tilde{f}_{(1+\epsilon)N}$ satisfies
(\ref{eq:Pr-A=1}) for $N$ large enough. Denote by $N'(x,s^a)$ the
number of occurrences of the pair $(x,s^a)$ in the last $\epsilon
N$ letters of the joint sequence $(\bx,\bs^a)$. Also denote by
$N'(s^a)$ the number of occurrences of $s^a$ in the last $\epsilon
|\calX|^{-1} \,N$ letters of the sequence $\bs^a$.

If $\calS^a \ne \emptyset$, we have
\[ \forall x,s^a ~: \quad Pr [ N'(x,s^a) \le \epsilon^2 N ]
    = Pr [ N'(s^a) \le \epsilon^2 N ] .
\]
Since $\eE[N'(s^a)] = N \epsilon |\calX|^{-1} p_{S^a}(s^a)$, the above probabilities
vanish exponentially with $N$ provided that
$\epsilon < |\calX|^{-1} \min_{s^a} p_{S^a}(s^a)$. Hence
\begin{eqnarray*}
   Pr[A=1] & = & Pr \left[ \min_{x,s^a} p_{\bx\bs^a}(x,s^a) \le \epsilon^2 \right] \\
        & \le & |\calX| \,|\calS^a| \,\max_{x,s^a}
                Pr[ p_{\bx\bs^a}(x,s^a) \le \epsilon^2] \\
        & \le & |\calX| \,|\calS^a| \,\max_{x,s^a}
                Pr [ N'(x,s^a) \le \epsilon^2 (1+\epsilon) N ] \\
        & = & |\calX| \,|\calS^a| \,\max_{s^a}
                Pr [ N'(s^a) \le \epsilon^2 (1+\epsilon) N ] .
\end{eqnarray*}
Thus (\ref{eq:Pr-A=1}) holds for
$\epsilon (1+\epsilon) < |\calX|^{-1} \min_{s^a} p_{S^a}(s^a)$
and $N$ large enough.

If $\calS^a = \emptyset$, straightforward changes in the above
derivation yield $Pr[A=1] = 0$, i.e., (\ref{eq:Pr-A=1}) holds
again. This concludes the proof. \hfill $\Box$

\Section{Discussion}
\label{sec:discussion}

In their landmark paper, Gel'fand and Pinsker \cite{Gel80} showed
that random binning achieves the capacity of a DMC with random
states known to the encoder. However their encoder was not
designed to provide positive error exponents at rates below
capacity. In this paper we have addressed this limitation and
proposed and optimized a new random-coding scheme. The codebook
consists of a stack of codeword-arrays indexed by the encoder's
state sequence type $\lambda$. The size of these arrays is
$2^{N\rho(\lambda)} \times 2^{NR}$, i.e., the number of rows is a
function of $\lambda$. The decoder is the Maximum Penalized Mutual
Information decoder (\ref{eq:MPMI-0}), where the penalty is the
same function $\rho(\lambda)$ that determines the array sizes.
This new MPMI decoder can be interpreted as an empirical
generalized MAP decoder.

The channel models studied in this paper generalize the original
Gel'fand-Pinsker setup in two ways. First, partial information
about the state sequence is available to the encoder, adversary,
and decoder. Second, both CDMC and CAM channel models are studied.

We have considered four combinations of maximum/expected cost
constraints for the transmitter and CDMC/CAM designs for the
adversary, and obtained the same capacity in all four cases. There
is thus no advantage (in terms of capacity) to the transmitter in
operating under expected-cost constraints instead of the stronger
maximum-cost constraints.

In terms of error exponents however, there is a definite advantage
to the adversary in choosing a CDMC rather than a CAM design of
the channel. This is because 1) arbitrary memory does not help the
adversary because randomly-modulated codes and a MMI-type decoder
are used, 2) the set of conditional types the adversary can choose
from is constrained in the CAM case but not in the CDMC case, and
3) the error exponents are determined by the worst types. The
random-coding exponent is always upper bounded by a straight line
with slope $-1$ at all rates below capacity. That upper bound is
achieved in the CAM case, when no side information is available to
the encoder.

Finally, neither the MMI nor the MPMI decoder is practical, and it
remains to be seen whether good, practical encoders and decoders
can be developed.

\bigskip
{\bf Acknowledgements.} The authors are grateful to M.~Haroutunian,
A.~Lapidoth, N.~Merhav, and P.~Narayan for helpful comments.

\newpage
\appendix
\renewcommand{\theequation}{\Alph{section}.\arabic{equation}}
\Section{Relation Between CAM and AVC Models}
\label{sec:CAM}
In this appendix, we detail the relation between a channel model
$p_{\bY|\bX}$, with maximum distortion constraint (\ref{eq:D2-as}),
and the AVC model in \cite{Csi81}. The AVC is a family
of conditional p.m.f.'s $W(y|x,\theta)$, where $\theta \in \Theta$
(finite set) is a ``channel state'' selected by the adversary.
A cost function $l~:~ \Theta \to \rR^+$ for the states is also defined.
The channel law is of the form
\begin{equation}
   p(\by|\bx,\btheta) = \prod_{i=1}^N W(y_i|x_i,\theta_i)
\label{eq:AVC-Csi}
\end{equation}
where the sequence $\btheta = \{ \theta_1, \cdots, \theta_N \}$
is arbitrary except for a maximum-cost constraint
\begin{equation}
   l^N(\btheta) \triangleq
    \frac{1}{N} \sum_{i=1}^N l(\theta_i) \le l_{\max} .
\label{eq:AVC-cost}
\end{equation}
In some formulations of the jamming problem, $\btheta$ must be selected
by the adversary before seeing $\bx$; in other formulations, $\theta_i$
is allowed to depend on $x_i$ but not on other samples of $\bx$
\cite{Csi88,Hug96}; yet in other formulations (the A*VC model \cite{Csi81}),
$\theta_i$ is allowed to depend on $x_j$ for all $j \le i$.

If $\btheta$ is allowed to depend on the entire sequence $\bx$ in
a noncausal manner (as opposed to the above formulations of the
AVC problem), the problem with maximum distortion constraint
(\ref{eq:D2-as}) may be formulated as (\ref{eq:AVC-Csi}) and
(\ref{eq:AVC-cost}) with state $\theta$, channel $W$, and cost $l$
defined below. Let $\theta = (\theta',\theta'')$ where $\theta'
\in \calX$ and $\theta'' \in \calY$, hence $\Theta = \calX \times
\calY$. Let
\[ l(\theta) = d(\theta', \theta''),
    \qquad W(y|x,\theta) = \iI\{y=\theta''\} .
\]
The maximum-cost constraint (\ref{eq:AVC-cost}) is then equivalent
to the maximum-distortion constraint (\ref{eq:D2-as}), with $l_{\max} = D_2$.
The sequence $\btheta''$ may be chosen deterministically or stochastically,
using an arbitrary distribution.

\Section{Error Exponents for Channels Without Side Information}
\label{sec:DMC}

This appendix summarizes some known results on random-coding error exponents.
\begin{description}
\item[Single DMC:] Let $p_{Y|X}$ and $p_X$ be the channel law and input
p.m.f., respectively. Referring to \cite[p.~165-166]{Csi81}, we have
\begin{eqnarray}
   E_r(R, p_X, p_{Y|X}) & = & \min_{\tp_{Y|X}} [ D(\tp_{Y|X} || p_{Y|X} \,|~p_X)
    + | \tilde{I}_{XY}(p_X, \tp_{Y|X}) - R |^+ ] ,  \label{eq:Er-DMC0}
\end{eqnarray}
\item[Compound DMC:] Here $p_{Y|X}$ belongs to a set $\calA$. We have
\begin{eqnarray}
   E_r(R,p_X,\calA) & = & \min_{p_{Y|X} \in \calA}
        E_r(R, p_X, p_{Y|X})            \nonumber \\
    & = & \min_{p_{Y|X} \in \calA} \,\min_{\tp_{Y|X}}
        [ D(\tp_{Y|X} || p_{Y|X} |~\,p_X)
        + | \tilde{I}_{XY}(p_X, \tp_{Y|X}) - R |^+
                ]       \label{eq:Er-compoundDMC}
\end{eqnarray}
which is zero if $R \ge \min_{p_{Y|X} \in \calA} \tilde{I}_{XY}(p_X, p_{Y|X})$.
\item[Private Watermarking:] the set $\calA$ is defined by the distortion
constraint (\ref{eq:D2-avg}). Then \cite{Som03}
\begin{eqnarray}
   E_r^{\CAM}(R,D_1,D_2) & = & \max_{p_{X|S}} \min_{p_{Y|XS} \in \tilde{\calA}}
        [ \tilde{I}_{SY|X}(p_S, p_{X|S}, p_{Y|XS})  \nonumber \\
        && \mbox{\hspace*{1in}}
        + |\tilde{I}_{XY|S}(p_S, p_{X|S}, p_{Y|XS}) - R |^+
                ]           \label{eq:Er-privateWM}
\end{eqnarray}
where $\tilde{\calA} \triangleq \{ p_{Y|XS} ~:~
\sum_s p_S(s) p_{X|S}(x|s) p_{Y|XS}(y|x,s) d(x,y) \le D_2 \}$.
The maximization over $p_{X|S}$ is also subject to a distortion constraint.
\item[Jamming] with channel state $S$ selected independently of input $X$
    \cite{Csi88,Hug96}. We have
\begin{eqnarray}
   E_r^{jam}(R) & = & \max_{p_X} \min_{p_S}
        \min_{\mbox{\small $\begin{array}{c}
        \tp_{YSX} ~: \\ \tp_X = p_X, \tp_S = p_S \end{array}$}}
        [ D(\tp_{YSX} || p_{Y|SX} p_X p_S)
        + |\tilde{I}_{XY}(p_X, \tp_{Y|X}) - R|^+ ]  \nonumber \\
        &&                          \label{eq:Er-jam}
\end{eqnarray}
\end{description}

\Section{Proof of Proposition~\ref{prop:pub-C}}
\label{sec:pub-C}

The set $\calA$, denoted here as $\calP_{Y|X}(D_2)$, is the set
of DMC's that introduce maximum Hamming distortion $D_2$.
Let the attack channel $p_{Y|X}^*$ be the BSC
with crossover probability $D_2$. Considering $p_{Y|X}^*$
may not be the worst channel, we have
\begin{eqnarray}
C^{\mathrm{\mathrm{pub}}} &=& \sup_L \max_{p_{XU|S} \in \calP_{XU|S}(L,D_1)}
    \min_{p_{Y|X} \in \calP_{Y|X}(D_2)}
    J_L(p_S \,p_{XU|S} \,p_{Y|X})           \nonumber\\
&\le& \sup_L \max_{p_{XU|S} \in \mathcal P_{XU|S}(L,D_1)}
    J_L(p_S \,p_{XU|S} \,p_{Y|X}^*)             \nonumber\\
&=& g^*(D_1,D_2),
\label{eq:CleG}
\end{eqnarray}
where the last step is derived in \cite{Bar03,Pra03}.
The function $g^*$ is defined in (\ref{eq:C-pub}).

Next we prove that $C^{\mathrm{\mathrm{pub}}}\ge g^*(D_1,D_2)$.
Consider $D_1=D_1^\prime \theta$,
where $D_1^\prime \in [0,\frac{1}{2}]$ and $\theta \in [0,1]$. Let
$p_{U|S}^*$ be the BSC with crossover probability $D_1^\prime$.
Furthermore, $X=U$ makes the distortion equal to $D_1^\prime$.
(Note that $L = |\calU| = 2$ in this case.)
Clearly,
\begin{eqnarray}
C^{\mathrm{\mathrm{pub}}}(D_1^\prime) &\ge & \min_{p_{Y|X} \in \calP_{Y|X}(D_2)}
    J_L(p_S \,p_{XU|S}^* \,p_{Y|X})\nonumber\\
&=& \min_{p_{Y|X} \in \calP_{Y|X}(D_2)} I(X;Y)
    -\underbrace{(1-h(D_1^\prime))}_{I(U;S)}\nonumber\\
&=& (1-h(D_2))-(1-h(D_1^\prime))\nonumber\\
&=& h(D_1^\prime)-h(D_2), 
\end{eqnarray}
where
\[ \min_{p_{Y|X} \in \calP_{Y|X}(D_2)} I(X;Y) = 1-h(D_2) \]
is achieved by $p^*_{Y|X}$.

Using time-sharing arguments, Barron {\em et al.} \cite{Bar03}
proved that capacity is a concave function of $D_1$
in the case $\calA = \{ p_{Y|X}^* \}$. It can be shown
that their result holds in the case $\calA = \calP_{Y|X}(D_2)$
considered here. Therefore we have
\[C^{\mathrm{\mathrm{pub}}}(D_1) = C^{\mathrm{\mathrm{pub}}}(D_1^\prime \theta)
    \ge \theta C^{\mathrm{\mathrm{pub}}}(D_1^\prime)
    \ge \theta \left(h(D_1^\prime)-h(D_2)\right) ,
    \quad \forall \theta \in [0,1] .
\]
It may be verified that
\[ \max_{0\le \theta \le 1}\theta \left(h(D_1^\prime)-h(D_2)\right)
    = g^*(D_1,D_2) .
\]
Therefore
\begin{equation}
C^{\mathrm{\mathrm{pub}}} \ge g^*(D_1,D_2).
\label{eq:CgeG}
\end{equation}

From (\ref{eq:CleG}) and (\ref{eq:CgeG}), we conclude that
$C^{\mathrm{\mathrm{pub}}}=g^*(D_1,D_2)$; also $|\calU| = 2$.
\hfill $\Box$
\Section{Proof of Proposition~\ref{prop:pub-Er-CAM}}
\label{sec:pub-Er-CAM}
From (\ref{eq:Er-CAM}), we have
\begin{eqnarray}
E_r^{\mathrm{CAM,pub}}(R) &=& \sup_L \,\min_{\tp_S}
    \max_{p_{XU|S} \in \calP_{XU|S}(L,D_1)}
    \min_{p_{Y|XUS} \in \calP_{Y|XUS}(D_2)} \Big[
    D(\tilde p_S||p_S)+I_{Y;US|X}(\tp_S \,p_{XU|S} \,p_{Y|XUS})
                                    \nonumber \\
& & \mbox{\hspace*{0.25in}}+|J_L(\tp_S \,p_{XU|S} \,p_{Y|XUS})-R|^+ \Big] .
\label{eq:CAM-RC-Public}
\end{eqnarray}

{\bf Step~1}.
First we prove that
\begin{eqnarray}
   F(D_1,D_2) & \triangleq & \sup_L \,\min_{\tp_S}
    \max_{p_{XU|S}\in \calP_{XU|S}(L,D_1)}
    \min_{p_{Y|XUS}\in \calP_{Y|XUS}(D_2)}
    J_L(\tilde p_S \,p_{XU|S} \,p_{Y|XUS})      \nonumber \\
    & = & C^{\mathrm{\mathrm{pub}}} ,
\label{eq:L2}
\end{eqnarray}
with equality if $\tp_S=p_S$.

Referring to (\ref{eq:C-pub}), we first consider the regime
in which time sharing is not needed:
$D_1 \ge \delta_2 = 1 - 2^{- \overline{h}(D_2)}$ and therefore
$C^{\mathrm{\mathrm{pub}}}=\overline{h}(D_1)-\overline{h}(D_2)$.
Letting $U=X$ and $p^*_{X|S}$ be the BSC with crossover probability $D_1$,
we obtain a lower bound on $F(D_1,D_2)$:
\begin{eqnarray}
F(D_1,D_2)
&\ge& \min_{\tp_S} \min_{p_{Y|XS}\in \calP_{Y|XS}(D_2)}
    J_L(\tp_S \,p_{X|S}^* \,p_{Y|XS})           \nonumber \\
&=& \min_{\tp_S} \min_{p_{Y|XS}\in \calP_{Y|XS}(D_2)}
    \Big[ \tilde I_{X;Y}(\tp_S \,p^*_{X|S} \,p_{Y|XS})
    - \tilde I_{X;S}(\tp_S \,p_{X|S}^*) \Big]       \nonumber \\
&=& \min_{\tp_S} \min_{p_{Y|XS}\in \calP_{Y|XS}(D_2)}
    \Big[ \tilde I_{X;Y}(\tilde p_S \,p^*_{X|S} \,p_{Y|XS})
    -\left( \overline{h}(D_1 \star p_0) - \overline{h}(D_1) \right)\Big]
\label{eq:lemma2-1}
\end{eqnarray}
where we use the shorthand $p_0 = \tp_S(0)$. Next, write $p_{Y|XS}$ as
\begin{center}
\begin{tabular}{c|c|c|c|c}
  $p_{Y|XS}$ & $XS=00$ & $XS=10$ & $XS=01$ & $XS=11$\\ \hline
  $Y=0$ & $1-e$ & $f$ & $1-g$ & $h$ \\
  $Y=1$ & $e$ & $1-f$ & $g$ & $1-h$ \\
 \end{tabular}.
\end{center}
The p.m.f. of $X$ induced by $\tp_S$ and $p^*_{X|S}$ is given by
\begin{eqnarray*}
   p_X & = & (p_{X0}, p_{X1}) \\
     & = & \left(p_0(1-D_1)+(1-p_0)D_1, \,p_0 D_1 +(1-p_0)(1-D_1)\right).
\end{eqnarray*}
We derive
\begin{eqnarray}
\lefteqn{\min_{p_{Y|XS}\in \calP_{Y|XS}(D_2)}
    \tilde I_{X;Y}(\tp_S \,p^*_{X|S} \,p_{Y|XS})}        \nonumber\\
&=& \min_{e,f,g,h: \atop p_0((1-D_1)e+D_1 f)+(1-p_0)((1-D_1)h+D_1 g)\le D_2}
    \overline{h}(p_{X0}(1-\alpha)+p_{X1} \beta)
    -p_{X0}\overline{h}(1-\alpha)-p_{X1}
    \overline{h}(\beta)\nonumber\\
&=& \overline{h}(D_1 \star p_0) - \overline{h}(D_2)
\label{eq:lemma2-2}
\end{eqnarray}
where
\[ \alpha=\frac{p_0(1-D_1)e+(1-p_0) D_1 g}{p_{X0}} \quad \mathrm{and}
    \quad \beta=\frac{p_0 D_1 f+(1-p_0)(1-D_1)h}{p_{X1}} .
\]
The minimum is achieved by
\[ \alpha^*=\frac{D_2}{p_{X0}}\frac{p_{X1}-D_2}{1-2D_2} , \quad
    \beta^*=\frac{D_2}{p_{X1}}\frac{p_{X0}-D_2}{1-2D_2}.
\]

Combining (\ref{eq:lemma2-1}) and (\ref{eq:lemma2-2}), we obtain
\begin{equation}
   F(D_1,D_2) \ge \overline{h}(D_1)-\overline{h}(D_2) = C^{\mathrm{\mathrm{pub}}} .
\label{eq:L2-lb}
\end{equation}

In the case $D_1 < \delta_2$, capacity is achieved using
time-sharing: $C^{\mathrm{\mathrm{pub}}} >
\overline{h}(D_1)-\overline{h}(D_2)$. Similarly to \cite{Bar03},
it can be shown that $F(D_1,D_2)$ is a nondecreasing concave
function of $D_1$. Hence,
\begin{equation}
   F(D_1,D_2) = F(D_1^\prime \theta,D_2)\ge \max_{0 \le \theta \le 1}
    \theta F(D_1^\prime,D_2) \ge \max_{0 \le \theta \le 1} \theta
    \left( \overline{h}(D_1^\prime)-\overline{h}(D_2) \right)=C^{\mathrm{\mathrm{pub}}}.
\label{eq:L2-TS}
\end{equation}

For all values of $D_1$, letting $\tp_S=p_S$ in (\ref{eq:L2}) and
further restricting the minimization over $p_{Y|XUS}$, we have
\begin{equation}
   F(D_1,D_2) \le \sup_L \max_{p_{XU|S}\in \calP_{XU|S}(L,D_1)}
    \min_{p_{Y|X}\in \calP_{Y|X}(D_2)} J_L(\tp_S \,p_{XU|S} \,p_{Y|X})
    = C^{\mathrm{\mathrm{pub}}}.
\label{eq:L2-ub}
\end{equation}
Combining (\ref{eq:L2-lb}), (\ref{eq:L2-TS}) and (\ref{eq:L2-ub}),
we obtain (\ref{eq:L2}).

{\bf Step~2}. The first two bracketed terms in
(\ref{eq:CAM-RC-Public}) are nonnegative. This yields a lower
bound on $E_r^{\mathrm{CAM,pub}}(R)$:
\begin{eqnarray}
E_r^{\mathrm{CAM,pub}}(R)
&\ge& \sup_L \min_{\tp_S} \max_{p_{XU|S} \in \calP_{XU|S}(L,D_1)}
    \min_{p_{Y|XUS} \in \calP_{Y|XUS}(D_2)}
    |J_L(\tp_S \,p_{XU|S} \,p_{Y|XUS})-R|^+     \nonumber\\
&=& |C^{\mathrm{\mathrm{pub}}}-R|^+
\label{eq:proposition 5-2}
\end{eqnarray}
where the equality is due to (\ref{eq:L2}).

{\bf Step~3}. If we fix $\tp_S=p_S=(\frac{1}{2},\frac{1}{2})$ and
restrict $p_{Y|SUX}$ to be of the form $p_{Y|X}$, we obtain an
upper bound on $E_r^{\mathrm{CAM,pub}}(R)$:
\begin{eqnarray}
E_r^{\mathrm{CAM,pub}}(R)
&\le& \sup_L \max_{p_{XU|S} \in \calP_{XU|S}(L,D_1)}
    \min_{p_{Y|X} \in \calP_{Y|X}(D_2)}
    |J_L(p_S \,p_{XU|S} \,p_{Y|X})-R|^+         \nonumber \\
&=& |C^{\mathrm{\mathrm{pub}}}-R|^+ .
\label{eq:proposition 5-1}
\end{eqnarray}

{\bf Step~4}. Combining (\ref{eq:proposition 5-1}) and
(\ref{eq:proposition 5-2}), we obtain
$E_r^{\mathrm{CAM,pub}}(R)=|C^{\mathrm{\mathrm{pub}}}-R|^+$.
\hfill $\Box$

\Section{Proof of Proposition~\ref{prop:deg-C}} \label{sec:deg-C}

We have
\begin{equation}
C^{\mathrm{deg}} = \max_{p_X \in \calP_X(D_1) }
    \min_{p_{Y|X} \in \calP_{Y|XS}(D_2)}I(X;Y).
\label{eq:CapacityDeg}
\end{equation}

Let $a = p_X(1)$, $e = p_{Y|X}(1|0)$, and $f = p_{Y|X}(0|1)$,
which satisfy the distortion constraints
\[ a \le D_1 , \quad (1-a)e + af \le D_2 . \]
Substituting these probabilities into (\ref{eq:CapacityDeg}), we obtain
\[
C^{\mathrm{deg}} = \max_{a \le D_1} \min_{(1-a)e+af \le D_2} \left
    [\overline{h}((1-a)(1-e)+af)
    -(1-a)\overline{h}(e)-a\overline{h}(f) \right].
\]

Solving the above max-min problem in the case $D_1 \ge \delta_2 =
1 - 2^{\overline{h}(D_2)}$, we obtain the optimal $p^*_X$ and
$p_{Y|X}^*$ from
\[
    a = D_1, \quad
    e = \frac{D_2(D_1-D_2)}{(1-D_1)(1-2D_2)} , \quad
    f = \frac{D_2(1-D_1-D_2)}{D_1(1-2D_2)} .
\]
After some algebraic simplifications, we obtain
$C^{\mathrm{deg}} = \overline{h}(D_1)-\overline{h}(D_2)$.
Applying the same time-sharing argument as in the proof of
Prop.~\ref{prop:pub-C}, we obtain $C^{\mathrm{deg}}=g^*(D_1,D_2)$, which is
the same as the capacity $C^{\mathrm{\mathrm{pub}}}$ for the public watermarking game.
\hfill $\Box$

\Section{Proof of \protect (\ref{eq:bound2})} \label{sec:bound2}

The inequality $1 - \prod_i (1-\alpha_i)^{t_i} \le 1$ being
trivial, it remains to prove that
\[ 1 - \prod_i (1-\alpha_i)^{t_i} \le \sum_i \alpha_i t_i  \]
or equivalently,
\[ \prod_i (1-\alpha_i)^{t_i} \ge 1 - \sum_i \alpha_i t_i  . \]
Define the $K$-vector $\one$ whose components are all equal to 1,
the $K$-vectors $\balpha$ and $\bt$ with components $\{ \alpha_i \}$
and $\{ t_i \}$, and $\Omega = [1, \infty)^K$, the domain of $\bt$.
Denote by $\nabla f$ the gradient vector of a function $f$
defined on $\rR^K$, and by $\mathbf{a} \cdot \mathbf{b}$ the dot product
of two vectors in $\rR^K$.
Define the functions
\begin{eqnarray}
   f(\bt) & = & \prod_i (1-\alpha_i)^{t_i} \label{eq:f} \\
   g(\bt) & = & f(\one) + (\bt - \one) \cdot \nabla f(\one) \label{eq:g} \\
   h(\bt) & = & 1 - \balpha \cdot \bt       \label{eq:h}
\end{eqnarray}
We need to prove that $f(\bt) \ge h(\bt)$ for all $\bt \in \Omega$
and $\balpha \in [0,1]^K$.
In Step~1 below we establish that $f(\bt) \ge g(\bt)$.
In Step~2, we prove that $g(\bt) \ge h(\bt)$.

{\bf Step~1}. The function $g(\bt)$ describes a hyperplane tangent
to the graph of $f(\bt)$ at $\bt=\one$. The function $f(\bt)$ may
be written as
\begin{equation}
   f(\bt) = \exp \left\{ \sum_i t_i \ln(1-\alpha_i) \right\} .
\label{eq:f2}
\end{equation}
It is convex and therefore $f(\bt) \ge g(\bt)$,
owing to the hyperplane separation theorem.

{\bf Step~2}. From (\ref{eq:g}) and (\ref{eq:h}), we have
\begin{equation}
   g(\bt) - h(\bt) = [f(\one)-h(\one)] + \sum_i (t_i-1) \left[
    \left. \frac{\partial f(\bt)}{\partial t_i} \right|_{\bt=\one}
    + \alpha_i \right] .
\label{eq:g-h}
\end{equation}
Observe that $f(\one) = \prod_i (1-\alpha_i)$ and
$h(\one) = 1 - \sum_i \alpha_i$;
therefore we have the well-known inequality $f(\one) \ge h(\one)$.
Next, since each term $t_i - 1$ in (\ref{eq:g-h}) is nonnegative
for $\bt \in \Omega$, it suffices to prove that
\begin{equation}
   \left. \frac{\partial f(\bt)}{\partial t_i} \right|_{\bt=\one}
    \ge - \alpha_i
\label{eq:ineq-gh0}
\end{equation}
to establish that $g(\bt) - h(\bt) \ge 0$ for all $\bt \in \Omega$.

From (\ref{eq:f2}), we obtain
\begin{equation}
   \left. \frac{\partial f(\bt)}{\partial t_i} \right|_{\bt=\one}
    = f(\one) \ln(1-\alpha_i) , \quad 1 \le i \le K , \quad \forall \bt ,
\label{eq:del-f}
\end{equation}
where $0 \le f(\one) \le 1-\alpha_i$.
Since $\ln(1-\alpha_i) \le 0$, this implies
\begin{equation}
   f(\one) \ln(1-\alpha_i) \ge (1-\alpha_i) \ln(1-\alpha_i) .
\label{eq:ineq-gh1}
\end{equation}
We now prove that
\begin{equation}
   (1-\alpha_i) \ln(1-\alpha_i) \ge - \alpha_i
\label{eq:ineq-gh2}
\end{equation}
which, combined with (\ref{eq:ineq-gh1}) and (\ref{eq:del-f}),
will establish (\ref{eq:ineq-gh0}).
Putting $x = 1-\alpha_i$, we apply the inequality
$\ln \frac{1}{x} \le \frac{1}{x} -1$ to claim that
\[ x \ln x = - x \ln \frac{1}{x} \ge -x \left(\frac{1}{x} -1\right) = x-1 \]
which proves (\ref{eq:ineq-gh2}).
The proof is complete.
\hfill $\Box$

\Section{Proof of Proposition~\ref{prop:cara-C}}
\label{sec:cara-C}

The variational distance for $p, p' \in \calP_Y$ is defined as
$\| p-p' \| = \sum_y |p(y)-p'(y)|$ and extended
to conditional pmf's $p, p' \in \calP_{Y|X}$ as
$\| p-p' \| = \max_x \sum_y |p(y|x)-p'(y|x)|$.

\begin{lemma} \cite[p.~33]{Csi81}.
For any $p, p' \in \calP_Y$, we have
\[ \| p-p' \| \le \theta \le \frac{1}{2} \quad \Rightarrow \quad
	|\tilde{H}_{p}(Y) - \tilde{H}_{p'}(Y)| \le \theta \log \frac{|\calY|}{\theta}
\]
\label{lem:csiszar1}
\end{lemma}

\begin{lemma}
For any $p_X \in \calP_X$ and $p, p' \in \calP_{Y|X}$, we have
\[ \| p-p' \| \le \theta \le \frac{1}{2} \quad \Rightarrow \quad
	|\tilde{H}_{p_X p}(Y|X) - \tilde{H}_{p_X p'}(Y|X)| 
		\le \theta \log \frac{|\calY|}{\theta} .
\]
\label{lem:csiszar2}
\end{lemma}
{\em Proof}:
\begin{eqnarray*}
   |\tilde{H}_{p_X p}(Y|X) - \tilde{H}_{p_X p'}(Y|X)| 
	& = & \left| \sum_x p_X(x) [\tilde{H}_{p}(Y|X=x) - \tilde{H}_{p'}(Y|X=x) ]
										\right| \\
	& \le & \max_x |\tilde{H}_{p}(Y|X=x) - \tilde{H}_{p'}(Y|X=x)| \\
	& \le & \theta \log \frac{|\calY|}{\theta}
\end{eqnarray*}
where the last inequality follows from Lemma~\ref{lem:csiszar1}. \\

\noindent
{\bf Proof of Proposition~\ref{prop:cara-C}.}

The upper bound is straightforward. We now derive the lower bound.

{\bf Step 1}.
Define a discretized set $\calA_l \subset \calA$ of attack channels as follows:
\[ \calA_l = \left\{ q \in \calA ~:~ 
	q(y|x,s^a) \in \{ 0, l^{-1}, 2\,l^{-1}, \cdots, 1 \} 
	\,\, \forall y,x,s^a \right\} . 
\]

{\bf Step 2}.
Any attack channel $p \in \calA$ satisfies
\[ \sum_x p_X(x) d_2(x) \le D_2 \]
where
\begin{equation}
   d_2(x,s^a) \triangleq \sum_y p(y|x,s^a) d(x,y)
\label{eq:d2}
\end{equation}
is the distortion introduced by $p$ when $X=x$ and $S^a=s^a$.

We construct an approximation $\hat{p}$ of $p$
with the following properties:
\begin{eqnarray}
   \sum_y \hat{p}(y|x,s^a) & = & 1		\label{eq:p-hat-property1} \\
   \sum_y \hat{p}(y|x,s^a) d(x,y) & \le & d_2(x,s^a) , \quad \forall x,s^a,
							\label{eq:p-hat-property2}
\end{eqnarray}
hence $\hat{p} \in \calA_l$.
The construction of $\hat{p}$ is as follows.
Define $p^+(y|x,s^a)$ as the least upper bound
on $p(y|x,s^a)$ in the set $\{ l^{-1}, 2\,l^{-1}, \cdots, 1 \}$.
Therefore $0 < p^+(y|x,s^a) - p(y|x,s^a) \le l^{-1}$. Define 
\begin{eqnarray}
   k & = & \sum_y (l\,p^+(y|x,s^a) - l\,p(y|x,s^a)) \nonumber \\
	& = & l \sum_y p^+(y|x,s^a) - l  
			\quad \in \{ 0, 1, \cdots, |\calY|-1 \} ,
\label{eq:k}
\end{eqnarray}
$d_k(x)$ as the sum of the $k$ largest values of $d(x,y)$ when $y$
ranges over $\calY$, and $\calY_k(x)$ as the set of $k$ corresponding
values of $y$, that is, we have $d_k(x) = \sum_{y \in \calY_k(x)} d(x,y)$.
Now let
\begin{equation}
   \hat{p}(y|x,s^a) = p^+(y|x,s^a) - \frac{1}{l} \,1_{\{y \in \calY_k\}} .
\label{eq:p-hat}
\end{equation}

We have the following properties:
\[ \sum_y \hat{p}(y|x,s^a) = \sum_y p^+(y|x,s^a) - \frac{k}{l} = 
	\sum_y p(y|x,s^a) = 1 ; \] 
therefore (\ref{eq:p-hat-property1}) holds.
Also
\begin{eqnarray*}
   \sum_y d(x,y) \hat{p}(y|x,s^a) = \sum_y d(x,y) p(y|x,s^a)
	&  & - \sum_y d(x,y) (p^+(y|x,s^a) - \hat{p}(y|x,s^a)) \\
	& & \qquad + \sum_y d(x,y) (p^+(y|x,s^a) - p(y|x,s^a)) .
\end{eqnarray*}
From (\ref{eq:d2}), the first sum in the right side is equal to $d_2(x,s^a)$.
Owing to (\ref{eq:p-hat}), the second sum is equal to
\[ \sum_y d(x,y) \frac{1}{l} \,1_{\{y \in \calY_k\}} = \frac{1}{l} d_k(x) . \]
The third sum is equal to
\[ \frac{1}{l} \sum_y d(x,y) (l\,p^+(y|x,s^a) - l\,p(y|x,s^a)) 
	\le  \frac{1}{l} d_k(x) \]
where the inequality holds because
\[ l\,p^+(y|x,s^a) - l\,p(y|x,s^a) \le 1 \quad
	\mathrm{and} \quad \sum_y (l\,p^+(y|x,s^a) - l\,p(y|x,s^a)) = k .
\]
Hence (\ref{eq:p-hat-property2}) holds as well, and $\hat{p} \in \calA_l$.
The cardinality of $\calA_l$ is at most
$(l+1)^{|\calY|\,|\calX|\,|\calS^a|}$.

By construction of $\hat{p}$, we have
\[
  \| p_{Y|XS^a} - \hat{p}_{Y|XS^a}\| \le \frac{|\calY|}{l} \triangleq \theta .
\]

{\bf Step~3}.
Consider an alphabet $\calU$ of arbitrarily large cardinality.
We have
\[ J_{|\calU|}(\cdot) = I(U;YS^d) - I(U;S^e) 
	= H(Y|S^d) - H(Y|US^d) + I(U;S^d) - I(U;S^e) , 
\]
hence
\begin{eqnarray}
   \lefteqn{|J_{|\calU|}(p_S \,p_{XU|S^e} \,p_{Y|XS^a}) 
		- J_{|\calU|}(p_S \,p_{XU|S^e} \,\hat{p}_{Y|XS^a})|} 	\nonumber \\
	& = &  |\tilde{H}_{p_{Y|XS^a}}(Y|S^d) - \tilde{H}_{p_{Y|XS^a}}(Y|US^d) 
			- \tilde{H}_{\hat{p}_{Y|XS^a}}(Y|S^d) 
			+ \tilde{H}_{\hat{p}_{Y|XS^a}}(Y|US^d)| 		\nonumber \\
	& \le & |\tilde{H}_{p_{Y|XS^a}}(Y|S^d) - \tilde{H}_{\hat{p}_{Y|XS^a}}(Y|S^d)|
			+ |\tilde{H}_{p_{Y|XS^a}}(Y|US^d) - \tilde{H}_{\hat{p}_{Y|XS^a}}(Y|US^d)|
											\nonumber \\
	& \le & 2 \,\theta \log \frac{|\calY|}{\theta} 
		= 2 |\calY| \,\frac{\log l}{l}
\label{eq:delta-J}
\end{eqnarray}
where the second inequality is obtained by application of Lemmas~\ref{lem:csiszar1}
and \ref{lem:csiszar2}.

{\bf Step~4}.
By application of Caratheodory's theorem, given a pmf $p_{XUS^e}$ where
$\calU$ has arbitrarily large cardinality, and given $L$ real-valued functionals
$f_i, 1 \le i \le L$ defined over the set $\calP_{XS^e}$, there exist $L$
elements $u_1, \cdots, u_L$ of $\calU$ and $L$ nonnegative numbers 
$\alpha_1, \cdots, \alpha_L$ summing to 1 such that
\[ \sum_{u \in \calU} p_U(u) f_i(p_{XS^e|U=u}) 
	= \sum_{u=1}^L \alpha_u f_i(p_{XS^e|U=u}) , \quad i=1,2, \cdots, L. \]


The payoff function in the mutual-information game takes the form
\begin{eqnarray*}
   J_{|\calU|}(p_S p_{XU|S^e} p_{Y|XS^a}) 
	& = & I(U;YS^d) - I(U;S^e) \\
	& = & \sum_{u \in \calU} p_U(u) [- H(YS^d|U=u) + H(S^e|U=u)] 
		+ H(YS^d) - H(S^e) .
\end{eqnarray*}
We apply Caratheodory's theorem to our problem by letting 
\begin{eqnarray*}
   f_i(p_{XS^e|U=u}) & = & p_{XS^e}(x,s^e) , \mbox{\hspace*{1.4in}}
				\quad 1 \le i(x,s^e) \le |\calX| \,|\calS^e| - 1, \\
   f_i(p_{XS^e|U=u}) & = & H(YS^d|U=u) - H(S^e|U=u) , 
				\quad |\calX| \,|\calS^e| \le i(p_{Y|XS^a}) \le
					|\calX| \,|\calS^e| + |\calA_l| - 1 .
\end{eqnarray*}
The first $|\calX| \,|\calS^e| - 1$ functions correspond to the marginals of
$p_{XS^e}$ except one, and the next $|\calA_l| = l^{|\calY|\,|\calX|\,|\calS^a|}$
functions are indexed by the attack channels $p_{Y|XS^a} \in \calA_l$.
Hence, defining $\calU' = \{ 1, \cdots, L \}$, there exist $L$ nonnegative
numbers $\alpha_1, \cdots, \alpha_L$ summing to 1 and a random variable
$U' \in \calU'$ such that
\begin{eqnarray*}
   p_{XU'S^e}(x,u',s^e) & = & p_{XS^e|U}(x,s^e|u_{u'}) \,\alpha_{u'} 
							\qquad \forall x, s^e \\
   J_L(p_S p_{XU'|S^e} p_{Y|XS^a}) & = & J_{|\calU|}(p_S p_{XU|S^e} p_{Y|XS^a}) 
							\quad \forall p_{Y|XS^a} \in \calA_l .
\end{eqnarray*}
Hence it suffices to consider
\[ |\calU| = L = |\calX|\,|\calS^e| + l^{|\calY|\,|\calX|\,|\calS^a|} - 1 , \]
as stated in (\ref{eq:L-cara-capacity}), to achieve the maximum in
\[ \max_{p_{XU|S^e} \in \calP_{XU|S^e}(|\calU|,D_1)} 
	\,\min_{p_{Y|XS^a} \in \calA_l} J_L(p_S p_{XU|S^e} p_{Y|XS^a}) . \]

{\bf Step~5}.
For any choice of $\calU$, we have
\begin{eqnarray}
   C_L & \stackrel{(a)}{=} & \max_{p_{XU|S^e} \in \calP_{XU|S^e}(L,D_1)}
		\min_{p_{Y|XS^a} \in \calA} J_L(p_S p_{XU|S^e} p_{Y|XS^a}) 		
											\nonumber \\
	& \stackrel{(b)}{\ge} & \max_{p_{XU|S^e} \in \calP_{XU|S^e}(L,D_1)}
		\min_{p_{Y|XS^a} \in \calA_l} J_L(p_S p_{XU|S^e} p_{Y|XS^a})
		- 2 |\calY| \,\frac{\log l}{l}				\nonumber \\
	& \stackrel{(c)}{=} & \max_{p_{XU|S^e} \in \calP_{XU|S^e}(|\calU|,D_1)} 
			\min_{p_{Y|XS^a} \in \calA_l}
			J_{|\calU|}(p_S p_{XU|S^e} p_{Y|XS^a})
			- 2 |\calY| \,\frac{\log l}{l} 			\nonumber \\
	& \stackrel{(d)}{\ge} & \max_{p_{XU|S^e} \in \calP_{XU|S^e}(|\calU|,D_1)} 
			\min_{p_{Y|XS^a} \in \calA}
			J_{|\calU|}(p_S p_{XU|S^e} p_{Y|XS^a})
			- 2 |\calY| \,\frac{\log l}{l}
\label{eq:delta-C}
\end{eqnarray}
where (a) is the definition of $C_L$;
(b) is because (\ref{eq:delta-J}) holds uniformly for all $p_{XU|S^e}$
and $p_{Y|XS^a} \in \calA$; and
(c) is a consequence of Caratheodory's theorem in Step~4; and
(d) holds because $\calA_l \subset \calA$.

The inequality (\ref{eq:delta-C}) holds in the limit as $|\calU| \to \infty$, hence
\[
   C_L \ge C - 2 |\calY| \,\frac{\log l}{l} ,
\]
which proves the claim.
\hfill $\Box$

\Section{Proof of Proposition~\ref{prop:cara-Er}}
\label{sec:cara-Er}

Define sets 
$\calP_Y(\epsilon) = \{ p_Y \in \calP_Y~:~ \min_y p(y) \ge \epsilon \}$
and similarly, $\calP_{Y|X}(\epsilon) = 
\{ p_{Y|X} \in \calP_{Y|X} ~:~ \min_{x,y} p(y|x) \ge \epsilon \}$,
for any $\epsilon \in [0, 1/|\calY|]$.
In preparation for the proof of the proposition, we define a log-uniform
quantizer $\Phi_l$ and present three lemmas.

{\bf Pmf quantization}.
Given $l \ge |\calY|$, we define $\epsilon = l^{-1}$ and
a pmf quantization mapping $\Phi_l ~:~ \calP_Y(\epsilon) \to \calP_Y(\epsilon)$
as follows. Define the log-uniform quantizer $Q_l~:~[\epsilon,1] \to \calQ_l$ 
with $l$ reproduction levels
\begin{equation}
  \calQ_l = \{ \epsilon = \epsilon^{l\epsilon}, \epsilon^{(l-1)\epsilon} , 
	\cdots, \epsilon^{2\epsilon}, \epsilon^\epsilon \} .
\label{eq:Ql}
\end{equation}
and quantization function
\begin{equation}
   Q_l(z) = \left\{ \begin{array}{ll}
			\epsilon & :~ z = \epsilon \\
			\epsilon^{i\epsilon} & :~ \epsilon^{i\epsilon} < z
								\le \epsilon^{(i-1)\epsilon} ,
								\quad i = 1, 2, \cdots, l .  
		\end{array} \right.
\label{eq:log-quantizer}
\end{equation}
Observe that the {\em ratio} between adjacent reproduction levels,
$\epsilon^\epsilon \uparrow 1$ as $l \to \infty$. Moreover the {\em difference}
between adjacent reproduction levels is upper-bounded by 
$1 - \epsilon^\epsilon \le \epsilon \ln \epsilon^{-1} = \frac{\ln l}{l}$.
Both notions of precision will be useful in the proof.

For any $p \in \calP_Y(\epsilon)$, define
\begin{equation}
   q(y) = Q_l(p(y)) ,
\label{eq:q}
\end{equation}
the sum $\sigma = \sum_y q(y)$, and the pmf $\hat{p}(y) = \frac{1}{\sigma} q(y)$.
Hence
\begin{equation}
   \hat{p}(y) = \Phi_l \,p(y) 
	   \triangleq \frac{Q_l(p(y))}{\sum_y Q_l(p(y))}, \quad y \in \calY .
\label{eq:Phi-l}
\end{equation}



\begin{lemma}
For any integer $l \ge |\calY| \ge 2$ and $p, \tp \in \calP_Y(l^{-1})$, we have
\[ |D(\tp\|p) - D(\Phi_l \,\tp\|\Phi_l \,p)| 
	\le 2 (|\calY|+1) \frac{\log^2 l}{l} . \]
\label{lem:D-gap2}
\end{lemma}

\noindent
{\em Proof}.
Let $\epsilon = l^{-1}$. Since $p(y) > \epsilon$, it follows from
(\ref{eq:log-quantizer}) and (\ref{eq:q}) that
\begin{equation}
   \epsilon^\epsilon p(y) \le q(y) < p(y) .
\label{eq:qp-ratio}
\end{equation}
Summing (\ref{eq:qp-ratio}) over $y \in \calY$, we obtain
\begin{equation}
   \epsilon^\epsilon \le \sigma = \sum_y q(y) \le 1 .
\label{eq:sigma}
\end{equation}
We have 
\begin{equation}
   \epsilon^\epsilon \le \frac{q(y)}{p(y)} \le 1
	\quad \mathrm{and} \quad
	1 \le \frac{1}{\sigma} = \frac{\Phi_l \,p(y)}{q(y)} \le \epsilon^{-\epsilon}
\label{eq:B1-bound-ratio}
\end{equation}

For each $y \in \calY$, we have
\[ |p(y) - \Phi_l \,p(y)| 
	\le |p(y) - q(y)| + |q(y) - \Phi_l \,p(y)| 
	\le \frac{\ln l}{l} + (1-\sigma) \Phi_l \,p(y)
	\le [1+\Phi_l \,p(y)] \frac{\ln l}{l} ,
\]
hence
\begin{equation}
   \|p - \Phi_l \,p \| \le (|\calY|+1) \frac{\ln l}{l} .
\label{eq:variation-bound}
\end{equation}

Multiplying the inequalities in (\ref{eq:B1-bound-ratio}) and taking logarithms,
we obtain
\begin{equation}
   \left| \log \frac{\Phi_l \,p(y)}{p(y)} \right| \le - \epsilon \log \epsilon 
	= \frac{\log l}{l} .
\label{eq:log-q-ratio}
\end{equation}
Similarly,
\begin{equation}
   \left| \log \frac{\Phi_l \,p(y)}{p(y)} \right| 
	\le \frac{\log l}{l} , \quad \forall y.
\label{eq:log-q-ratio2}
\end{equation}
Hence
\begin{eqnarray*}
   D(\tp\|p) - D(\Phi_l \,\tp\|\Phi_l \,p) 
	& = & \sum_y \tp(y) \,\log \frac{\tp(y)}{p(y)} 
		- \sum_y \Phi_l \,\tp(y) \log \frac{\Phi_l\,\tp(y)}{\Phi_lp(y)} \\ 
	& = & \sum_y \tp(y) \left( \log \frac{\tp(y)}{p(y)} 
			- \log \frac{\Phi_l\,\tp(y)}{\Phi_l \,p(y)} \right)
		+ \sum_y (\tp(y) - \Phi_l \tp(y)) \log \frac{\Phi_l\,\tp(y)}{\Phi_l\,p(y)} \\
	& = & \sum_y \tp(y) \left( \log \frac{\Phi_l \,p(y)}{p(y)} 
			- \log \frac{\Phi_l\,\tp(y)}{\tp(y)} \right)
		+ \sum_y (\tp(y) - \Phi_l \tp(y)) \,\log \frac{\Phi_l\,\tp(y)}{\Phi_l\,p(y)} .
\end{eqnarray*}
Hence
\begin{eqnarray*}
   |D(\tp\|p) - D(\Phi_l \,\tp\|\Phi_l \,p)| 
 	& \le & \max_y \left| \log \frac{\Phi_l \,p(y)}{p(y)} \right|
			+ \max_y \left| \log \frac{\Phi_l \,\tp(y)}{\tp(y)} \right|
		+ \|\tp - \Phi_l \,\tp \|
			\,\max_y \left| \log \frac{\Phi_l \,\tp(y)}{\Phi_l \,p(y)} \right| \\
	& \le & 2 \frac{\log l}{l} + (|\calY|+1) \frac{\ln l}{l} \log l \\
	& = & 2 \frac{\log l}{l} \left[ 1 + (|\calY|+1) \frac{\ln 2}{2} \log l \right] \\
	& \le & 2 (|\calY|+1) \frac{\log^2 l}{l}
\end{eqnarray*}
where the second inequality follows from (\ref{eq:variation-bound}),
(\ref{eq:log-q-ratio}), and (\ref{eq:log-q-ratio2}), and the third inequality
from the fact that $l \ge |\calY| \ge 2$.
\hfill $\Box$

The following lemma establishes a bound on the variation in conditional
Kullback-Leibler divergence $D(\tp_{Y|X}\|p_{Y|X} |\tp_X)$ when the
mapping $\Phi_l$ is applied to each pmf $p_{Y|X}(\cdot |x) \in \calP_Y(\epsilon)$.
The resulting pmf is denoted by 
$\Phi_l \,p_{Y|X} = \{ \Phi_l \,p_{Y|X}(\cdot|x) ,\,x \in \calX\}$.

\begin{lemma}
For any integer $l \ge |\calY| \ge 2$, 
$\tp_{XY} = \tp_X \tp_{Y|X} \in \calP_{XY}$,
and $p_{Y|X}, \tp_{Y|X} \in \calP_{Y|X}(\epsilon)$, we have
\[ |D(\tp_{Y|X}\|p_{Y|X} |\tp_X) - D(\Phi_l \,\tp_{Y|X}\|\Phi_l \,p_{Y|X} |\tp_X)| 
	\le 2 (|\calY|+1) \frac{\log^2 l}{l} . \]
\label{lem:D-gap3}
\end{lemma}

{\em Proof}. 
\begin{eqnarray*}
   \lefteqn{|D(\tp_{Y|X}\|p_{Y|X} |\tp_X) 
		- D(\Phi_l \,\tp_{Y|X}\|\Phi_l \,p_{Y|X} |\tp_X)|} \\
	& = & \left| \sum_x \tp_X(x) [D(\tp_{Y|X=x}\|p_{Y|X=x}) 
			- D(\Phi_l \,\tp_{Y|X=x}\|\Phi_l \,p_{Y|X=x}) ] \right| \\
	& \le & \max_x |D(\tp_{Y|X=x}\|p_{Y|X=x}) 
			- D(\Phi_l \,\tp_{Y|X=x}\|\Phi_l \,p_{Y|X=x})| \\
	& \le & 2 (|\calY|+1) \frac{\log^2 l}{l}
\end{eqnarray*}
where the last inequality follows from Lemma~\ref{lem:D-gap2}. 
\hfill $\Box$ \\

\noindent
{\bf Proof of Proposition~\ref{prop:cara-Er}}.

The lower bound is straightforward. We now derive the upper bound.
The class of attack channels under expected distortion constraint $D_2$
is denoted by $\calA(D_2)$; the dependency on $p_X$ is not explicitly
indicated.

Define $\calU_L = \{ 1, \cdots, L\}$, where $L$ is given in (\ref{eq:L-cara}).
Let $\calU$ have arbitrarily cardinality, possibly larger than $L$.
Define the shorthands
\[ \tilde{\calA} \triangleq \calP_{YS^aS^d|XS^e}
	\quad \mathrm{and} \quad 
	\tilde{\calA}_{\calU} \triangleq \calP_{YS^aS^d|XUS^e}
\]
and the functionals (with a little abuse of notation)
\begin{eqnarray*}
   \lefteqn{E_{r,|\calU|}(\tp_{S^e}, p_{XU|S^e}, \tp_{YS^aS^d|XUS^e}, p_{Y|XS^a})} \\
	& \triangleq & D(\tp_{S^e} p_{XU|S^e} \tp_{YS^aS^d|XUS^e} 
		\| p_S p_{XU|S^e} p_{Y|XS^a})
		+ |J_{|\calU|}(\tp_{S^e} p_{XU|S^e} \tp_{YS^aS^d|XUS^e}) - R |^+
\end{eqnarray*}
and
\begin{eqnarray}
   E_{r,|\calU|}(\tp_{S^e}, \tilde{\calB}_{\calU}, \calB)
	& \triangleq & \max_{p_{XU|S^e} \in \calP_{XU|S^e}(|\calU|,D_1)}
		\,\min_{\tp_{YS^aS^d|XUS^e} \in \tilde{\calB}_{\calU}} 
			\,\min_{p_{Y|XS^a} \in \calB} 			\nonumber \\
	& & \quad E_{r,|\calU|}(\tp_{S^e}, p_{XU|S^e}, \tp_{YS^aS^d|XUS^e},
			p_{Y|XS^a}) , \quad \forall \tilde{\calB}_{\calU} 
				\subseteq \tilde{\calA}_{\calU}, \,\calB \subseteq \calA .
\label{eq:ErU-b}
\end{eqnarray}
Hence
\begin{eqnarray*}
   E_{r,|\calU|}(D_2) & = & \min_{\tp_{S^e}} 
			E_{r,|\calU|}(\tp_{S^e}, \tilde{\calA}_{\calU}, \calA(D_2)) \\
   E_r(D_2) & = & \lim_{|\calU| \to \infty} E_{r,|\calU|}(D_2) .
\end{eqnarray*}

Let $\epsilon = 1/l$ and 
\begin{eqnarray}
   D_2' & = & D_2 (1+ \epsilon \ln \epsilon) < D_2 	\label{eq:D2'} \\
   D_2'' & = & D_2' - \epsilon \,|\calY||\,|\calS^a|\,|\calS^d| \overline{D}
										\nonumber \\
	   & = & D_2 - \epsilon (|\calY||\,|\calS^a|\,|\calS^d| \overline{D}
				+ D_2 \ln \epsilon^{-1}) .		\label{eq:D2''} 
\end{eqnarray}

The proof consists of the following steps:
\begin{enumerate}
\item ({\em Pmf lifting step}).
	Define the subsets
	\begin{eqnarray*}
	   \calA(D_2;\epsilon) & \triangleq & \left\{ p_{Y|XS^a} \in \calA(D_2) 
		~:~ \min_{y,x,s^a} p_{Y|XS^a}(y|x,s^a) \ge \epsilon \right\} \\
	   \tilde{\calA}(\epsilon) & \triangleq & \left\{ \tp_{YS^aS^d|XS^e}
		\in \tilde{\calA} ~:~ \min_{y,x,s} \tp_{YS^aS^d|XS^e}(y,s^a,s^d|x,s^e)
		\ge \epsilon \right\}
	\end{eqnarray*}
	of $\calA(D_2)$ and $\tilde{\calA}$ in which the conditional pmf's
	$p_{Y|XS^a}$ and $\tp_{YS^aS^d|XS^e}$ are lower-bounded by $\epsilon$.
 	Also define $\tilde{\calA}_{\calU}(\epsilon)$ as the set of all pmf's
	$\tp_{YS^aS^d|XUS^e}$ whose conditional marginals $\tp_{YS^aS^d|XS^e,U=u}$
	are in $\tilde{\calA}(\epsilon)$ for all $u \in \calU$.

	For any $\tp_{S^e}$, $p_{XU|S^e} \in \calP_{XU|S^e}(|\calU|, D_1)$,
	$\tp_{YS^aS^d|XUS^e} \in \tilde{\calA}_{\calU}$
	and $p_{Y|XS^a} \in \calA(D_2'')$, we show there exist 
	$\tp'_{YS^aS^d|XUS^e} \in \tilde{\calA}_{\calU}(\epsilon)$
	and $p'_{Y|XS^a} \in \calA(D_2';\epsilon)$ such that
	\begin{eqnarray}
	   \lefteqn{E_{r,|\calU|}(\tp_{S^e}, p_{XU|S^e}, \tp_{YS^aS^d|XUS^e}, 
		\,p_{Y|XS^a})}							\nonumber \\
	   & \ge & E_{r,|\calU|}(\tp_{S^e}, p_{XU|S^e}, \tp'_{YS^aS^d|XUS^e}, 
			\,p'_{Y|XS^a})| - \frac{4}{l} |\calY|\,|\calS^a|\,|\calS^d|
			\,\log \frac{l}{2|\calS^a|^{5/4} |\calS^d|^{1/4} c^{1/4}}
										\nonumber \\	
	\label{eq:ErU-bound-lift}
	\end{eqnarray}
	where the constant $c$ was defined in the statement of the proposition.

\item ({\em Pmf quantization step}).
	We define finite nets 
	$\tilde{\calA}_l(\epsilon) \subset \tilde{\calA}(\epsilon)$
	and $\calA_l(D_2;\epsilon) \subset \calA(D_2;\epsilon)$ 
	whose cardinalities are at most $l^{|\calY|\,|\calX|\,|\calS|}$
	and $l^{|\calY|\,|\calX|\,|\calS^a|}$, respectively.
	Also define $\tilde{\calA}_{l,\calU}(\epsilon)$ as the set of all pmf's
	$\tp_{YS^aS^d|XUS^e}$ whose conditional marginals
	$\tp_{YS^aS^d|XS^e,U=u}$ are in $\tilde{\calA}_l(\epsilon)$
	for all $u \in \calU$.

	For any $\tp_{S^e}$, $p_{XU|S^e}$,
	$\tp'_{YS^aS^d|XUS^e} \in \tilde{\calA}_{\calU}(\epsilon)$
	and $p_{Y|XS^a}' \in \calA(D_2';\epsilon)$, we show there exist 
	$\hat{p}_{YS^aS^d|XUS^e} \in \tilde{\calA}_{l,\calU}(\epsilon)$
	and $\hat{p}_{Y|XS^a} \in \calA_l(D_2;\epsilon)$ such that
	\begin{eqnarray}
	   \lefteqn{E_{r,|\calU|}(\tp_{S^e}, p_{XU|S^e}, \tp'_{YS^aS^d|XUS^e},
			\,p'_{Y|XS^a})}						\nonumber \\
		& \ge & E_{r,|\calU|}(\tp_{S^e}, p_{XU|S^e}, \hat{p}_{YS^aS^d|XUS^e}, 
			\,\hat{p}_{Y|XS^a}) 
			- 5 |\calY|\,|\calS^a|\,|\calS^d| \,\frac{\log^2 l}{l} .
	\label{eq:ErU-bound-Q}
	\end{eqnarray}
	From (\ref{eq:ErU-bound-lift}) and (\ref{eq:ErU-bound-Q}), for
	\begin{equation}
	   l \ge \,\exp_2 \left[ 1 + \sqrt{ \left| - \frac{1}{2}
		\log (8 |\calS^a|^5 |\calS^d| c) \right|^+ } \,\right]
	\label{eq:lmin-0}
	\end{equation}
	we obtain
	\begin{eqnarray}
	   \lefteqn{E_{r,|\calU|}(\tp_{S^e}, p_{XU|S^e}, \tp_{YS^aS^d|XUS^e}, 
		\,p_{Y|XS^a})} 							\nonumber \\
		& \ge & E_{r,|\calU|}(\tp_{S^e}, p_{XU|S^e}, \hat{p}_{YS^aS^d|XUS^e}, 
			\,\hat{p}_{Y|XS^a}) - \Delta(l)
	\label{eq:ErU-bound}
	\end{eqnarray}
	where
	\begin{equation}
	   \Delta(l) \triangleq  7 |\calY|\,|\calS^a|\,|\calS^d| \,\frac{\log^2 l}{l} .
	\label{eq:Delta}
	\end{equation}

\item By application of Caratheodory's theorem, we show that for each $\tp_{S^e}$,
	the supremum of the function $E_{r,|\calU|}(\tp_{S^e}, p_{XU|S^e},
	\tilde{A}_{l,\calU}(\epsilon), \calA_l(D_2;\epsilon))$
	over $p_{XU|S^e}$ is achieved for $|\calU|=L$ given in (\ref{eq:L-cara}).

\item Combining the results above, we show that
	$E_{r,L}(D_2'') \ge E_{r,|\calU|}(D_2) - \Delta(l)$.
	Taking the limit as $|\calU| \to \infty$ proves the claim.
\end{enumerate}

{\bf Step 1}. ({\em Pmf lifting}).
Denote by
\[ \mu_Y(y) = \frac{1}{|\calY|} \quad \mathrm{and} \quad
	 \mu_{S^aS^d}(s^a,s^d) = \frac{1}{|\calS^a|\,|\calS^d|} \]
the uniform distributions over $\calY$ and $\calS^a \times \calS^d$,
respectively. The average distortion for the hypothetical attack channel
$p_{Y|XS^a} = \mu_Y$ satisfies
\begin{equation}
   \eE d(X,Y) \le \max_x \eE d(x,Y) 
	= \max_x \frac{1}{|\calY|} \sum_y d(x,y) = \overline{D} .
\label{eq:distortion-r}
\end{equation}
To each $p_{Y|XS^a} \in \calA(D_2'')$ and
$\tp_{YS^aS^d|XUS^e} \in \tilde{\calA}$, associate the conditional pmf's
\begin{eqnarray}
   p'_{Y|XS^a}(y|x,s^a) & = & (1-\epsilon \,|\calY|\,|\calS^a|\,|\calS^d|)
	\,p_{Y|XS^a}(y|x,s^a) + \epsilon \,|\calY|\,|\calS^a|\,|\calS^d| \,\mu_Y(y)
											\nonumber \\
   \tp'_{YS^aS^d|XUS^e}(y,s^a,s^d|x,u,s^e)
	& = & (1-\epsilon \,|\calY|\,|\calS^a|\,|\calS^d|)
		\,\tp_{YS^aS^d|XUS^e}(y,s^a,s^d|x,u,s^e)			\nonumber \\
	& & \quad + \epsilon \,|\calY|\,|\calS^a|\,|\calS^d|
		\,\mu_{Y}(y) \,\mu_{S^aS^d}(s^a,s^d) , \quad \forall y,x,u,s
\label{eq:tp'}
\end{eqnarray}
which are slight modifications of $p_{Y|XS^a}$ and $\tp_{YS^aS^d|XUS^e}$
and are lower-bounded by $\epsilon$.
We refer to the mappings in (\ref{eq:tp'}) as ``pmf lifting''.

Since average distortion is a linear functional of the attack channel,
the average distortion for $p'_{Y|XS^a}$ is upper bounded by
\[ (1-\epsilon \,|\calY|\,|\calS^a|\,|\calS^d|) D_2''
	+ \epsilon \,|\calY|\,|\calS^a|\,|\calS^d| \,\overline{D} \le D_2' . 
\]
Hence $p'_{Y|XS^a} \in \calA(D_2';\epsilon)$.

Since $E_{r,|\calU|}(\cdot)$ takes the form $D(\cdot) + |J_{|\calU|}(\cdot) - R|^+$,
the variation in the error exponent due to the above pmf lifting operations
satisfies the upper bound:
\begin{eqnarray}
   E_{r,L}(\tp_{S^e}, p_{XU|S^e}, \,\tp'_{YS^aS^d|XUS^e}, \,p'_{Y|XS^a})
		- E_{r,L}(\tp_{S^e}, p_{XU|S^e}, \,\tp_{YS^aS^d|XUS^e}, \,p_{Y|XS^a})
											\nonumber \\
	\le \Delta D + \Delta J_{|\calU|}
\label{eq:ErU-lift}
\end{eqnarray}
where
\begin{eqnarray*}
   \Delta J_{|\calU|}
	& = & J_{|\calU|}(\tp_{S^e} \,p_{XU|S^e} \,\tp'_{YS^aS^d|XUS^e})
		- J_{|\calU|}(\tp_{S^e} \,p_{XU|S^e} \,\tp_{YS^aS^d|XUS^e})
\end{eqnarray*}
and
\begin{eqnarray*}
   \Delta D
	& = & D(\tp_{S^e} p_{XU|S^e} \,\tp'_{YS^aS^d|XUS^e} 
			\| p_S p_{XU|S^e} \,p'_{Y|XS^a}) 
		- D(\tp_{S^e} p_{XU|S^e} \,\tp_{YS^aS^d|XUS^e} 
			\| p_S p_{XU|S^e} \,p_{Y|XS^a}) \\
	& = & D(\tp'_{YS^aS^d|XUS^e} 
		\| p_{S^aS^d|S^e} \,p'_{Y|XS^a} ~| \,\tp_{S^e} p_{XU|S^e}) 
		- D(\tp_{YS^aS^d|XUS^e} 
		\| p_{S^aS^d|S^e} \,p_{Y|XS^a} ~| \,\tp_{S^e} p_{XU|S^e}) .
\end{eqnarray*}
The last equality follows from
the chain rule for Kullback-Leibler divergence,
\begin{eqnarray*}
\lefteqn{D(\tp_{S^e} p_{XU|S^e} \tp_{YS^aS^d|XUS^e} \| p_S p_{XU|S^e} \,p'_{Y|XS^a})} \\
	& = & D(\tp_{S^e} \| p_{S^e}) + D(\tp_{YS^aS^d|XUS^e} 
		\| p_{S^aS^d|S^e} \,p'_{Y|XS^a} ~| \,\tp_{S^e} p_{XU|S^e}) .
\end{eqnarray*}

The effect of pmf lifting on $\Delta D$ is as follows.
By convexity of conditional Kullback-Leibler divergence \cite{Csi81}, 
from (\ref{eq:tp'}) we have
\begin{eqnarray*}
  \lefteqn{D(\tp'_{YS^aS^d|XUS^e} 
		\| p_{S^aS^d|S^e} \,p'_{Y|XS^a} ~| \,\tp_{S^e} p_{XU|S^e})} \\
	& \le & (1-\epsilon \,|\calY|\,|\calS^a|\,|\calS^d|) D(\tp_{YS^aS^d|XUS^e} 
		\| p_{S^aS^d|S^e} \,p_{Y|XS^a} ~| \,\tp_{S^e} p_{XU|S^e}) \\
	& & \qquad + \epsilon \,|\calY|\,|\calS^a|\,|\calS^d| D(\mu_Y \,\mu_{S^aS^d} 
		\| p_{S^aS^d|S^e} \,\mu_Y ~| \,\tp_{S^e} p_{XU|S^e}) \\
	& \le & D(\tp_{YS^aS^d|XUS^e} 
		\| p_{S^aS^d|S^e} \,p_{Y|XS^a} ~| \,\tp_{S^e} p_{XU|S^e}) 
		+ \epsilon \,|\calY|\,|\calS^a|\,|\calS^d|
		\, \max_s \log \frac{\mu_{S^aS^d}(s^a,s^d)}{p_{S^aS^d|S^e}(s^a,s^d|s^e)} \\
	& \le & D(\tp_{YS^aS^d|XUS^e} 
		\| p_{S^aS^d|S^e} \,p_{Y|XS^a} ~| \,\tp_{S^e} p_{XU|S^e}) 
		+ \epsilon \,|\calY|\,|\calS^a|\,|\calS^d|
			\log \frac{1}{|\calS^a|\,|\calS^d| \,c}
\end{eqnarray*}
where the constant $c$ was defined in the statement of the proposition.
Hence
\begin{equation}
   \Delta D	\le \epsilon \,|\calY|\,|\calS^a|\,|\calS^d|
			\log \frac{1}{|\calS^a|\,|\calS^d| \,c} .
\label{eq:D-gap}
\end{equation}

The effect of pmf lifting on $J_{|\calU|}(\cdot) = H(YS^d) - H(YS^d|U) - I(U;S^e)$
is as follows. From (\ref{eq:tp'}), we have
\begin{eqnarray}
   \| \tp_{YS^aS^d|XUS^e} - \tp'_{YS^aS^d|XUS^e} \| 
	& = & \| \,\epsilon |\calY|\,|\calS^a|\,|\calS^d| \,p_{YS^aS^d|XUS^e} 
		- \epsilon \,|\calY|\,|\calS^a|\,|\calS^d| \,\mu_Y \,\mu_{S^aS^d} \|
											\nonumber \\
	& \le & 2 \epsilon \,|\calY|\,|\calS^a|\,|\calS^d|
	\triangleq \theta .
\label{eq:delta-tp}
\end{eqnarray}
Analogously to (\ref{eq:delta-J}), we have
\begin{eqnarray}
   |\Delta J_{|\calU|}|
	& = & \left| \tilde{H}_{\tp_{YS^aS^d|XUS^e}}(YS^d) 
			- \tilde{H}_{\tp_{YS^aS^d|XUS^e}}(YS^d|U)	\right.	\nonumber \\
	& & \qquad \left. - \tilde{H}_{\tp'_{YS^aS^d|XUS^e}}(YS^d) 
			+ \tilde{H}_{\tp'_{YS^aS^d|XUS^e}}(YS^d|U) \right|	\nonumber \\
	& \le & \left| \tilde{H}_{\tp_{YS^aS^d|XUS^e}}(YS^d) 
			- \tilde{H}_{\tp'_{YS^aS^d|XUS^e}}(YS^d) \right|	\nonumber \\
	& & \qquad + \left| \tilde{H}_{\tp_{YS^aS^d|XUS^e}}(YS^d|U)
			+ \tilde{H}_{\tp'_{YS^aS^d|XUS^e}}(YS^d|U) \right|	\nonumber \\
	& \le & 2 \theta \log \frac{|\calY|\,|\calS^d|}{\theta}		\nonumber \\
	& = & 4 \epsilon \,|\calY|\,|\calS^a|\,|\calS^d|
			\log \frac{1}{2\epsilon \,|\calS^a|}
\label{eq:J-gap}
\end{eqnarray}
where the last inequality follows from (\ref{eq:delta-tp}) and 
Lemmas~\ref{lem:csiszar1} and \ref{lem:csiszar2}.

Combining (\ref{eq:ErU-lift}), (\ref{eq:D-gap}), and (\ref{eq:J-gap}),
we obtain (\ref{eq:ErU-bound-lift}).

{\bf Step 2}. ({\em Pmf quantization}).
Consider $p'_{Y|XS^a} \in \calA(D_2';\epsilon)$.
For each value of $(x,s^a)$, apply the quantization mapping 
$\Phi_l ~:~ \calP_Y(\epsilon) \to \calP_Y(\epsilon)$ defined in (\ref{eq:Phi-l})
to the pmf $p'_{Y|XS^a}(\cdot|x,s^a)$ 
and denote by $\Phi_l \,p'_{Y|XS^a} \in \calP_{Y|XS^a}(\epsilon)$
the resulting conditional pmf. Also let
\[ \sigma(x,s^a) = \sum_y Q_l(p'_{Y|XS^a}(y|x,s^a)) \ge \epsilon^\epsilon \]
where the inequality is obtained as in (\ref{eq:sigma}).
Similarly, given $\tp'_{YS^aS^d|XUS^e} \in \tilde{\calA}_{\calU}(\epsilon)$,
define the quantized conditional pmf $\Phi_l \,\tp'_{YS^aS^d|XUS^e}$ 
which belongs to the finite set $\tilde{\calA}_{l,\calU}(\epsilon)$.

The average distortion associated with $\Phi_l \,p'_{Y|XS^a}$ is
\begin{eqnarray*}
   \lefteqn{\sum_{y,x,s^a} d(x,y) \,\Phi_l \,p'_{Y|XS^a}(y|x,s^a)
		\,p_{XS^a}(x,s^a)} \\
	& = & \frac{1}{\sigma(x,s^a)} \sum_{y,x,s^a} d(x,y) 
		\,Q_l(p'_{Y|XS^a}(y|x,s^a)) \,p_{XS^a}(x,s^a) \\
	& \le & \frac{1}{\sigma(x,s^a)} \sum_{y,x,s^a} d(x,y) 
		\,p'_{Y|XS^a}(y|x,s^a) \,p_{XS^a}(x,s^a) \\
	& \le & \epsilon^{-\epsilon} D'_2 \\
	& = & \epsilon^{-\epsilon} (1 + \epsilon \ln \epsilon) D_2 \\
	& \le & \epsilon^{-\epsilon} e^{\epsilon \ln \epsilon} D_2 \\
	& = & D_2 .
\end{eqnarray*}
Therefore $\Phi_l \,p'_{Y|XS^a} \in \calA_l(D_2;\epsilon)$.

For any choice of $\tp_{S^e}$, $p_{XU|S^e}$, 
$\tp'_{YS^aS^d|XUS^e} \in \tilde{\calA}_{\calU}(\epsilon)$ and
$p'_{Y|XS^a} \in \calA(D_2';\epsilon)$, we now bound
the effect of $\Phi_l$ on the error exponent as follows:
\begin{eqnarray}
   E_{r,|\calU|}(\tp_{S^e}, p_{XU|S^e}, \,\tp'_{YS^aS^d|XUS^e}, \,p'_{Y|XS^a})
		- E_{r,|\calU|}(\tp_{S^e}, p_{XU|S^e}, \Phi_l \,\tp'_{YS^aS^d|XUS^e}, 
				\Phi_l \,p'_{Y|XS^a})				\nonumber \\
	\le \Delta D + \Delta J_{|\calU|} .
\label{eq:ErU-Q}
\end{eqnarray}
Here
\[
   \Delta J_{|\calU|}
	= J_{|\calU|}(\tp_{S^e} \,p_{XU|S^e} \,\tp'_{YS^aS^d|XUS^e})
		- J_{|\calU|}(\tp_{S^e} \,p_{XU|S^e} \,\Phi_l \,\tp'_{YS^aS^d|XUS^e})
\]
and
\begin{eqnarray}
   |\Delta D|
	& = & \left| D(\tp_{S^e} p_{XU|S^e} \,\tp'_{YS^aS^d|XUS^e} 
			\| p_S p_{XU|S^e} \,p'_{Y|XS^a}) \right.			\nonumber \\ 
	& & \qquad \left. - D(\tp_{S^e} p_{XU|S^e} \Phi_l \,\tp'_{YS^aS^d|XUS^e} 
			\| p_S p_{XU|S^e} \Phi_l \,p'_{Y|XS^a}) \right|		\nonumber \\
	& = & \left| D(\tp'_{YS^aS^d|XUS^e} 
			\| p'_{Y|XS^a} \,p_{S^aS^d|S^e} \,|\,\tp_{S^e} p_{XU|S^e}) \right.
												\nonumber \\
	& & \qquad \left. - D(\Phi_l \,\tp'_{YS^aS^d|XUS^e} 
			\| \Phi_l \,p'_{Y|XS^a} \,p_{S^aS^d|S^e} \,|\,\tp_{S^e} p_{XU|S^e})
										\right|	\nonumber \\
	& \le & 2 (|\calY| \,|\calS^a| \,|\calS^d| + 1) \frac{\log^2 l}{l}  \nonumber \\
	& \le & 3 |\calY| \,|\calS^a| \,|\calS^d| \frac{\log^2 l}{l}
\label{eq:deltaD-Q}
\end{eqnarray}
where the first inequality is obtained by application of Lemma~\ref{lem:D-gap3},
and the second because $|\calY| \ge 2$.
Next, we have
\begin{eqnarray*}
   \| \tp'_{YS^aS^d|XUS^e}) - \Phi_l \,\tp'_{YS^aS^d|XUS^e} \|
	\le (|\calY|\,|\calS^a|\,|\calS^d| + 1) \,\frac{\ln l}{l} \triangleq \theta .
\end{eqnarray*}
where the inequality is a straightforward generalization of (\ref{eq:variation-bound}).
Similarly to (\ref{eq:J-gap}), we obtain
\begin{eqnarray}
   |\Delta J_{|\calU|}| \le 2 \theta \log \frac{|\calY|\,|\calS^d|}{\theta}
	& = & 2 (|\calY|\,|\calS^a|\,|\calS^d| + 1) \,\frac{\ln l}{l} 
		\log \frac{l}{|\calS^a| \ln l} \nonumber \\
	& \le & 2 |\calY|\,|\calS^a|\,|\calS^d| \,\frac{\log^2 l}{l}
\label{eq:deltaJ-Q}
\end{eqnarray}
where the last inequality holds because $\log e < \frac{2}{3}$
and $|\calS^a| \ln l > 1$.

Combining (\ref{eq:ErU-Q}), (\ref{eq:deltaD-Q}) and (\ref{eq:deltaJ-Q}),
we obtain
\begin{eqnarray*}
   |E_{r,|\calU|}(\tp_{S^e}, p_{XU|S^e}, \,\tp_{YS^aS^d|XUS^e}, \,p'_{Y|XS^a})
		- E_{r,|\calU|}(\tp_{S^e}, p_{XU|S^e}, \Phi_l \,\tp_{YS^aS^d|XUS^e}, 
				\Phi_l \,p'_{Y|XS^a})| 				\\
	\le |\Delta D| + |\Delta J_{|\calU|}| 
	\le 5 |\calY|\,|\calS^a|\,|\calS^d| \,\frac{\log^2 l}{l} ,
\end{eqnarray*}
which establishes (\ref{eq:ErU-bound-Q}).

Combining (\ref{eq:ErU-b}), (\ref{eq:ErU-bound-lift}) and 
(\ref{eq:ErU-bound-Q}), for any choice of $\tp_{S^e}$ we obtain
\begin{eqnarray}
   \lefteqn{E_{r,|\calU|}(\tp_{S^e}, \tilde{\calA}_{\calU}, \calA(D_2''))
	- E_{r,|\calU|}(\tp_{S^e}, \tilde{\calA}_{l,\calU}, \calA_l(D_2;\epsilon))}
											\nonumber \\
	& \ge & - 5 |\calY|\,|\calS^a|\,|\calS^d| \,\frac{\log^2 l}{l} 
		- \frac{4}{l} |\calY|\,|\calS^a|\,|\calS^d|
			\,\log \frac{l}{2|\calS^a|^{5/4} |\calS^d|^{1/4} c^{1/4}}
											\nonumber \\
	& = & - \Delta(l)	+ \frac{2}{l} |\calY|\,|\calS^a|\,|\calS^d| \left[ \log^2 l
			- 2 \log \frac{l}{2|\calS^a|^{5/4} |\calS^d|^{1/4} c^{1/4}}
			\right]							\nonumber \\
	& = & - \Delta(l)	+ \frac{2}{l} |\calY|\,|\calS^a|\,|\calS^d| \left[ \log^2 l
			- 2 \log l + \frac{1}{2} \log(16 |\calS^a|^5\,|\calS^d|\,c)
			\right]							\nonumber \\
	& \ge & - \Delta(l)
\label{eq:delta-l}
\end{eqnarray}
where $\Delta(l)$ was defined in (\ref{eq:Delta}).
The term in brackets is positive when (\ref{eq:lmin-0}) is satisfied.

{\bf Step~3}. ({\em Caratheodory}).
Define
\begin{eqnarray*}
   \lefteqn{\tilde{J}_{|\calU|}(\tp_{S^e}, p_{X|S^e,U=u}, \tp_{YS^aS^d|XS^e,U=u})} \\
	& \triangleq & H(YS^d) - H(YS^d|U=u) - H(S^e) + H(S^e|U=u) , 
		\quad \forall u \in \calU
\end{eqnarray*}
where the quintuple $(S^eS^aS^dXY)$, conditioned on $U=u$, is distributed as \\
$\tp_{S^e} \,p_{X|S^e,U=u} \,\tp_{YS^aS^d|XS^e,U=u}$. 
Likewise, we view the conditional divergence 
\[ D(\tp_{YS^aS^d|XS^e,U=u} \| p_{Y|XS^a} \,p_{S^aS^d|S^e} 
	~| p_{X|S^e,U=u} \,\tp_{S^e}) 
\]
as a function of $\tp_{S^e}$, $p_{X|S^e,U=u}$, $\tp_{YS^aS^d|XS^e,U=u}$,
and $p_{Y|XS^a}$.
We may thus write
\begin{eqnarray*}
   \lefteqn{D(\tp_{YS^aS^d|XUS^e} \| p_{Y|XS^a} p_{S^aS^d|S^e} 
		~| p_{X|US^e} \tp_{S^e})} \\
	& = & \sum_{u \in \calU} p_U(u) D(\tp_{YS^aS^d|XS^e,U=u} 
		\| p_{Y|XS^a} \,p_{S^aS^d|S^e} ~| p_{X|S^e,U=u} \,\tp_{S^e}) \\
   J_{|\calU|}(\tp_{S^e}, p_{XU|S^e}, \tp_{YS^aS^d|XUS^e})
	& = & \sum_{u \in \calU} p_U(u) 
		\tilde{J}_{|\calU|}(\tp_{S^e}, p_{X|S^e,U=u}, \tp_{YS^aS^d|XS^e,U=u})
\end{eqnarray*}
and
\begin{eqnarray*}
   \lefteqn{E_{r,|\calU|}(\tp_{S^e}, p_{XU|S^e}, \tp_{YS^aS^d|XUS^e}, p_{Y|XS^a})} \\
	& = & D(\tp_{S^e} \| p_{S^e}) + \sum_{u \in \calU} p_U(u) D(\tp_{YS^aS^d|XS^e,U=u} 
		\| p_{Y|XS^a} \,p_{S^aS^d|S^e} ~| p_{X|S^e,U=u} \,\tp_{S^e}) \\
	& & \qquad \qquad \qquad + \left| \sum_{u \in \calU} p_U(u) 
		\tilde{J}_{|\calU|}(\tp_{S^e}, p_{X|S^e,U=u}, \tp_{YS^aS^d|XS^e,U=u}) - R 
			\right|^+ .
\end{eqnarray*}

The cardinality of the discretized set $\calA_l(D_2;\epsilon)$ of attack channels
$\hat{p}_{Y|XS^a}$ is less than $l^{|\calY|\,|\calX|\,|\calS^a|}$. Likewise
the cardinality of the discretized set $\tilde{\calA}_l(\epsilon)$ of channels
$\hat{p}_{YS^aS^d|XS^e}$ is less than $l^{|\calY|\,|\calX|\,|\calS|}$.

We now define the following $L$ functionals over $\calP_{SXY}$ (recall that
$S = (S^e,S^a,S^d)$):
\begin{eqnarray*}
   f_i(p_{SXY|U=u}) & = & \tp_{S^e}(s^e) \,p_{X|S^e}(x|s^e), \mbox{\hspace*{1.4in}}
				\quad 1 \le i(x,s^e) \le |\calX| \,|\calS^e| - 1, \\
   f_i(p_{SXY|U=u}) & = & D(\hat{p}_{YS^aS^d|XS^e,U=u} 
		\| \,\hat{p}_{Y|XS^a} \,p_{S^aS^d|S^e} ~| p_{X|S^e,U=u} \,\tp_{S^e}) \\
	& & \qquad |\calX| \,|\calS^e| \le i(\hat{p}_{YS^aS^d|XS^e}, \hat{p}_{Y|XS^a})
			\le |\calX| \,|\calS^e| + |\tilde{\calA}_l(\epsilon)| 
				\,|\calA_l(D_2;\epsilon)| - 1, \\
   f_i(p_{SXY|U=u}) & = & {\tilde{J}_L
		(\tp_{S^e}, p_{X|S^e,U=u}, \hat{p}_{YS^aS^d|XS^e,U=u})} \\
	& & \qquad |\calX| \,|\calS^e| + |\tilde{\calA}_l| \,|\calA_l|
			\le i(\hat{p}_{YS^aS^d|XS^e}) \\
	& & \mbox{\hspace*{1.5in}}
			\le |\calX| \,|\calS^e| + |\tilde{\calA}_l(\epsilon)| 
				\,(1+|\calA_l(D_2;\epsilon)|) - 1.
\end{eqnarray*}
The first $|\calX| \,|\calS^e| - 1$ functions correspond to the marginals of
$p_{XS^e}$ except one, and the next
\[ |\tilde{\calA}_l(\epsilon)| \,(1+|\calA_l(D_2;\epsilon)|) 
		\le l^{|\calY|\,|\calX|\,(|\calS| + |\calS^a|)}
\]
functions are indexed by the channels $\hat{p}_{Y|XS^a} \in \calA_l(D_2,\epsilon)$
and $\hat{p}_{YS^aS^d|XS^e} \in \tilde{\calA}_l(\epsilon)$.
Hence, applying Caratheodory's theorem, we conclude there exist $L$ nonnegative
numbers $\alpha_1, \cdots, \alpha_L$ summing to 1 and
a random variable $U' \in \calU_L$ such that
\begin{eqnarray*}
   p_{SU'XY}(s,u',x,y) & = & p_{XS^e|U}(x,s^e|u_{u'}) \,\alpha_{u'} 
					\,\hat{p}_{YS^aS^d|XUS^e}(y,s^a,s^d|x,u_{u'},s^e) \\
	& & \qquad \forall s,u',x,y , \\
   E_{r,L}(\tp_{S^e}, p_{XU'|S^e}, \hat{p}_{YS^aS^d|XU'S^e}, \hat{p}_{Y|XS^a})
	& = & E_{r,|\calU|}(\tp_{S^e}, p_{XU|S^e}, \hat{p}_{YS^aS^d|XUS^e},
			\hat{p}_{Y|XS^a}) ,							\\
	& & \qquad \forall \hat{p}_{Y|XS^a} \in \calA_l(D_2;\epsilon) ,\,
			\hat{p}_{YS^aS^d|XUS^e} \in \tilde{\calA}_{l,\calU}(\epsilon) .
\end{eqnarray*}
Hence, given any $\tp_{S^e}$, it suffices to consider
\[ |\calU| = L = |\calX|\,|\calS^e| 
	+ l^{|\calY|\,|\calX|\,(|\calS| + |\calS^a|)} - 1 \]
to achieve
\[ \max_{p_{XU|S^e} \in \calP_{XU|S^e}(|\calU|,D_1)} 
	\,\min_{\hat{p}_{Y|XS^a} \in \calA_l(D_2;\epsilon)}
	\,\min_{\hat{p}_{YS^aS^d|XUS^e} \in \tilde{\calA}_l(\epsilon)}
	\,E_{r,|\calU|}(\tp_{S^e}, p_{XU'|S^e}, \hat{p}_{YS^aS^d|XUS^e},
		\hat{p}_{Y|XS^a}) .
\]
Hence 
\begin{equation}
   E_{r,|\calU|}(\tp_{S^e}, \tilde{\calA}_{l,\calU}(\epsilon), \calA_l(D_2;\epsilon))
	= E_{r,L}(\tp_{S^e}, \tilde{\calA}_{l,\calU_L}(\epsilon), \calA_l(D_2;\epsilon)) .
\label{eq:ErL-cara}
\end{equation}

{\bf Step~4}.
Let $\tp_{S^e}^*$ achieve the minimum in 
\[ E_{r,L}(D_2'') = \min_{\tp_{S^e}}
	E_{r,L}(\tp_{S^e}, \tilde{\calA}_{\calU_L}, \calA(D_2'')) . \]
It follows from the previous steps that
\begin{eqnarray}
   E_{r,L}(D_2'') 
	& \stackrel{(a)}{=} & E_{r,L}(\tp_{S^e}^*, \tilde{\calA}_{\calU_L},
			\calA(D_2'')) 						\nonumber \\
	& \stackrel{(b)}{\ge} & E_{r,L}(\tp_{S^e}^*, \tilde{\calA}_{l,\calU_L}
			(\epsilon), \calA_l(D_2;\epsilon)) - \Delta(l)	\nonumber \\
	& \stackrel{(c)}{=} & E_{r,|\calU|}(\tp_{S^e}^*, \tilde{\calA}_{l,\calU}
			(\epsilon), \calA_l(D_2;\epsilon)) - \Delta(l)	\nonumber \\
	& \stackrel{(d)}{\ge} & E_{r,|\calU|}(\tp_{S^e}^*, \tilde{\calA}_{\calU},
			\calA(D_2)) - \Delta(l)					\nonumber \\
	& \ge & \min_{\tp_{S^e}} E_{r,|\calU|}(\tp_{S^e}, \tilde{\calA}_{\calU},
			\calA(D_2)) - \Delta(l)					\nonumber \\
	& = & E_{r,|\calU|}(D_2) - \Delta(l)	
\label{eq:ErL-bound1}
\end{eqnarray}
where 
(a) follows from the definition of $\tp_{S^e}^*$,
(b) follows from (\ref{eq:delta-l}) with $\calU = \calU_L$,
(c) follows from (\ref{eq:ErL-cara}), and 
(d) holds because $\tilde{\calA}_{l,\calU}(\epsilon) \subset \tilde{\calA}_{\calU}$
and $\calA_l(D_2;\epsilon) \subset \calA(D_2)$.

Equation (\ref{eq:ErL-bound1}) holds for any $\calU$, hence
\begin{eqnarray*}
    E_{r,L}(D_2'') & \ge & \lim_{|\calU| \to \infty} E_{r,|\calU|}(D_2)
			- \Delta(l) \\
	& = & E_r(D_2) - \Delta(l) ,
\end{eqnarray*}
which concludes the proof.
\hfill $\Box$

\newpage

\begin{center}
\huge \textbf{Authors' Biographies}
\end{center}
\bigskip
\textbf{Pierre Moulin} received his
doctoral degree from Washington University in St. Louis in 1990,
after which he joined at Bell Communications Research in
Morristown, New Jersey, as a Research Scientist. In 1996, he
joined the University of Illinois at Urbana-Champaign, where he is
currently Professor in the Department of Electrical and Computer
Engineering, Research Professor at the Beckman Institute and the
Coordinated Science Laboratory, and affiliate professor in the
Department of Statistics.

His fields of professional interest include image and video
processing, compression, statistical signal processing and
modeling, media security, decision theory, and information theory.

Dr. Moulin has served on the editorial boards of the IEEE
Transactions on Information Theory and the IEEE Transactions on
Image Processing.  He is co-founding Editor-in-Chief of the IEEE
Transactions on Information Forensics and Security.  He has served
IEEE in various other capacities and is currently a member of the
IEEE Signal Processing Society Board of Governors.

He received a 1997 Career award from the National Science
Foundation and an IEEE Signal Processing Society 1997 Senior Best
Paper award. He is also co-author (with Juan Liu) of a paper that
received an IEEE Signal Processing Society 2002 Young Author Best
Paper award. He was 2003 Beckman Associate of UIUC's Center for
Advanced Study. He is an IEEE Fellow, recipient of UIUC's 2005
Sony Faculty award, and plenary speaker for ICASSP 2006.

\bigskip
\noindent\textbf{Ying Wang} received her Ph.D. degree in
electrical engineering from the University of Illinois at
Urbana-Champaign in 2006. She is now with QUALCOMM Flarion
Technologies, Bedminster, New Jersey, as a senior engineer. Her
research interests include information hiding, multimedia security
and forensics, statistical signal processing, image processing,
information theory, detection and estimation theory, and wireless
communications.


\newpage
\small
\begin{thebibliography}{99}
\bibitem{Gel80} S. I. Gel'fand and M. S. Pinsker, ``Coding for Channel
        with Random Parameters,'' {\em Problems of Control and
        Information Theory}, Vol.~9, No.~1, pp. 19---31, 1980.
\bibitem{Hee83} C.~Heegard and A. A. El Gamal, ``On the Capacity of
        Computer Memory with Defects,'' {\em IEEE Trans. Information Theory},
        Vol.~29, No.~5, pp.~731---739, Sep. 1983.
\bibitem{Cos83} M.~Costa, ``Writing on Dirty Paper,''
        {\em IEEE Trans. Information Theory}, Vol.~29, No.~3, pp.~439---441, May 1983.
\bibitem{Ere01} U.~Erez and R.~Zamir, ``Error Exponents of Modulo-Additive Noise
    Channels with Side Information at the Transmitter,''
    {\em IEEE Trans. Information Theory}, Vol. 47, No. 1, pp. 210---218, Jan.~2001.
\bibitem{Ere04} U. Erez and R. Zamir, ``Achieving $\frac{1}{2} \log (1+\mathrm{SNR})$
    on the AWGN Channel with Lattice Encoding and Decoding,
    {\em IEEE Trans. on Information Theory}, Vol.~50, pp.~2293---2314,
    Oct.~2004.
\bibitem{Liu06} T. Liu, P. Moulin and R. Koetter,
    ``On Error Exponents of Modulo Lattice Additive Noise Channels,''
    {\em IEEE Trans. on Information Theory}, Vol.~52,  No.~2, pp.~454-471,
    Feb.~2006.
\bibitem{Cov02} T. M. Cover and M. Chiang,
        ``Duality Between Channel Capacity and Rate Distortion with
    Side Information,''
        {\em IEEE Trans. Information Theory}, Vol.~48, No.~6, pp.~1629---1638,
        June 2002.
\bibitem{Bar03}R. J. Barron, B. Chen, G. W. Wornell,
``The duality between information embedding and source coding
with side information and some applications",
\emph{IEEE Trans. on Information Theory},
Vol.~49, No.~5, pp. 1159-1180, May 2003.
\bibitem{Pra03}  S. S. Pradhan, J. Chou and Ramchandran,
        ``Duality Between Source Coding and Channel Coding
    and its Extension to the Side Information Case,''
    {\em IEEE Trans. on Information Theory},
    Vol.~49, No.~5, pp. 1181-1203, May 2003.
\bibitem{Wil00} F. M. J. Willems, ``An Information Theoretical Approach
        to Information Embedding,''
        {\em Proc. 21st Symp. Information Theory in the Benelux}, pp.~255---260,
        Wassenaar, The Netherlands, May 2000.
\bibitem{Che01} B. Chen and G. W. Wornell,
        ``Quantization Index Modulation Methods: A Class of Provably
        Good Methods for Digital Watermarking and Information Embedding,''
        {\em IEEE Trans. Information Theory}, Vol.~47, No.~4, pp.~1423---1443, May 2001.
\bibitem{Mou03}P. Moulin and J. A. O'Sullivan,
    ``Information-theoretic analysis of information hiding",
    \emph{IEEE Trans. on Information Theory},
    Vol. 49, No.3, pp. 563--593, March 2003.
\bibitem{Coh02} A. S. Cohen and A. Lapidoth,
    ``The Gaussian Watermarking Game,''
        {\em IEEE Trans. Information Theory}, Vol.~48, No.~6, pp.~1639---1667,
        June 2002.
\bibitem{Som03} A. Somekh-Baruch and N. Merhav,
        ``On the Error Exponent and Capacity Games of Private
    Watermarking Systems,''
    \emph{IEEE Trans. on Information Theory},
    Vol. 49, No.3, pp. 537--562, March 2003.
\bibitem{Som04} A. Somekh-Baruch and N. Merhav,
        ``On the Capacity Game of Public Watermarking Systems,''
    \emph{IEEE Trans. on Information Theory},
    Vol. 50, No.3, pp. 511--524, Mar. 2004.
\bibitem{Csi81} I. Csisz\'{a}r and J. K\"{o}rner, {\em Information Theory:
        Coding Theory for Discrete Memoryless Systems},
        Academic Press, NY, 1981.
\bibitem{Csi88} I. Csisz\'{a}r and P. Narayan, ``Arbitrarily
    Varying Channels with Constrained Inputs and States,''
    {\em IEEE Trans. Information Theory},
       Vol.~31, No.~1, pp.~42---48, Jan.~1988.
\bibitem{Lap98} A. Lapidoth and P. Narayan, ``Reliable Communication
        Under Channel Uncertainty,'' {\em IEEE Trans. Information Theory},
        Vol.~44, No.~6, pp.~2148---2177, Oct.~1998.
\bibitem{Eri85} T. Ericson,
    ``Exponential Error Bounds for Random Codes
    in the Arbitrarily Varying Channel,''
    \emph{IEEE Trans. on Information Theory},
    Vol. 31, No.1, pp. 42--48, Jan. 1985.
\bibitem{Hug96} B. L. Hughes and T. G. Thomas,
    ``On Error Exponents for Arbitrarily Varying Channels,''
    \emph{IEEE Trans. on Information Theory},
    Vol. 42, No.1, pp. 87--98, Jan. 1996.
\bibitem{Ahl86} R. Ahlswede, ``Arbitrarily Varying Channels
    with States Sequence Known to the Sender,''
    \emph{IEEE Trans. on Information Theory},
    Vol. 32, No.5, pp. 621--629, Sep. 1986.
\bibitem{Har88} E. A. Haroutunian and M. E. Haroutunian,
``E-Capacity Upper Bound for a Channel with Random Parameter,''
{\em Problems of Control and Information Theory},
Vol.~17, No.~2, pp.~99--105, 1988.
\bibitem{Har91} M. E. Haroutunian, ``Bounds on E-Capacity of a Channel
with a Random Parameter,'' {\em Probl. Info. Transmission},
Vol.~27, No.~1, pp.~14--23, 1991.
\bibitem{Har00} M. E. Haroutunian, ``Letter to the Editor,''
{\em Probl. Info. Transmission}, Vol.~36, No.~4, p.~140, 2000.
\bibitem{Har01} M. E. Haroutunian, ``New Bounds on E-Capacity of
Arbitrarily Varying Channel and Channels with Random Parameters,''
{\em Trans. IIAP NAS RA and YSU} (Inst. for Informatics and Automation
Problems of the Nat. Acad. Sci. Rep. Armenia and Yerevan State U.),
{\em Mathematical Problems of Computer Science}, Vol.~22, pp.~44--59, 2001.
Available from {\tt http://ipia.sci.am/itas/Mariam/mariam.htm}.
\bibitem{Har04} M. E. Haroutunian and S. A. Tonoyan,
``Random Coding Bound of Information Hiding E-Capacity,'' {\em
Proc. IEEE Int. Symp. Info. Theory}, p.~536, Chicago, IL,
June-July 2004.
\bibitem{Som04b} A. Somekh-Baruch and N. Merhav,
``On the Random Coding Error Exponents of the Single-User and the
Multiple-Access Gel'fand-Pinsker Channels,'' {\em Proc. IEEE Int.
Symp. Info. Theory}, p.~448, Chicago, IL, June-July 2004.
\bibitem{Mou04} P. Moulin and Y. Wang,
``Error Exponents for Channel Coding With Side Information,''
presented at the Recent Results session,
{\em IEEE Int. Symp. Info. Theory}, Chicago, IL, June 2004.
\bibitem{Csi98} I. Csisz\'{a}r, ``The Method of Types,''
    {\em IEEE Trans. Information Theory},
       Vol.~44, No.~6, pp.~2505---2523, Oct.~1998.
%
\bibitem{Mou04b} P.~Moulin and Y.~Wang,
    ``New Results on Steganographic Capacity,''
    {\em Proc. CISS'04}, Princeton, NJ, March 2004.
\bibitem{Michalewicz} Z. Michalewicz, \emph{Genetic Algorithms + Data
Structures = Evolution Programs}. New York: Springer, 1996.
\bibitem{Rudolph} G. Rudolph, \emph{Convergence Properties of Evolutionary
Algorithms}. Verlag Dr. Kovac: Hamburg, Germany, 1997.
\end{thebibliography}
\normalsize
\end{document}